\documentclass[structabstract]{aa}
\usepackage{txfonts}
\usepackage{graphicx}
\usepackage{natbib}
\bibpunct{(}{)}{;}{a}{}{,} % to follow the A&A style
\def\her{\mbox{\sc Heros}}
\def\fer{\mbox{\sc Feros}}

\def\Rstar{\mbox{${\rm R}_{\star}$}} 
\def\Rsun{\mbox{${\rm R}_{\odot}$}}

\def\HeI{\ion{He}{i}}
\def\BrG{Br\,$\gamma$}
\def\zetaTau{$\zeta$~Tau}

\begin{document}
\title{Cyclic variability of the circumstellar disk of the Be star $\zeta$~Tau}
\subtitle{I. Long-term monitoring observations
    \thanks{Based partly on observations collected at the
    European Southern Observatory, Chile (Prop.\ Nos. 073.D-0234,
    074.D-0240, 078.D-0542, and 081.D-2005; as well as archival data from programs
    074.D-0573 and 076.B-0055)}}
   \author{S.~\v{S}tefl \inst{1}
             \and 
             Th.~Rivinius \inst{1}
             \and  
           A.~C.~Carciofi \inst{2}
             \and 
            J.-B.~Le~Bouquin \inst{1}
             \and
            D.~Baade \inst{3}
              \and  
            K.S.~Bjorkman \inst{4} $^{\star \star}$
               \and 
            E.~Hesselbach \inst{4} $^{\star \star}$
               \and
            C.~A.~Hummel \inst{3}
               \and  
            A.~T.~Okazaki \inst{5} 
              \and 
            E.~Pollmann \inst{6}
               \and
            F.~Rantakyr\"o \inst{7} 
               \and 
            J.P.~Wisniewski \inst{8}
   \thanks{
   %IRTF_affiliation_note:- J Wisniewski, K Bjorkman, E Hesselbach should have the added affiliation tag of   $^a$ 
    Visiting Astronomer at the Infrared Telescope Facility, which is operated by the 
    University of Hawaii under cooperative agreement NNX08AE38A with the National Aeronautics and Space 
    Administration, Science Mission Directorate, Planetary Astronomy Program.
    }
}
\offprints{S.\ \v{S}tefl, \email{sstefl@eso.org}}
\institute{European Organisation for Astronomical Research in the Southern
           Hemisphere, Casilla 19001, Santiago 19, Chile 
           \and
           Instituto de Astronomia, Geof{\'\i}sica e Ci{\^e}ncias
           Atmosf{\'e}ricas, Universidade de S\~ao Paulo,
           Rua do Mat\~ao 1226, Cidade Universit\'aria, S\~ao Paulo, SP 05508-900, Brazil 
           \and
           European Organisation for Astronomical Research in the
           Southern Hemisphere, Karl-Schwarzschild-Str.~2,
           85748 Garching bei M\"unchen, Germany 
           \and
           University of Toledo, Department of Physics \& Astronomy,
           MS111 2801 W. Bancroft Street Toledo, OH 43606, USA 
           \and
           Faculty of Engineering, Hokkai-Gakuen University,
           Toyohira-ku, Sapporo 062-8605, Japan 
           \and
           Emil-Nolde-Str.12, 51375 Leverkusen, Germany
           \and
           Gemini Observatory,  Southern Operations Center, c/o AURA, Casilla 603, La Serena, Chile
           \and
           NSF Astronomy \& Astrophysics Postdoctoral Fellow,
           Department of Astronomy, University of Washington, Box
           351580, Seattle, WA 98195, USA 
}
\date{Received: $<$date$>$; accepted: $<$date$>$; \LaTeX ed: \today}  
\authorrunning{\v{S}tefl et al.}
\titlerunning{Cyclic disk variations of the Be star $\zeta$~Tau}
\abstract
  %Context, Optional
{Emission lines formed in decretion disks of Be stars often undergo 
long-term cyclic variations, especially in the violet-to-red ($V/R$) ratio of 
their primary components.  The underlying structural and dynamical 
variations of the disks are only partly understood.  From observations of 
the bright Be-shell 
star $\zeta$ Tau, the possibly broadest and longest data set illustrating 
the prototype of this behaviour was compiled from our own and archival observations.   
It comprises optical and infrared spectra, broad-band polarimetry, and
interferometric observations.}
 % aims heading (mandatory)
{The dense, long-time monitoring permits a better separation of repetitive 
and ephemeral variations.  The broad wavelength coverage includes lines 
formed under different physical conditions, i.e.\ different locations in 
the disk, so that the dynamics can be probed throughout much of the disk.  
Polarimetry and interferometry constrain the spatial structure.  All together, 
the objective is a better understand the dynamics and life cycle 
of decretion disks.}
  % methods heading (mandatory)
{Standard methods of data acquisition, reduction, and analysis were applied.}
  % results heading (mandatory){}
{From 3 $V/R$ cycles between 1997 and 2008, a mean cycle length in 
H$\alpha$ of 1400\,-\,1430\,days was derived.  After each minimum in $V/R$, 
the shell absorption weakens and splits into two components, leading 
to 3 emission peaks.  This phase may make the strongest contribution to the 
variability in cycle length.  There is no obvious connection between the 
$V/R$ cycle and the 133-day orbital period of the not otherwise 
detected companion.  $V/R$ curves of different lines are shifted in 
phase. Lines formed on average closer to the central star are ahead 
of the others.  The shell absorption lines fall into 2 categories 
differing in line width, ionization/excitation potential, and variability of 
the equivalent width.  They seem to form in separate regions 
of the disk, probably crossing the line of sight at different times.  
The interferometry has resolved the continuum and 
the line emission in Br$\gamma$ and HeI\,2.06.  The phasing of the 
Br$\gamma$ emission shows that the photocenter of the line-emitting 
region lies within the plane of the disk but is offset from the 
continuum source.  The plane of the disk is constant throughout the observed 
$V/R$ cycles.  The observations lay the foundation for the fully 
self-consistent, one-armed, disk-oscillation model developed in Paper II.  
}
{}
%}
\maketitle
%
%%%%%%%%%%%%%%%%%%%%%%%%%%%%%%%%%%%%%%%%%%%%%%%%%%%%%%%%%%%%%%%%%%%%%%%%

\section{Introduction}
\label{intro}

%\subsection{Cyclic long-term {\it V/R} variations}
Circumstellar disks of classical Be-stars have been known 
for decades as the
place of origin of their characteristic Balmer emission
lines.  Polarimetric \citep{1978A&A....69..291M} and combined
interferometric and spectro-polarimetric observations
\citep{1997ApJ...479..477Q, 1998A&A...335..261V} have brought 
the long discussion of the geometry of the disks to a definitive
conclusion. These observations confirm directly that the disks are
geometrically thin with low opening angles of about $5-15$ degrees.
The outer radius of the H$\alpha$ emission was estimated at typically
$10-20$ stellar radii.

The detailed dynamics of Be-star disks, however, is not as clearly
known. Various hypotheses exist, but the observations have not so far 
yielded fully unambiguous results. The relations between (a) stellar 
$v$ sin $i$ \citep{1989Ap&SS.161...61H} and (b) disk size 
\citep{1997ApJ...479..477Q}, and separation of the emission peaks, 
details of circumstellar line shapes \citep{2000A&A...359.1075H}, 
and the occurrence of central quasi-emission peaks in shell stars
\citep{1999A&A...348..831R} are all consistent with  Kepler-like rotation, 
in which the rotation velocity varies as $r^{-j}$, where $r$ is the distance 
from the star. For a strictly Keplerian disk with circular orbits the exponent 
$j$ is  equal to 0.5.
Analyses of spectral line profiles \citep{2000A&A...359.1075H} suggests that 
$j~<~0.65$.

\citet{1994Ap&SS.216...99H,2000ASPC..214..518H} and
\citet{1994IAUS..162..399W} showed that any radially outward directed 
velocities in the disk must be lower than a few km/s.  However, the angular
momentum in a Keplerian disk increases with $r^{1/2}$, Be disks
are formed from matter outflowing from the star, and at least part of the 
disk eventually escapes from the star's gravity.  Therefore, 
there must be a net 
angular momentum transfer from the star to the disk and then outwards through the 
disk.  The mechanism for this momentum transfer, albeit not yet conclusively  
identified, is most likely related to dynamical viscosity.  A  
viscous decretion disk model \citep{1991MNRAS.250..432L,1996MNRAS.280L..31P, 
2001PASJ...53..119O} has recently been successfully applied to the 
circumstellar disk of the Be star $\delta$ Sco \citep{2006ApJ...652.1617C} and, 
at present, is the strongest candidate to describe the structure of  
Be-star disks.

Purely circular Keplerian disks are also unable to explain the complex
variability of the Balmer emission lines, notably the so-called
$V/R$ (the violet-to-red flux ratio of the emission peaks) variations.
Most observations of $V/R$ variations concern lower Balmer lines,  
but they can also be seen in virtually all other emission lines.  
Often, $V/R$ variations are cyclic with 
timescales of $5 - 10$ years \citep{1991PASJ...43...75O}.  The amplitudes 
can be large to spectacular.  Nevertheless, many $V/R$ cycles look 
strongly perturbed because of concomitant large variations in the 
emission-line strength relative to the continuum and, presumably, 
the unsteady character of the mass loss from the central star.  

The first  models for the $V/R$ variations, a precessing elliptical ring
\citep{1973ApJ...183..541H}, spheroidal/ellipsoidal variable-mass loss 
\citep{1987pbes.coll..384D}, and  a variable stellar wind causing expansions and 
contractions of the disk \citep{1991A&A...241..159M} were not physically 
self-consistent. They were finally made obsolete by Okazaki's model of 
one-armed density oscillations in the disk 
\citep{1991PASJ...43...75O,1997A&A...318..548O}.  
It is based on work  by \citet{1983PASJ...35..249K}, who showed
that the only possible global oscillation mode in an often 
thin Keplerian disk has $m=1$.  The idea of density waves
was developed further by \citet{1992A&A...265L..45P},
\citet{2000ASPC..214..409O}, and \citet{2006A&A...456.1097P}.
A few scattered interferometric observations detected decentered density 
enhancements in circumstellar disks of $\zeta$~Tau \citep{1998A&A...335..261V} 
and $\gamma$~Cas \citep{1999A&A...345..203B}. 

The wide acceptance of the model is based mainly on qualitative 
comparisons.  Its detailed application to specific sets of $V/R$ curves
and emission line profiles has been pending to date.  A strong reason 
is the poor repetitiveness of $V/R$ cycles, which makes it difficult to 
decide what the model can be expected to reproduce.  Furthermore, 
the typically long 
duration of the cycles has kept the statistics at a low level.  
Also, detailed comparisons between model and observations have only 
become feasible with the advent of three-dimensional radiative transfer 
codes  \citep{2006ApJ...639.1081C}.  The generic theoretical modeling 
combining the dynamics of $m$=1 disk oscillations with a radiative-transfer 
code was successfully pioneered by \citet{1997A&A...320..852H}. 

\citet{2007ASPC..361..274S} analyzed the $V/R$ variations in five Be-shell 
stars with a companion star.  They find that in some stars the $V/R$ 
variability is phase-locked to the orbital motions.  In others it is not.  
A peculiarity seemingly limited to binary Be-shell stars (but not exhibited by 
all of them) is the appearance of triple-peak H$\alpha$ emission profiles 
(which may also be described by a doubling of the self-absorption 
in the disk).  $\zeta$\,Tau is one of the examples presented by 
\citet{2007ASPC..361..274S}.  The occurrence of such profiles seems 
restricted to a narrow phase interval at the beginning of the transition 
from $V \ll R$ to $V \sim R$. 
They are not reproduced by present versions of the disk-oscillation model.

%%%%%%%%%%%%%%%%%%%%%%%%%%%%%%%%%%%%%%%%%%%%%%%%%%%%%%%%%%%%%%%%%
\begin{table*}[t]
\begin{minipage}{175mm}
\begin{center}
\caption[]{\label{datasets} Spectroscopic datasets in the visual range}
\begin{tabular}{cc@{\,--\,}cc@{\,--\,}cllrrrr}
\hline\noalign{\smallskip}
\hline
\rule{0ex}{2.5ex} Data        & \multicolumn{2}{c}{Observing} & \multicolumn{2}{c}{JD}         &                                     &                                    & 
 \multicolumn{1}{c}{No.\ of} & \multicolumn{1}{c}{Resolving} & \multicolumn{1}{c}{Spectral}   &                              \\
 
set                           & \multicolumn{2}{c}{season}    & \multicolumn{2}{c}{2400000+}   & \raisebox{1.5ex}[1.5ex]{Telescope}  & \raisebox{1.5ex}[1.5ex]{Instrument} & 
\multicolumn{1}{c}{spectra}   &\multicolumn{1}{c}{power}      & \multicolumn{1}{c}{range [\AA]}& \raisebox{1.5ex}[1.5ex]{Ref.} \\
\hline
A & \multicolumn{2}{c}{1991}    & \multicolumn{2}{c}{48\,347} & Heidelberg 0.9m  & FLASH          &   1     & 20\,000 & 4050-6780 & o \\  
B & 1992      & 2007          & 49\,049    & 51\,486        & Ond\v{r}ejov 2m  & slit spectr.   &  43     &  8\,500 & 6300-6700 & 1 \\ 
C & \multicolumn{2}{c}{1994\,\&\,2000}  &  \multicolumn{2}{c}{49\,592\,\&\,51\,572}  & OHP 1.93m          & {\sc Elodie}     &   3  & 45\,000 & 3850-6800 & 2 \\
D & 2000 & 2003 & 51\,045 & 52\,726 & Ond\v{r}ejov 2m     & \her             &  23  & 20\,000 & 3700-8600 & 1 \\
E & 2005 & 2008 & 53\,399 & 54\,714 & ESO/MPI 2.2m       & \fer             &  25  & 48\,000 & 3700-9000 & 1,o \\
F & 2000 & 2008 & 51\,850 & 54\,174 & Schmidt-Cassegrain & Slitless-Grating & 129  & 14\,000 & 6500-6700 & 3 \\
G & 2006 & 2007 & 54\,005 & 54\,515 & SCT 0.3m           & LHIRES           &  12  & 13\,425 & 6520-6700 & 4 \\ 
H &  \multicolumn{2}{c}{2006}  &  \multicolumn{2}{c}{54\,387\,\&\,54\,083}    & Celestron 11       & LHIRES           &   3  & 17\,000 & 6520-6620 & 5 \\ 
I &  \multicolumn{2}{c}{2006}  &  \multicolumn{2}{c}{54\,169\,\&\,54\,200} & Takahashi TSC225   & LHIRES           &   7  & 17\,000 & 3000-7000 & 6 \\ 
J & 2007 & 2008 & 54\,162 & 54\,550 & Schmidt-Cassegrain & Slitless-Grating &  32  & 14\,000 & 6500-6700 & 7 \\ 
K & \multicolumn{2}{c}{2007\,\&\,2008}  & \multicolumn{2}{c}{54\,442\,\&\,54\,527}  & Newton             & LHIRES           &   2  &  7\,000 & 6490-6700 & 8 \\
L & \multicolumn{2}{c}{2008}  & \multicolumn{2}{c}{54\,532}           & Celestron C14      & LHIRES           &   2  & 17\,000 & 6520-6610 & 9 \\ 
M & 2007 & 2008 & 54\,406 & 54\,515 & Meade LX200        & LHIRES           &  10  & 17\,000 & 6520-6610 & 10 \\ 
N & \multicolumn{2}{c}{2008} & 54\,106 & 54\,359 & C11                & LHIRES           &   3  & 17\,000 & 4000-8000 & 11 \\ 
O & \multicolumn{2}{c}{2007} &         \multicolumn{2}{c}{54\,331}   & OHP ED120          & LHIRES           &   1  & 17\,000 & 6400-6800 & 12 \\ 
P & 2006 & 2008 & 54\,021 & 54\,714 & ESO/MPG 2.2m       & \fer             &  18  & 48\,000 & 3700-9000 & o \\
Q & 1999 & 2007 & 51\,489 & 54\,433 & Ritter 1m          & echelle          &  61  & 26\,000 & 4600-6700 & o \\ 
R & \multicolumn{2}{c}{2007}  &        \multicolumn{2}{c}{54\,384}    & OPD/LNA            & ECASS            &   1  & 16\,000 & 6500-6620 & o \\ 
\hline\noalign{\smallskip} \hline
\end{tabular}
\end{center}
\end{minipage}
\begin{minipage}{175mm}
\vspace*{.5cm}
References: 1 -- \citet{2006A&A...459..137R}, 
            2 -- BeSS,observer: C.~Neiner,
            3.-- E.~Pollmann, data used in \citet{2008IBVS.5813....1P}, private communication 
            4 -- BeSS observer: B.~Mauclaire,
            5 -- BeSS observer: C.~Buil,
            6 -- BeSS observer: E.~Barbotin,
            7 -- BeSS observer: E.~Pollmann,
            8 -- BeSS observer: J.~Guarro Fl\'{o},
            9 -- BeSS observer: J.~Ribeiro,
           10 -- BeSS observer: J.-N.~Terry,
           11 -- BeSS observer: O.~Thizy, 
           12 -- BeSS observer: V.~Desnoux, 
            o -- this paper
\end{minipage}
%\label{datasets}
\end{table*}
%%%%%%%%%%%%%%%%%%%%%%%%%%%%%%%%%%%%%%%%%%%%%%%%%%%%%%%%%%%%%%%%%
%\subsection{The Be-shell star $\zeta$\,Tau}

The star $\zeta$\,Tau (123 Tau, HR 1910, HD 37\,202; B2\,IV) is among the most
suitable targets for an in-depth observational and theoretical test of the 
disk-oscillation model. The present emission state, which started in 1990
\citep[][their Fig. 2]{1995A&AS..112..201G}, 
shows very stable $V/R$ variations with a cycle length of about 1500 days
\citep{2006A&A...459..137R}.  Compared to other $V/R$-variable Be stars, 
secular and ephemeral variations have recently (see below) been small 
enough that observations obtained in different cycles can be fairly safely 
assembled into one picture.  

For decades, $\zeta$~Tau has been known as a spectroscopic
binary.  A comprehensive set of radial velocity measurements was 
compiled and analyzed by \citet{1984BAICz..35..164H}, who derived 
a period of 132.9735\,d.

Being a bright object reachable from both northern and southern
latitudes, $\zeta$~Tau is one of the most observed Be stars. 
At the distance of 417 light years (corresponding to the
Hipparcos parallax of 7.82\,mas) not only the circumstellar
disk can be resolved with present-day interferometers but also
spectro-interferometry extends this resolution to the disk dynamics.
The limb-darkened photospheric diameter is estimated at 0.4\,mas
\citep[see e.g.][]{2004AJ....127.1194T, 2007ApJ...654..527G}.  At the
Hipparcos distance, this corresponds to a radius of 5.5 to 6\,\Rsun .

The major axis of the H$\alpha$ emitting disk was measured by
\citet{1997ApJ...479..477Q} to 4.53\,mas, corresponding to a radius of
11.3\,\Rstar  \  and by \citeauthor{2004AJ....127.1194T} to 3.14\,mas 
(almost 8\,\Rstar).  \citet{2007ApJ...654..527G} derived the major axis 
of 1.99\,mas (about 5\,\Rstar) for the K-band continuum  emitting
disk.  These values were derived from the full width at half maximum (FWHM) 
of a Gaussian fit.  The differences  can be reconciled by assuming that 
the disk was less fully developed during the later observations as is also 
suggested by a lower H$\alpha$ emission strength (see below). In the K-band 
continuum, \citeauthor{2007ApJ...654..527G} measured the 
semi-major axis of the disk to 1.79\,mas or 
4.5\,\Rstar (Gaussian FWHM).  All these values are lower than the
estimated Roche-lobe radius for the primary of the system, which
\citeauthor{2004AJ....127.1194T} compute as 5.3\,mas
(26.5\,\Rstar) at a binary separation of 9.2\,mas (46\,\Rstar).

%\subsection{Structure of this work}
This first paper of the series summarizes the extended observational
material (see Sect.~\ref{data}) and focuses on a detailed
phenomenological description of the variability corresponding to the
$V/R$ cycle.  The variations in the visible part of the spectrum are
analyzed in Sect.~\ref{VtoR}.  The variability in emission lines in
the $JHK$-bands is described in Sect.~\ref{resIRspect}.
Sect.~\ref{resinterfer} is devoted to the VLTI/AMBER
spectro-interferometry and contemporaneous optical and IR
spectroscopy, and Sect.~\ref{respolar} deals with new and previously
published polarimetric observations.  The last two sections discuss
and summarize the observational results.

The second paper (Carciofi et al., this volume) presents a detailed 
self-consistent physical model of the observations.  It combines 
the 2D global  oscillation model of \citet{1997A&A...318..548O} 
and the 3D radiative-transfer code HDUST of 
\citep{2006ApJ...639.1081C,2008ApJ...684.1374C}. 

\section{Observations}
\label{data}

\subsection{Visual Spectroscopy} 
\label{dataspectr}

In order to maximize the coverage of the $V/R$ H$\alpha$ cycles, new 
observations were combined with data from the literature, the ESO 
Science Archive, and from the BeSS database (see below).  
Table~\ref{datasets} summarizes the spectroscopic data sets used for this
study.  The description of the \her \ and \fer \ spectrographs can be
found in \citet{1998RvMA..11...177K} and \citet{1999Msngr..95...8K},
respectively.  The {\sc Flash} instrument is an earlier version of the \her\
spectrograph with only one spectral arm (covering the range from 
4050 to 6700\,\AA ) and a somewhat lower resolution due to the use 
of a larger fiber, which determines the effective slit width.

The OPD/LNA observation was made with a Cassegrain spectrograph,  
equipped with a 1200 grooves/mm grating blazed at 6562\,$\AA$ and a  
1024x1024 pixel CCD.  Observations at the Ritter observatory were done with 
an echelle spectrograph in the Cassegrain focus of the 1-m 
telescope, using the old camera equipped with a
1200x800 pixel CCD (cf.\ Sect.\,\ref{datapolar}). 

In recent years, the technological progress has brought bright stars
like $\zeta$\,Tau within reach of quite a few amateur telescopes and
spectrographs.  Spectra of many Be stars are regularly deposited in,
and conveniently available from, BeSS (http://basebe.obspm.fr).  As
Table\,\ref{datasets} shows, they account for a large fraction
of the total optical spectroscopy used in this paper.
  
\subsection{Infrared Spectroscopy}
\label{dataIRspectr} 
Fourteen infrared (IR) spectra were obtained between March
5, 2004 and October 9, 2007 using the SpeX spectrograph at the 3.0-m
NASA Infrared Telescope Facility (IRTF) on Mauna Kea
\citep{2003PASP..115..362R}.  With a 0$\farcs$3 x 15$\farcs$0
slit in the short-wavelength cross-dispersed (SXD) observing mode, a
resolving power of R$\sim$2000 spectroscopy was achieved from 0.8 to
2.4 $\mu$m.  The observing and data reduction techniques mirrored
those used in previous SpeX programs to study classical Be stars (see,
e.g., \citealt{2005PhDT........21W},
\citealt{2007ApJ...656L..21W}).  Observations of $\zeta$~Tau were
immediately followed by observations of a nearby A0V star located at a
similar air mass, to facilitate optimal telluric correction
\citep{2003PASP..115..389V}, along with a series of
quartz-tungsten-halogen flat field and argon arc lamp exposures.  The
IDL-based Spextool software was used to perform the basic data
reduction and spectral extraction using the techniques described in
\citet{2004PASP..116..362C}.

%The temporal distribution of the data is well seen in
%Fig.\,\ref{IR_VtoR}, Julian dates are written as labels in Fig.\
%\ref{IR_lp}.
%%%%%%%%%%%%%%%%%%%%%%%%%%%%%%%%%%%%%%%%%%%%%%%%%%%%%%%%%%%%%%%%%
  \begin{figure*}[t]
   \centering
   \includegraphics[viewport=50 73 553 790,angle=-90,width=18.8cm]{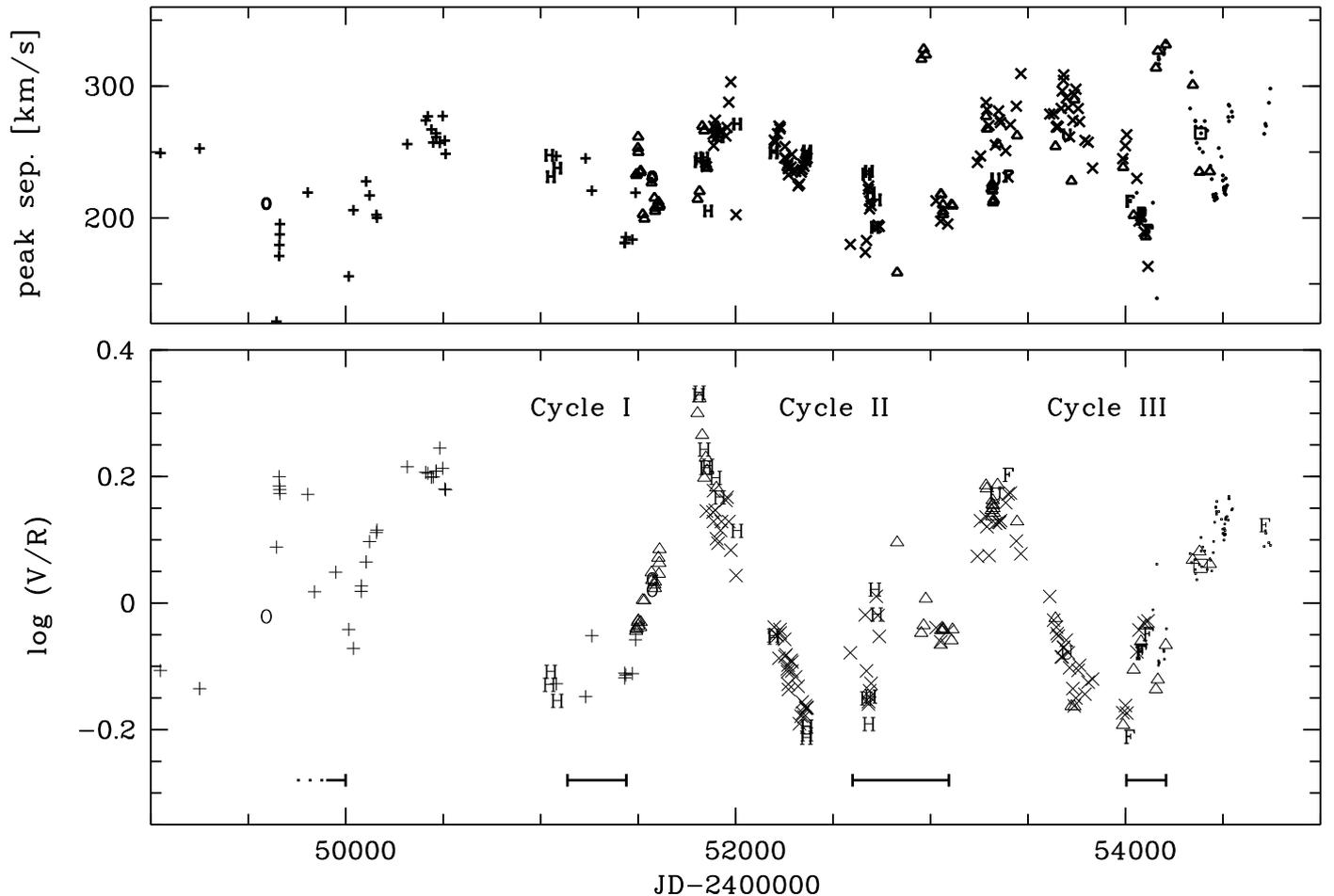}
   \caption{H$\alpha$ $V/R$ variations (lower panel) and emission peak separation 
     (upper panel)of $\zeta$~Tau in the period
     1992-2008 showing the detailed character of more than 3 $V/R$
     cycles since 1996. The symbols correspond to the following data
     sets: (+) Ond\v{r}ejov slit spectrograph, (O) OHP 1.93m, (x)
     Pollmann, (H) \her , (F) \fer , ($\bigtriangleup$) Ritter
     Observatory, ($\cdot$) BeSS database, ($\square$)
     A.~Carciofi, see also Table\,\ref{datasets}.  Phases with triple-peaked emission line profiles are 
    indicated by solid horizontal bars }
   \label{figvr}
    \end{figure*}
%%%%%%%%%%%%%%%%%%%%%%%%%%%%%%%%%%%%%%%%%%%%%%%%%%%%%%%%%%%%%%%%%%

\subsection{Interferometry}
\label{datainterfer}

Our near-IR interferometric observations with {\sc AMBER}
\citep{2007A&A...464....1P} and three 8-m telescopes (UT\,1, 2, and 4)
of ESO's VLTI interferometer were made during the night of December
12, 2006.  Interferometric fringes from all 3 pairs of telescopes
(hereafter called baselines) were measured across the K-band, at
medium spectral resolution of $R \approx 1500$.  Because fringe
tracking was not supported at the time, it was not possible to
properly record the full wavelength range.  The observations were,
therefore, taken around the
Br$\gamma$ (2.16\,$\mu$m) and \ion{He}{i} (2.06\,$\mu$m) lines.  The
\zetaTau{} observations were interlaced with observations of the
calibration stars HD\,39\,699 and HD\,59\,686, for which 
diameters of 1.03 mas and 1.30 mas, respectively 
\citep{2005A&A...433.1155M}, were adopted.  

In principle, three distinct quantities can be extracted for each  
spectral channel of the AMBER spectro-interferometer: the fringe  
visibility (measuring the spatial extension of the emitting material),  
the fringe phase (measuring the position of emitting material), and  
the closure-phase (the sum of the 3 phases obtained for the 3  
interferometric arms, sensitive to asymmetries in the source  
brightness distribution). In {\sc AMBER} data sets, the signal-to-noise ratio  
on the closure-phase is very low and only visibilities and phases were used
for the analysis.

The data reduction was done using the \texttt{amdlib-2.2} package, which 
employs the P2VM algorithm \citep{2007A&A...464...29T}.  In a first step, 
the  instrumental calibration matrix with the standard calibration files  
provided by ESO was computed. This is mandatory in order to properly convert the  
raw individual frames into raw interferometric visibilities and phases  
(about thousand measurements per observation). The next steps consist  
of averaging these different frames into individual measurements,  
and performing the final calibration. Based on our past experience of  
spectro-interferometry with AMBER, two different  strategies were adopted.  
In the following they are distinguished as \emph{absolute  reduction} and 
\emph{differential reduction}. Only a brief description is included here, 
as both methods are now considered to be standard data reduction steps for AMBER.

\paragraph{Absolute data reduction:} The objective of this method is  
to provide an absolutely calibrated estimation of the fringe  
visibility (the fringe phase cannot be determined in an absolute  
manner). It requires the multiplicative effect of the atmospheric  
turbulence (generally called transfer function) to be properly  
estimated.
\begin{itemize}
\item For each observation, the top 20\% of the frames  with the best 
signal-to-noise ratio were averaged. This ratio provides the most  
stable transfer function, see discussion in \citet{2007A&A...464...29T} 
and \citet{2007arXiv0705.1635M}.
\item The transfer function was estimated by averaging all observations  
of the calibration stars. Their scatter is a measure for the  
uncertainty of the transfer function. The resulting stability is very  
poor, with temporal fluctuations as large as 30\%, partially due to  
the atmospheric turbulence and to the Unit Telescope infrastructures,  
which generate non-stationary vibrations.
\item Absolutely calibrated quantities were derived by dividing --  
separately in all spectral channels -- the raw visibilities of  
\zetaTau{} by the derived transfer function. Error bars on the  
transfer function completely dominate the final uncertainties.
\item Finally, all spectral channels were averaged  (avoiding  
the \HeI{} and \BrG{} lines) to provide a single visibility across the  
band for each observation. Formally it introduces a small amount of  
wavelength smearing, which is however negligible at the level of  
precision.
\end{itemize}

It is worth noting that this observation strategy has been optimized for  
differential interferometry, not for absolute calibration of the  
fringe visibility. Only a single calibration star has been observed  
for each AMBER instrument setup, and the observations of the  
calibration stars do not bracket properly the observations of the  
science target. Therefore it is hard to asses the reliability of the  
absolute calibration (usually done by comparing values obtained using  
different calibration stars). Consequently, the fringe visibility  
derived with this method should be interpreted carefully.

\paragraph{Differential data reduction:} This method aims mainly at a  
precise measurement of the differential quantities (visibility and  
phase) in the \BrG{} and \HeI{} lines with respect to the adjacent  
continuum. Compared with the previous method, a significantly higher SNR  
can be obtained because the atmosphere affects all spectral channels  
in very nearly the same way. The main drawback is that the continuum  
level is obviously lost.
\begin{itemize}
\item First, as many as possible individual frames were averaged. To do so, all  
consecutive observations with the same instrument setup were merged  
together \emph{before} averaging the 70\% of the frames with the best  
signal-to-noise. It was tested that this method provides the best final accuracy on the  
differential quantities.
\item The here presented  {\sc AMBER} data were corrupted by  
ripples with low beat frequency in the spectral direction (few cycles  
along the complete K-band spectrum) \footnote{The interferences  
between the light beams reflected inside the AMBER polarizers created  
``Moir\'{e} fringes'' in the spectral direction. The optics responsible  
for these artifacts was identified and properly replaced in October  
2008.}). These artifacts mainly affect the phases and not the  
visibilities, explaining why they are not of a big concern regarding  
the absolute data reduction. To correct them, a high-pass filter  
along the spectral direction was applied to the phases.  
Different methods (high-pass filter, optimal filtering at the  
corrupted frequencies, manual fit of the continuum ripples) were tested
but  differences in the resulting \HeI{} and \BrG{} phase profiles could be 
seen.
\item At that stage, the continuum levels for both the phase and the  
visibility are arbitrary. The brightness distribution of the K-band  
continuum emitting region has already been constrained by  
\citet{2007ApJ...654..527G}. Because they used longer baselines and more precise  
absolute calibration, their results are significantly better than the {\sc AMBER}  
measurements for the continuum (see previous paragraph). Therefore  
the {\sc AMBER} continuum level around the \BrG{} and \HeI{}  lines was forced 
to to match the CHARA model of \citet{2007ApJ...654..527G}, which is displayed 
in Fig.~10 .
\end{itemize}

%%%%%%%%%%%%%%%%%%%%%%%%%%%%%%%%%%%%%%%%%%%%%%%%%%%%%%%%%%%%%%%%%
  \begin{figure}[t]
   \includegraphics[viewport=50 61 562 776,angle=270,width=9cm]{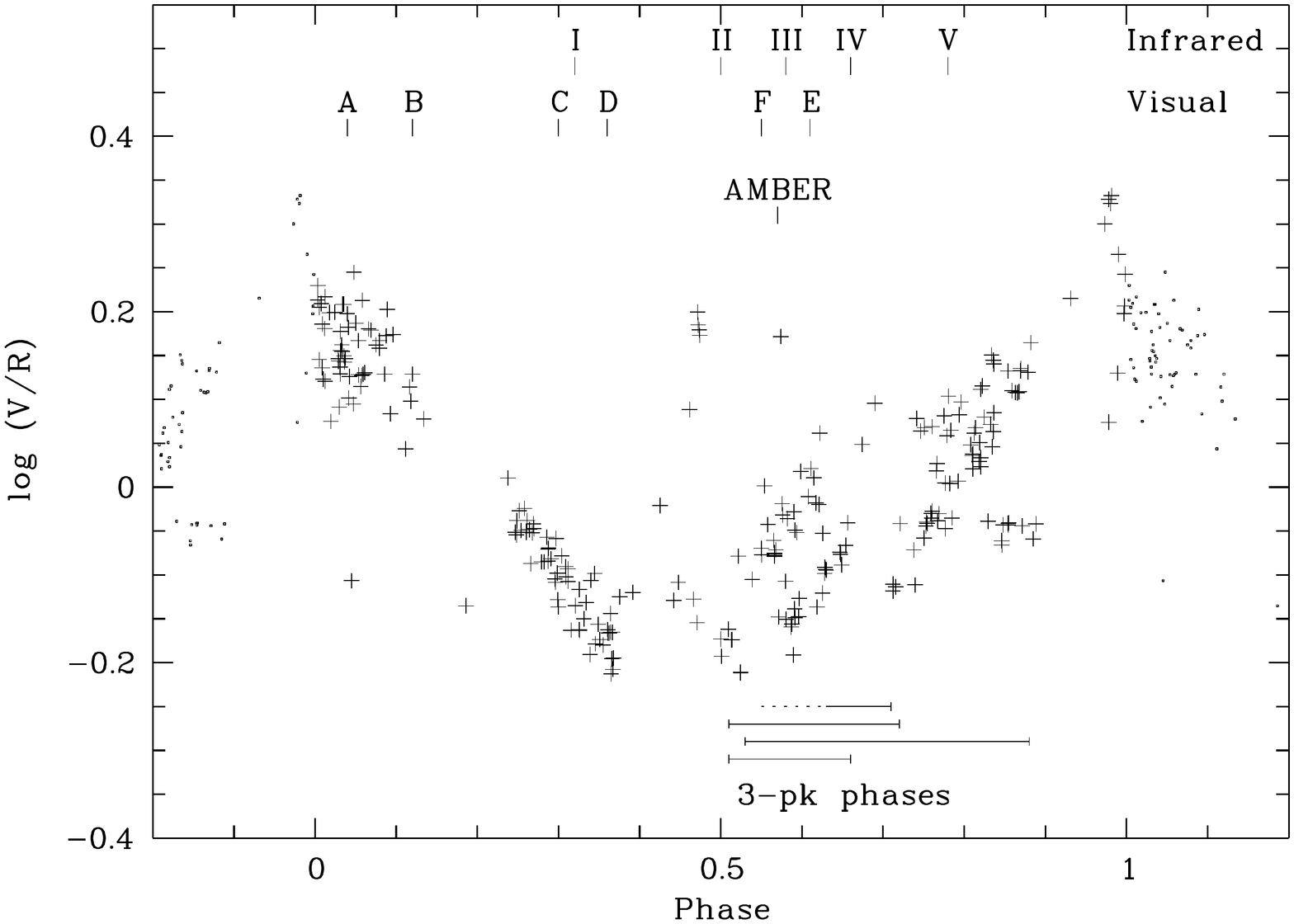}
   \caption{$V/R$ data from Fig.\ \ref{figvr}, phase binned with
     $JD\,50414 + 1429 \times E$.  Horizontal lines indicate the triple-peak phases
     (3-pk) from Table \ref{vtor_cycles}, which introduce a higher scatter in the 
     corresponding phase interval because they make the definition of the $V$ 
     and $R$ emission components doubtful.  The letters in the upper part of the figure 
     mark the phases of the {\sc AMBER} observation and of repesentative visual and 
     infrared spectra used in Sects.\ \ref{cycledesc} and \ref{resIRspect} (see also 
     Tables \ref{vtor-spectra} and \ref{vtor-IRspectra})}
    \label{fig_cycle}
   \end{figure}
%%%%%%%%%%%%%%%%%%%%%%%%%%%%%%%%%%%%%%%%%%%%%%%%%%%%%%%%%%%%%%%%%

\subsection{Polarimetry}
\label{datapolar}

The polarimetric data are from observations made with the
Halfwave Polarimeter   \citep[HPOL;][]{1996ASPC...97..100N}.  
The instrument was used primarily on the
0.9-m telescope at the Pine Bluff Observatory (PBO), operated by the
University of Wisconsin (UW).  The data are available through the
HPOL web site (http://www.sal.wisc.edu/HPOL) developed by M.R.\ Meade and B.L. Babler,
and data prior to 1995 are also available through the NASA MAST archive.
The database includes a wide range of hot stars, the polarimetric variability 
of which has been discussed in \citet{2005ASPC..343..406B}.

Data from 1989-1994 were obtained using a dual Reticon array detector,
which provided spectropolarimetry over a wavelength range of
3200-7600\,\AA, with a spectral resolution of 25\,\AA\
\citep{1996AJ....111..856W}.  In 1995, the HPOL detector was upgraded to a
$400\times 1200$ pixel CCD camera, and data since then have covered a
more extended wavelength range from 3200-10\,500\,\AA, with an improved
spectral resolution of 7\,\AA\ below 6000\,\AA\ and 10\,\AA\ above
this point \citep{1996ASPC...97..100N}.  For further details about
HPOL, see \citet{1990PhDT.........2N} and \citet{1996AJ....111..856W}.
As described by \citet{1996AJ....111..856W}, the data were processed
and analyzed using REDUCE, a specialized spectropolarimetric software
package developed at UW.  Instrumental polarization is monitored
regularly as part of the observing program at PBO, and all data are
fully calibrated for instrumental effects to an accuracy of 0.025\% in
the $V$ band.  

For the purposes of this study of $\zeta$\,Tau, the 
spectropolarimetric data were binned to approximate
broad-band $UBVRI$ results.   The
individual uncertainty of an HPOL measurement was estimated by a statistics
over a series of data of polarized standards: It is about 0.01\,\% for
the polarization degree and $1\,\deg$ for the polarization angle in $BVRI$, 
and about two to three times as high in the $U$-band.  

%%%%%%%%%%%%%%%%%%%%%%%%%%%%%%%%%%%%%%%%%%%%%%%%%%%%%%%%%%%%%%%%%
 \begin{figure*}[t]
\parbox{18.8cm}{%
\parbox{4.6cm}{\includegraphics[viewport=25 44 576 781,width=4.6cm,angle=0,clip]
{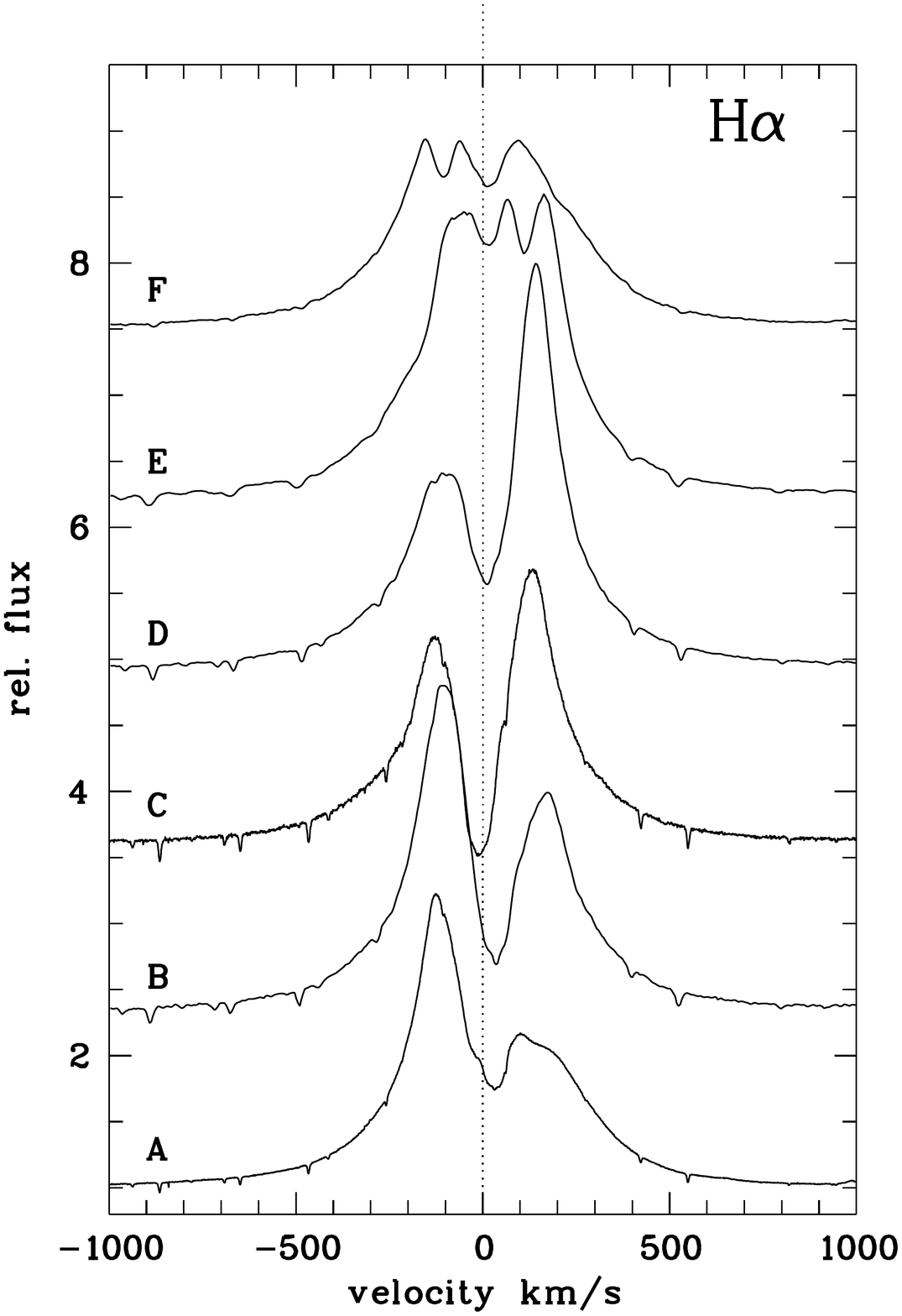}}%
\parbox{4.6cm}{\includegraphics[viewport=25 44 576 781,width=4.6cm,angle=0,clip]
{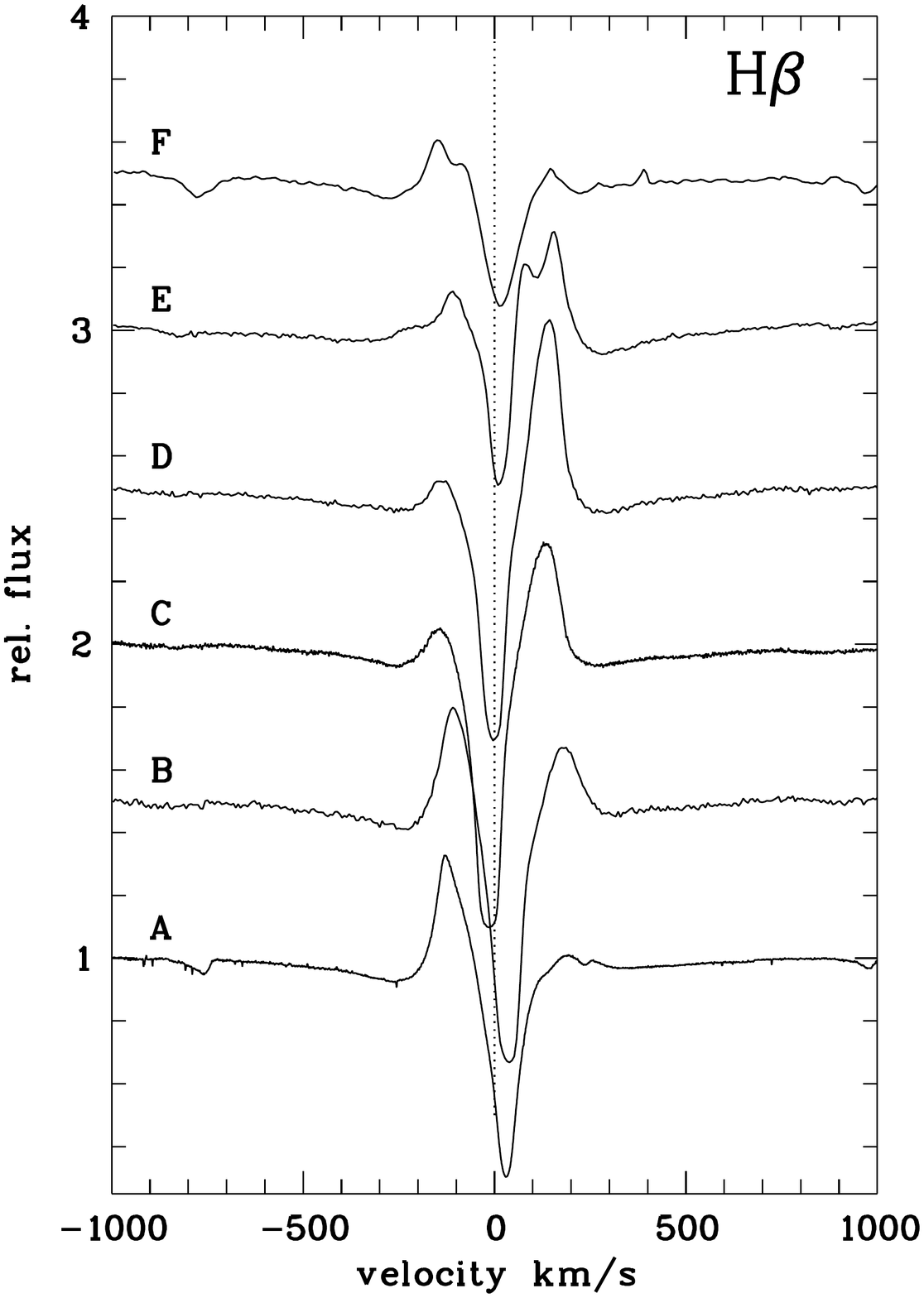}}%
\parbox{4.6cm}{\includegraphics[viewport=25 44 576 781,width=4.6cm,angle=0,clip]
{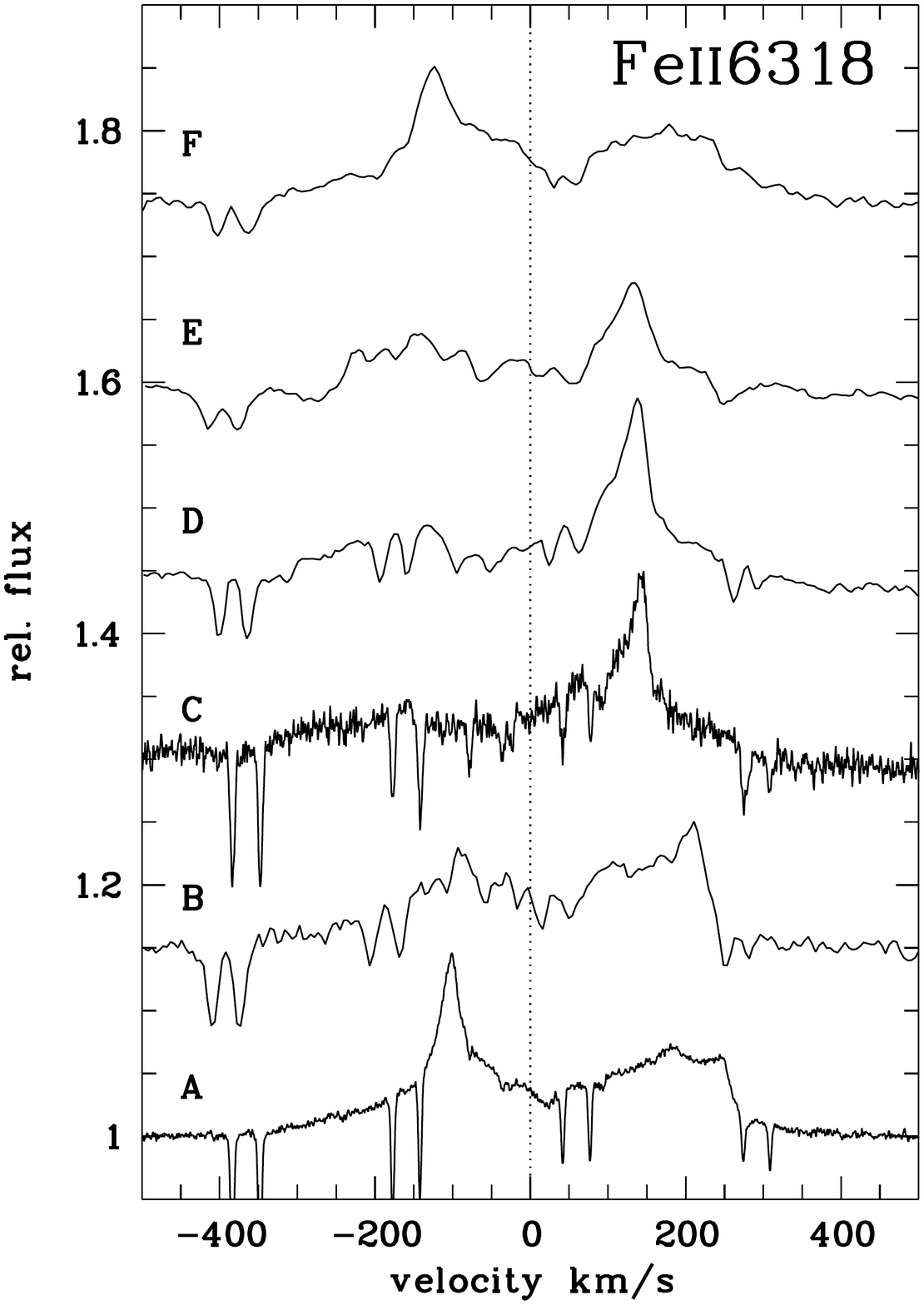}}%
\parbox{4.6cm}{\includegraphics[viewport=25 44 576 781,width=4.6cm,angle=0,clip]
{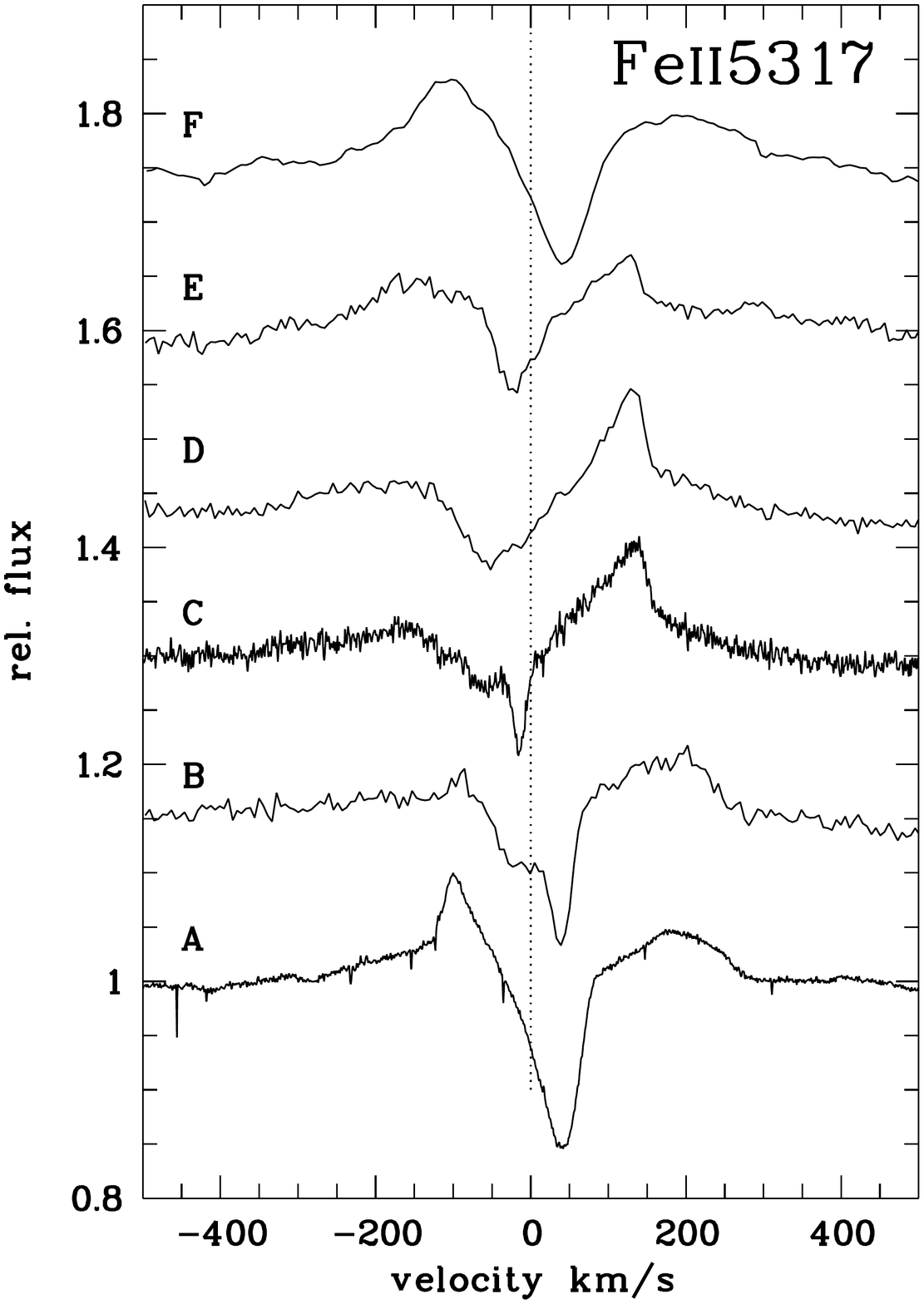}}%

\parbox{4.6cm}{\includegraphics[viewport=25 44 576 781,width=4.6cm,angle=0,clip]
{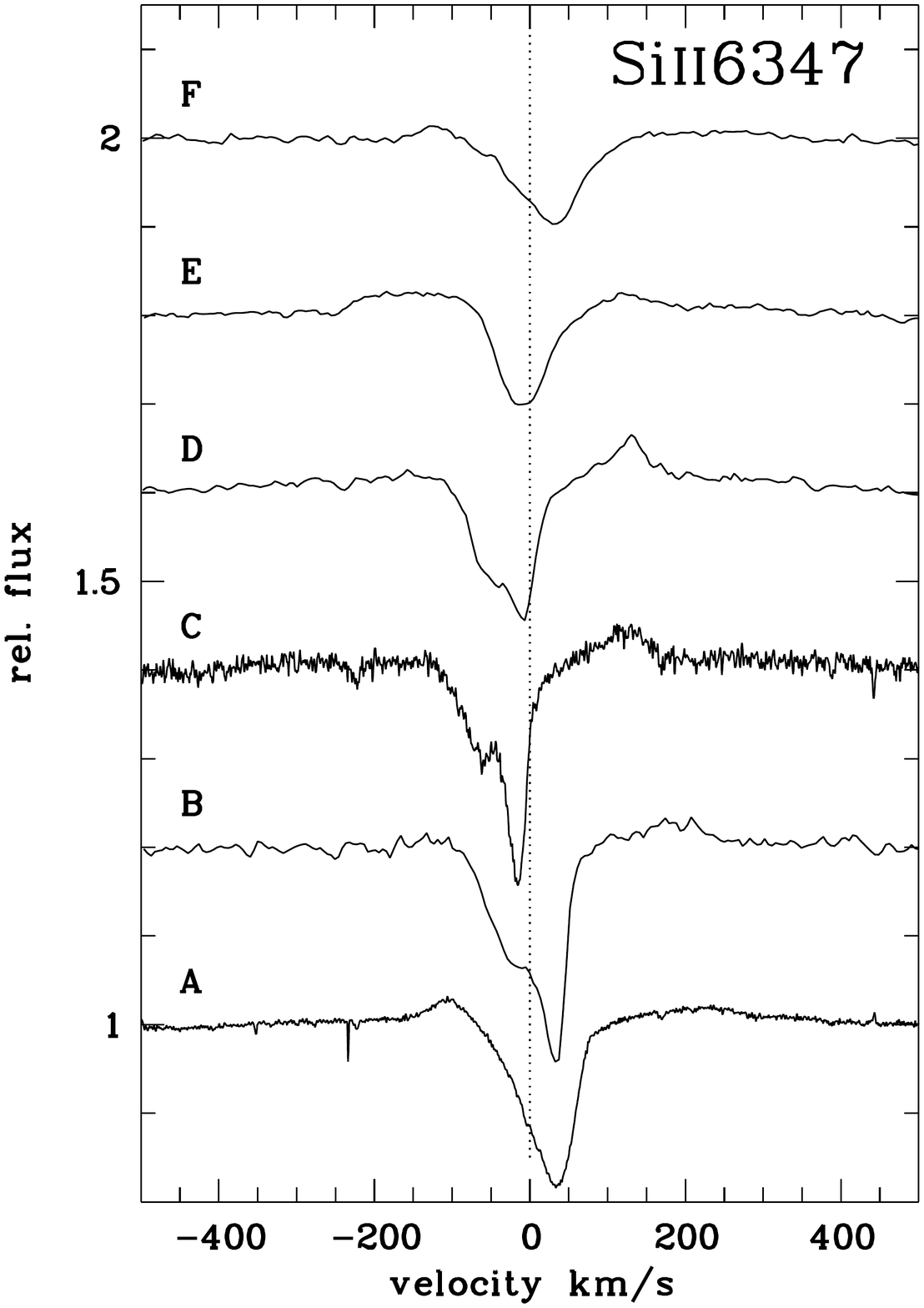}}%
\parbox{4.6cm}{\includegraphics[viewport=25 44 576 781,width=4.6cm,angle=0,clip]
{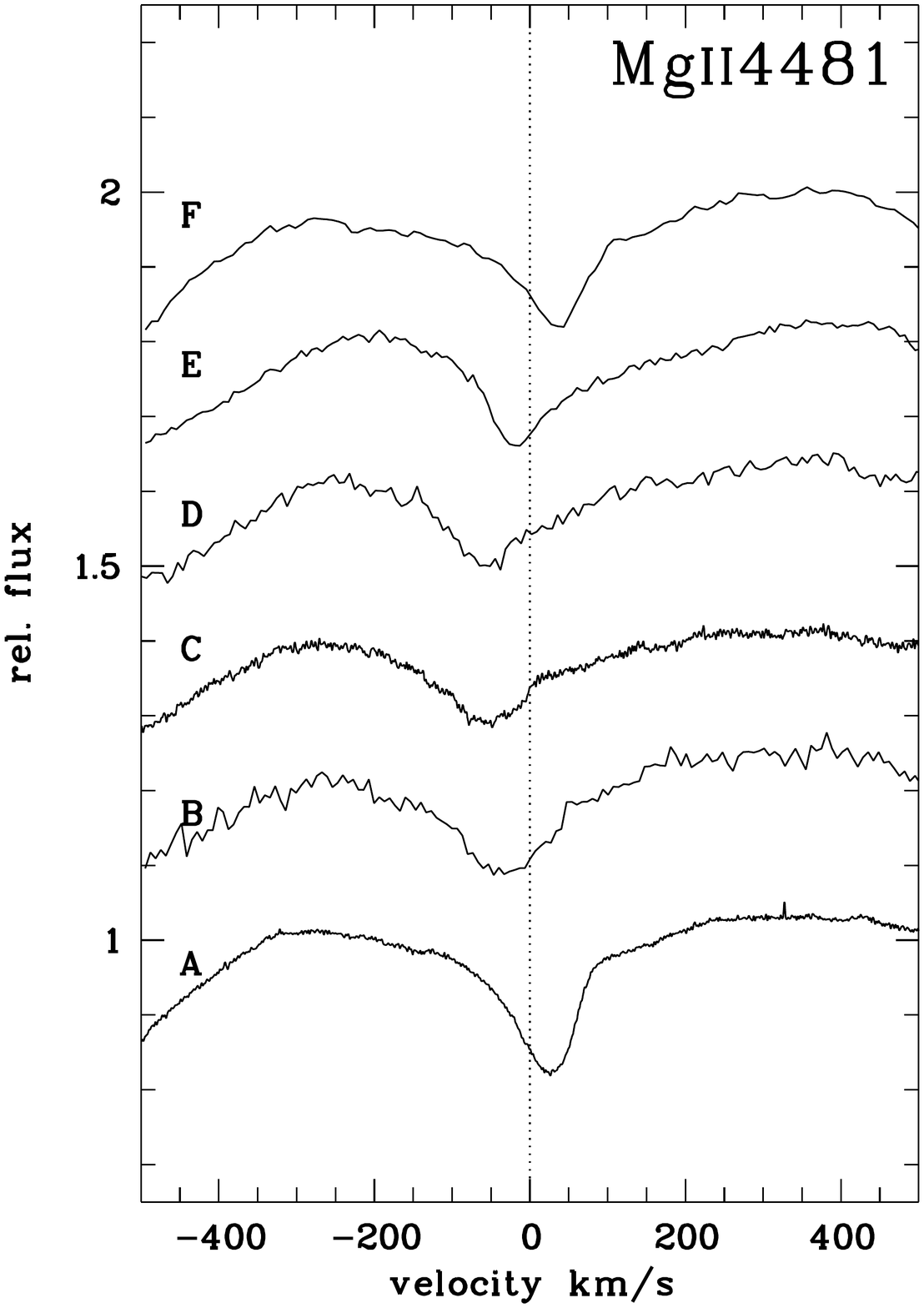}}%
\parbox{4.6cm}{\includegraphics[viewport=25 44 576 781,width=4.6cm,angle=0,clip]
{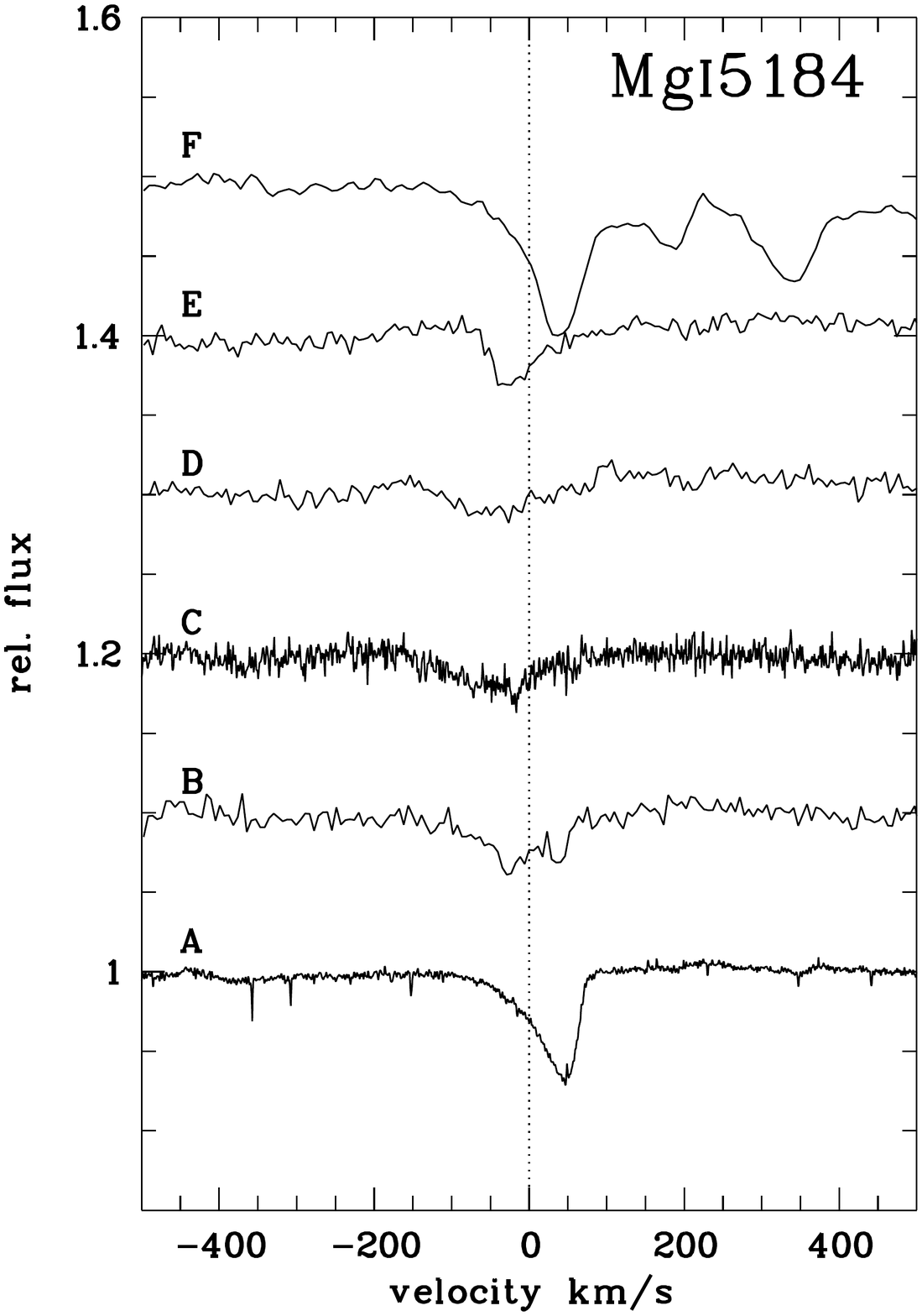}}%
\parbox{4.6cm}{\includegraphics[viewport=25 44 576 781,width=4.6cm,angle=0,clip]
{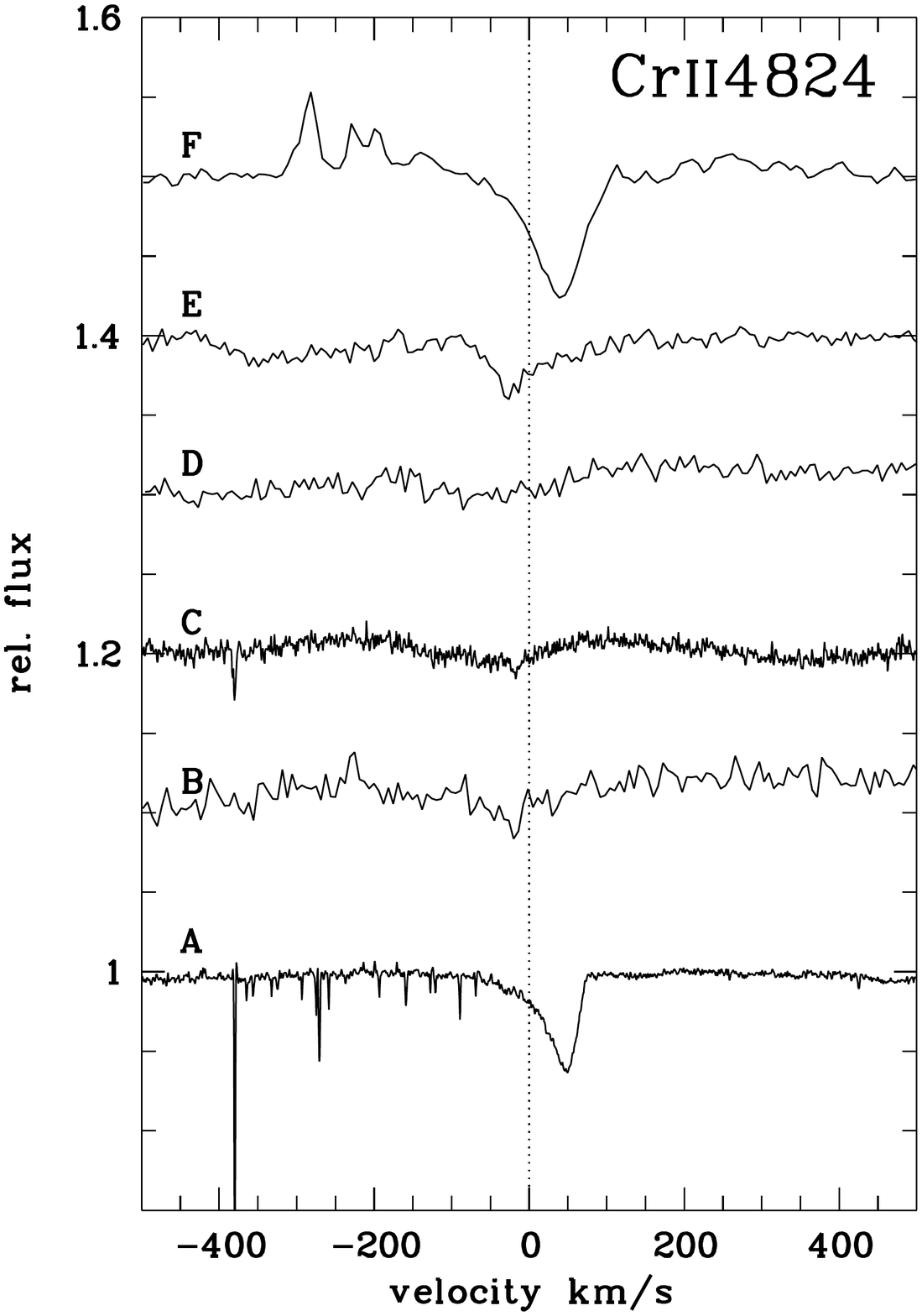}}%

\parbox{4.6cm}{\includegraphics[viewport=25 44 576 781,width=4.6cm,angle=0,clip]{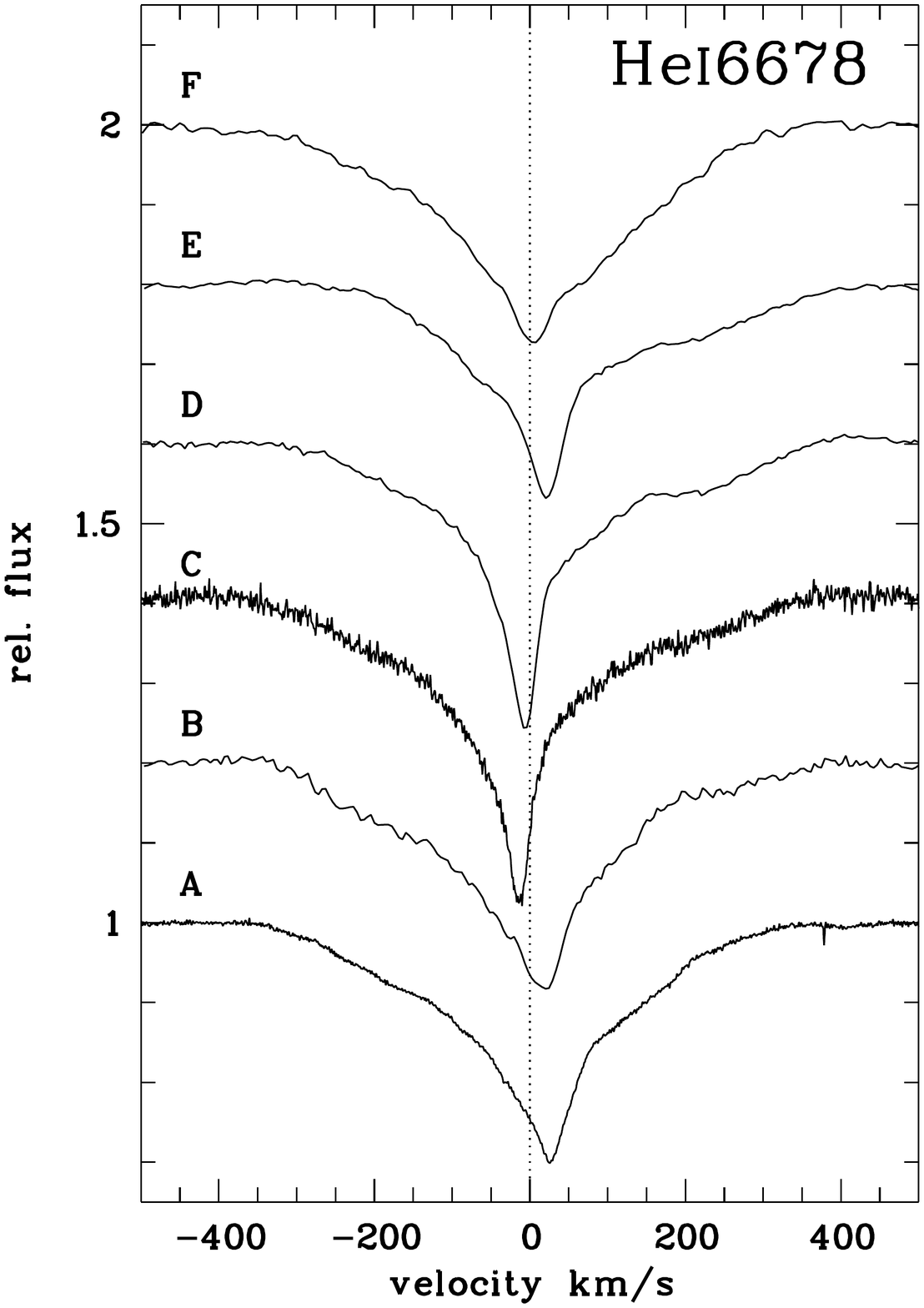}}%
\parbox{4.6cm}{\includegraphics[viewport=25 44 576 781,width=4.6cm,angle=0,clip]{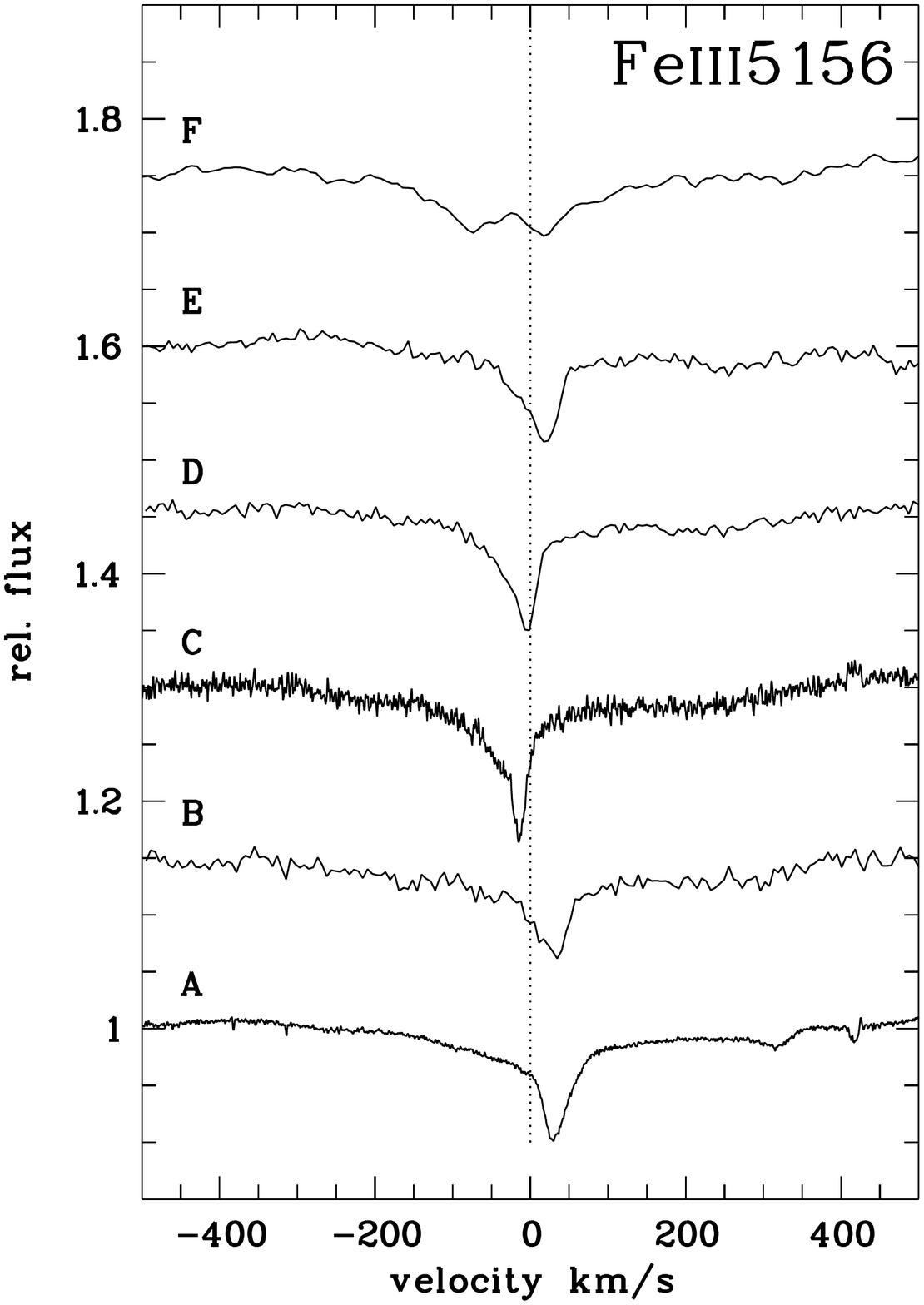}}%
\parbox{4.6cm}{\includegraphics[viewport=25 44 576 781,width=4.6cm,angle=0,clip]{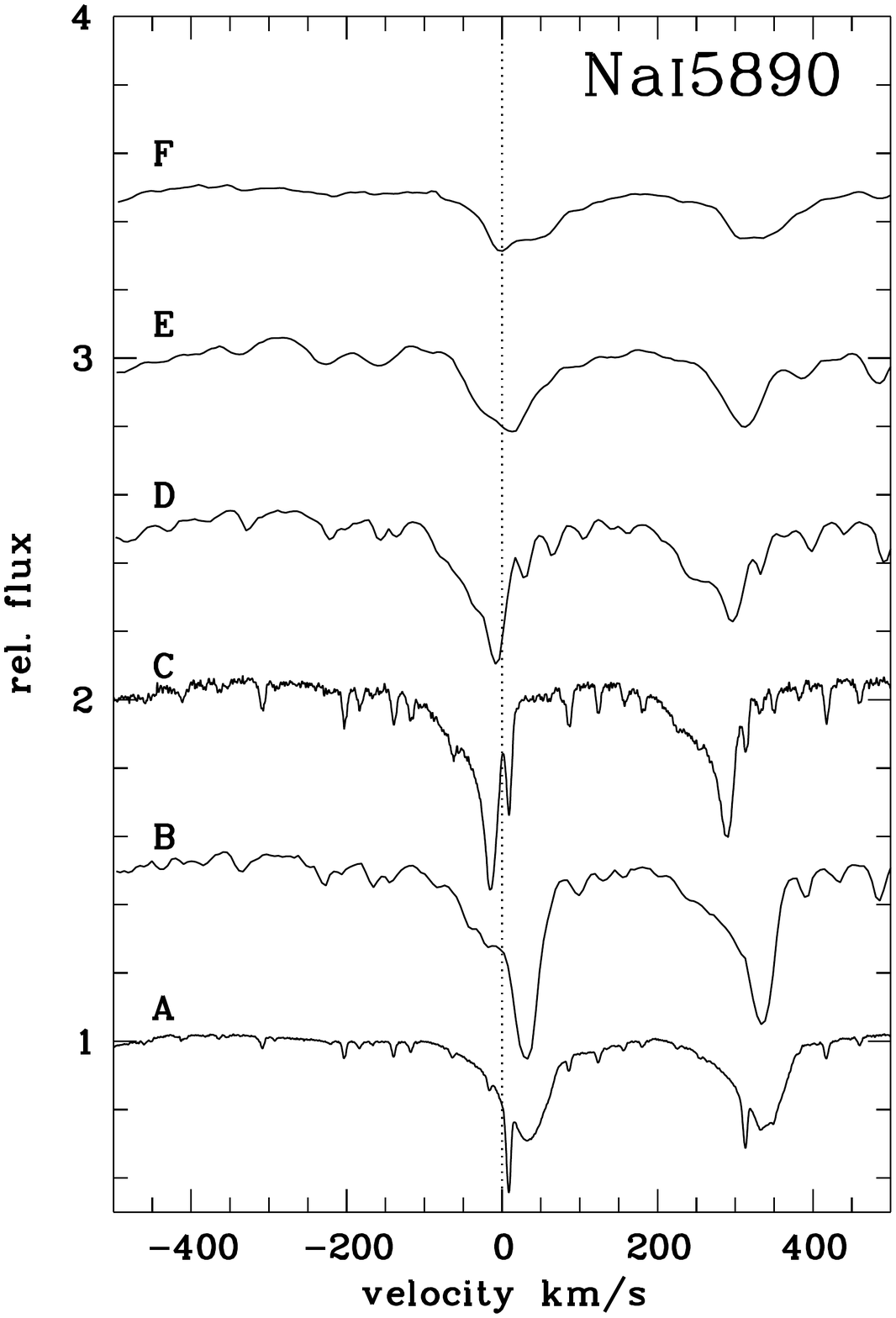}}%
\parbox{4.6cm}{\includegraphics[viewport=25 44 576 781,width=4.6cm,angle=0,clip]{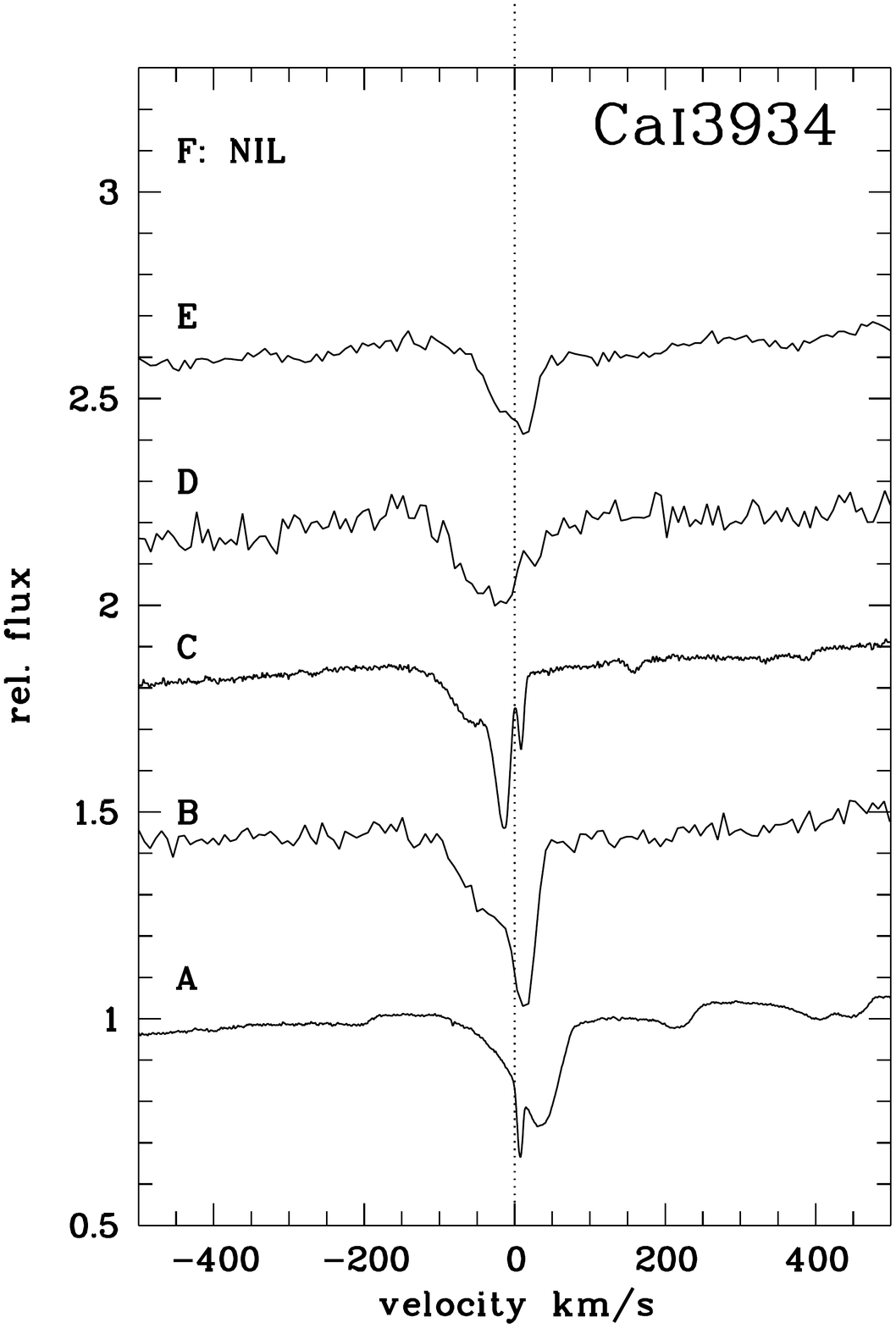}}%
}
\caption[xx]{\label{splines}Evolution of representative lines in the
  visible spectrum during the $V/R$ cycle. See
  Table~\ref{vtor-spectra} for the properties of spectra ``A'' to
  ``F''. The H$\alpha$ $V/R$ can bbe read off Fig.\,\ref{fig_cycle}
}
\end{figure*}
%%%%%%%%%%%%%%%%%%%%%%%%%%%%%%%%%%%%%%%%%%%%%%%%%%%%%%%%%%%%%%%%%

\section{Visual spectral line variability}
\label{VtoR}

\subsection{Long-term evolution of the $\zeta$~Tau circumstellar disk and its $V/R$ variations}
\label{diskev}

\citet[][his Fig.\ 2]{1984BAICz..35..164H} showed that the present 
well-pronounced $V/R$ variations were absent in earlier decades:  
At least from 1920 to about 1950 there was no $V/R$ cyclicity.  A 
$V/R$-variable phase started in the mid-1950s and lasted for about three 
 cycles until 1980.  From 1980 to 1990 the star was again
stable with $V\approx R$.  From 1980 to 1985 the equivalent width (EW) of 
the H$\alpha$ emission decreased from about $-23$ to $-12$\,\AA\ while the
star brightened by 0.3 to 0.4 magnitudes in the $UBV$ passbands, also
getting slightly bluer.  This behavior is typical of Be-shell stars as
described observationally by \citet{1983HvaOB...7...55H} and predicted 
before by \citet[][their Fig.\ 33]{1978ApJS...38..229P} for the case of 
decreasing base density of the disk.  

From 1985 to 1990 the disk was in a low-density state \citep[][their Figs.\ 2 
and 3]{1995A&AS..112..201G}.  
The $V/R$ variability was not well sampled observationally.  But short-term 
(days) variability probably dominated \citep[][their Fig.\ 5]{1995A&AS..112..201G}.  
It would, then, have resembled a phase observed in $\mu$\,Cen by 
\citet{1993A&A...274..356H}, in which rapid $V/R$ variations were due 
mainly to discrete mass loss events and the subsequent circularization of 
the ejected matter. 

Around 1990, the emission strength began to increase again to a
high level of almost $-30$\,\AA, together with brightness and colour
changes as expected for an edge-on disk with increasing base density.  
At the same time, the disk entered a phase of $V/R$ variations.  Since this
replenishing of the disk, the emission lines  became weaker (H\,$\alpha$ EW
between $-15$ and $-20$\,\AA\ ), but the $V/R$
variations are still ongoing.  The AAVSO database \footnote{American Association 
of Variable Star Observers; htpp://www.aavso.org/data} indicates that the 
magnitude was stable over the last 8 years ($<m_V>~\sim~3.\!^m02 \pm 0.\!^m07$).

%%%%%%%%%%%%%%%%%%%%%%%%%%%%%%%%%%%%%%%%%%%%%%%%%%%%%%%%%%%%%%%%%%%%%%%%%%%%%%%%
\begin{table}
\caption[]{Properties of individual $V/R$ cycles 
}
\begin{tabular}{ccccc@{\,--\,}cc}
  \hline\noalign{\smallskip}
  \hline
  Cycle  & \multicolumn{2}{c}{Max.~$V/R$}  & Cycle         &    \multicolumn{3}{c}{Triple peak}     \\
  number &      JD\,24     &   V/R     & length [d]    &    \multicolumn{2}{c}{[JD\,24 range]} & [d] \\
  \hline 
  I     &      50\,414      &   1.62       &   1419        &    51\,137 & 51\,440   &  303    \\
  II    &      51\,833      &   1.58       &   1527        &    52\,580 & 53\,094   &  494    \\
  III   &      53\,360      &   1.44       &   1230        &    54\,005 & 54\,208   &  203    \\
  IV    &      54\,590      &   1.40       &    N/A        &     N/A    &  N/A      &   NA    \\
  \hline\noalign{\smallskip}
\end{tabular}
\label{vtor_cycles}
\end{table}
%%%%%%%%%%%%%%%%%%%%%%%%%%%%%%%%%%%%%%%%%%%%%%%%%%%%%%%%%%%%%%%%

\subsection{Variations of the total H$\alpha$ equivalent width}
\label{haew}

Equivalent widths were measured by means of an automatic {\sc MIDAS}
\citep{1990apaa.conf..111G} procedure in the interval $6546 - 6581$\,\AA. 
Except for a few outliers, all total EWs between 1992 and 2008 are 
in the interval $-11.5$ to $-20$\,\AA, with a
mean of ($-15.5\pm 0.2$)\,\AA.  After an initial increase of
emission lasting from about 1991 to 1994 the value remained rather
stable.  This is a further indicator of the validity of the assumption 
that the individual cycles are comparable to each other.

In view of the uncertainties in the normalization of the
echelle spectra with imperfectly corrected wiggles and systematic
differences between spectrographs with very different spectral
resolution, a more detailed analysis of the measured total equivalent 
widths and their temporal variations was not attempted. 

\subsection{Shell absorption lines} 
\label{shellabs}
In the visual (and higher-resolution) spectra, two types of shell 
lines (dubbed ``narrow-'' and ``broad-line group'', NLG and BLG) 
and their associated cyclic behavior can be distinguished.
Pure NLG-type shell absorption is found in lines of
\ion{Fe}{ii} (near-UV lines only), \ion{Fe}{iii}, \ion{Ni}{ii},
\ion{O}{i}, \ion{Na}{i}, and \ion{He}{i}.  The cores of the Balmer lines, too, 
fall into the NLG category.  Pure BLG-type shell absorption is seen in
some lines of \ion{Si}{ii}, \ion{Fe}{ii} (visual wavelengths),
\ion{Cr}{ii}, \ion{Ti}{ii}, \ion{Mg}{i}, and \ion{O}{i}. This group
also includes \ion{Mg}{ii}\,4481, which is among the easiest lines to
excite in a Be circumstellar environment.  The NLG/BLG classification
of the most important spectral lines in the optical spectrum is marked
in Table~\ref{table-lines}.

%%%%%%%%%%%%%%%%%%%%%%%%%%%%%%%%%%%%%%%%%%%%%%%%%%%%%%%%%%%%%%%%%
% Table on Line-IDs
%%%%%%%%%%%%%%%%%%%%%%%%%%%%%%%%%%%%%%%%%%%%%%%%%%%%%%%%%%%%%%%%%
\begin{table}
\begin{center}
  \caption[xx]{\label{table-lines} Selected spectral lines in the optical spectrum 
and their classification into broad-line (BLG) and narrow-line group 
(NLG), respectively, as described in Sect.~\ref{cycledesc}
}
\begin{tabular}{ccl}
\hline\noalign{\smallskip}
\hline
 Wavelength &  Ion   &  Comment  \\
\hline 
3187.746 &  \ion{He}{i}     &  NLG, no phot. \\
3465.642 &  \ion{Ni}{ii}    &  NLG  \\
3468.678 &  \ion{Fe}{ii}    &  NLG  \\
3471.386 &  \ion{Ni}{ii}    &  NLG  \\
3513.987 &  \ion{Ni}{ii}    & NLG, strong \\
3530.505 &  \ion{He}{i}     & pure phot. \\
3554.412 &  \ion{He}{i}     & pure phot. \\
3587.268 &  \ion{He}{i}     & pure phot. \\
3634.238 &  \ion{He}{i}     & pure phot. \\
3968.469   &  \ion{Ca}{ii}&  BLG + NLG   \\
3994.997 &  \ion{N}{ii}     & pure phot. \\
4002.592    &  \ion{Si}{ii}   &  BLG   \\
4067.031   &  \ion{Ni}{ii}&  BLG + NLG   \\
4128.054   &  \ion{Si}{ii}&  BLG + NLG   \\
4130.872   &  \ion{Si}{ii}&  BLG + NLG   \\
4258.154    &  \ion{Fe}{ii}   &  BLG   \\
4294.099    &  \ion{Ti}{ii}   &  BLG   \\
4481.126    &  \ion{Mg}{ii}   &  BLG   \\
4824.127    &  \ion{Cr}{ii}   &  BLG   \\
5015.678 &  \ion{He}{i}     &  NLG + weak phot. \\
5100.727    &  \ion{Fe}{ii}   &  BLG   \\
5127.387 &  \ion{Fe}{iii}   &  NLG  \\
5156.111 &  \ion{Fe}{iii}   &  NLG  \\
5169.033   &  \ion{Fe}{ii}&  BLG + NLG   \\
5172.684    &  \ion{Mg}{i}    &  BLG   \\
5183.604    &  \ion{Mg}{i}    &  BLG   \\
5197.577   &  \ion{Fe}{ii}&  BLG + NLG   \\
5226.543    &  \ion{Ti}{ii}   &  BLG   \\
5316.615   &  \ion{Fe}{ii}&  BLG + NLG   \\
5679.558 &  \ion{N}{ii}     & pure phot. \\
5875.625 &  \ion{He}{i}     &  NLG + phot. \\
5889.951    &  \ion{Na}{i}    &  NLG   \\
5895.924    &  \ion{Na}{i}    &  NLG   \\
6147.741    &  \ion{Fe}{ii}   &  BLG   \\
6149.258    &  \ion{Fe}{ii}   &  BLG   \\
6156.755    &  \ion{O}{i}     &  BLG   \\
6158.187    &  \ion{O}{i}     &  BLG   \\
6347.109   &  \ion{Si}{ii}&  BLG + NLG   \\
6371.371   &  \ion{Si}{ii}&  BLG + NLG   \\
6678.154 &  \ion{He}{i}     &  NLG + phot. \\
7774.166 &  \ion{O}{i} & NLG  \\
\hline\noalign{\smallskip}
\end{tabular}
\end{center}
\end{table}
%%%%%%%%%%%%%%%%%%%%%%%%%%%%%%%%%%%%%%%%%%%%%%%%%%%%%%%%%%%%%%%%%

The assignment of a line to BLG or NLG is not in all cases and not at
all epochs unique.  For instance, \ion{Fe}{ii}\,5137 and
\ion{Si}{ii}\,6347 normally show the BLG characteristics, but may have
additional NLG components in certain phases of the $V/R$ cycle.  Other BLG lines, too,
exhibit additional temporary NLG components, especially around $V/R$ phase
$\tau\approx 0.25$ (see Sect.~\ref{cycledesc}). Typical representatives are the strongest shell
lines due to
\ion{Si}{ii}, \ion{Ca}{ii}, \ion{Ni}{ii}, and \ion{Fe}{ii}.

Besides the obvious property of their widths, the two groups also differ 
in their variations of radial velocity and symmetry as described in 
Sect.\,\ref{BNLGdiscuss}

\subsection{Length of the $V/R$ cycles}
 \label{qper}

Some confusion may occur due to different definitions of the $V/R$ ratio in the literature. 
They differ in the reference flux level, with respect  to which the peak maxima are measured. 
This paper uses the more common definition ($V/R~=~ F_{V}/F_{R}$, where $F_{V}$ and $F_{R}$ 
are the relative fluxes at the violet and red emission peak maxima, respectively), according 
to which $V/R$ 
is computed without prior correction for the underlying continuum level.  
Therefore, the results are less influenced by the spectrum normalization than in the other definition, 
which uses only the flux above the continuum level.  Consequently, this approach enables to combine 
reasonably well V/R values even in 
heterogeneous databases. Since the photometric variability has been small (Sect.~\ref{diskev}),
while  the H\,$\alpha$ line emission was strong, the continuum variability should have
little effect on the $V/R$ values measured in this way. The $V/R$ values obtained for 
the interval 1993-2008 are visualized in Fig.\,\ref{figvr}.

The observations before JD\,2\,450\,500 are insufficient to conclude
whether the $V/R$ cycle was not yet stabilized or whether the sampling
was too sparse to deduce a more regular $V/R$ variability.  However,
the subsequent observations define nicely three complete cycles marked
I, II, and III in Fig.\,\ref{figvr}.  Cycle IV just commenced about JD
2\,454\,590.  Starting Julian Dates and lengths of the cycles are
summarized in Table~\ref{vtor_cycles}.  The cycle-averaged $V/R$ curve
is displayed in Fig.\,\ref{fig_cycle}.

The mean $V/R$ curve is smooth and roughly symmetric except for 
some time shortly after the minimum.   During this ascending branch, 
the H$\alpha$ emission is split into three peaks (or the self-absorption 
is split into 2 components), making the definition of `the' V 
and `the' R component ambiguous (see
Sect.~\ref{3pp} and Fig.\ref{3ppevol}).  An increased scatter in $V/R$ 
values is the result (see Figs.~\ref{figvr} and  \ref{fig_cycle}).  Obviously, 
the disk variability cannot be characterized by a single parameter 
such as $V/R$ alone.  

By contrast, the $V/R$ maxima are very well defined and can be used 
as fiducial marks for measuring the length of $V/R$ cycles.  Fourth-order 
polynomials fitted to the peaks over different baselines indicate an error in 
their position of 5-8 days. 

In spite of the systematic uncertainty in $V/R$, Fig.\,\ref{figvr} seems to 
suggests a secondary $V/R$ maximum that coincides with the 
triple-peak phases and follows the 
main minimum.   Apparently, such secondary maxima in the cyclic $V/R$ 
variations were not so far reported for any Be star.  Their height and widths 
vary from cycle to cycle and the strongest secondary peak can be 
recognized in Cycle II at JD\,2\,452\,700~--~2\,453\,000.  

A formal Fourier frequency analysis was only 
applied to H$\alpha$ $V/R$ values not affected by a third component 
(see Table\,\ref{vtor_cycles} for the corresponding dates). Such a 
restriction is suggested by the sensitivity of most period-search methods
to non-sinusoidal perturbations and to a lower accuracy of 
the V/R values during the triple-peak phase. The CLEAN method
of the sinusoid fitting \citep{1996A&A...305..887K} gives  periods 
$P_1\,=\,(685 \pm 5)$~d and 
$P_2\,=\,(1405 \pm 26)$~d, obviously the former one is the harmonic produced 
due to variable cycle lengths.  Scargle's method \citep{1982ApJ...263..835S} 
is less likely to return harmonics and suggests 1429\,days.  

This mean duration of the $V/R$ cycles is shorter than the
values derived by \citet{2008IBVS.5813....1P}, 1475\,days, and by
\citet{2006A&A...459..137R}, 1503\,days.  The differences are 
well explained by the different data sets used.  
In particular, the earlier studies could not include the 
shorter-than-average triple-peak phase of Cycle III.  
Even more different and variable cycle length can be identified during the 
1955~-~1980 V/R variable phase. The compiled data presented by
\citet{1984BAICz..35..164H} show three maxima, defining -in the same sense as 
in this paper- two complete cycles of about 2290 and 1290 days.

Radial-velocity (RV) measurements suffer from the inhomogeneity of 
the database in a similar fashion as the absolute emission strength 
does. The impact of this inhomogeneity is reduced in the RV separation of 
the red and violet emission peaks (Fig.\,\ref{figvr}.  
The separation follows the V/R cycle, but with a possible phase shift and
a considerably higher scatter, which not surprisingly is most prominent 
when the triple-peak profiles are present.  The dates of maxima of peak 
separation are delayed by 100~-~200 days with respect to the $V/R$ maxima.
 
%%%%%%%%%%%%%%%%%%%%%%%%%%%%%%%%%%%%%%%%%%%%%%%%%%%%%%%%%%%%%%%%%
\begin{table}[b]
\begin{center}
  \caption[]{\label{vtor-spectra}  Optical spectra representative of the
    $V/R$-cycle. {\sc Flash} is the predecessor instrument of HEROS,
    i.e.\ the same instrument before the extension with a blue
    channel}
\begin{tabular}{cccc}
\hline\noalign{\smallskip}
\hline
  Spectrum &  [JD\,24...]  &  Instrument & Comment  \\
\hline 
A & 53\,333.8  &  UVES/ESO Archive    &  nightly average \\
B & 52\,009.3  &  {\sc Heros}/Data Set C   &  single spectrum  \\
C & 53\,700.8  &  UVES/ESO Archive         &  night average \\
D & 52\,364.3  &  {\sc Heros}/Data Set C   &  single spectrum  \\
E & 52\,725.3  &  {\sc Heros}/Data Set C   &  single spectrum  \\
F & 48\,347.6  &  {\sc Flash}/LSW Archive     & single spectrum   \\
\hline\noalign{\smallskip}
\end{tabular}
\end{center}
\end{table}
%%%%%%%%%%%%%%%%%%%%%%%%%%%%%%%%%%%%%%%%%%%%%%%%%%%%%%%%%%%%%%%%%

%%%%%%%%%%%%%%%%%%%%%%%%%%%%%%%%%%%%%%%%%%%%%%%%%%%%%%%%%%%%%%%%%
   \begin{figure}[t]
   \includegraphics[viewport=42 38 562 755,angle=270,width=9cm]{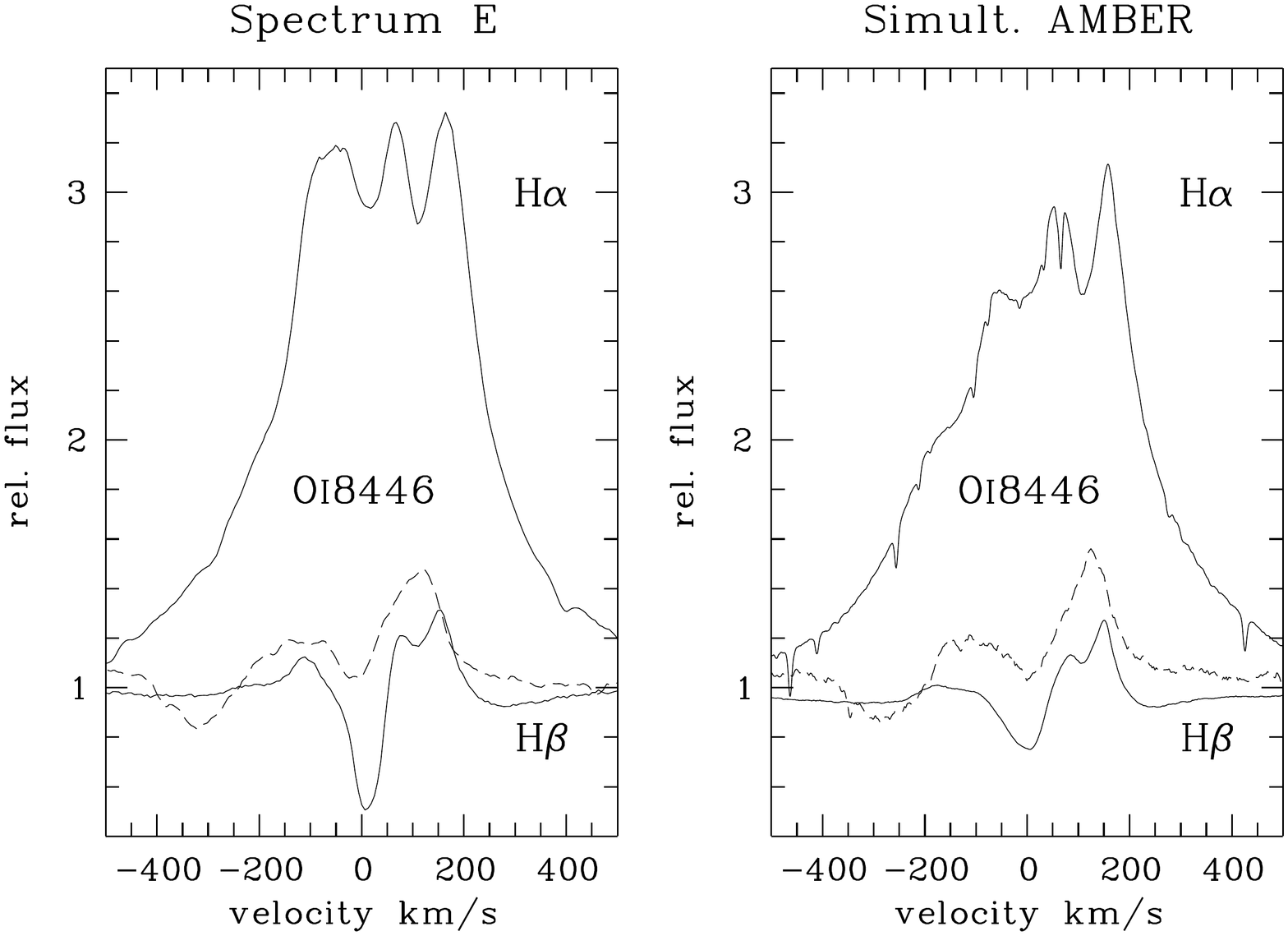}
   \caption{H$\alpha$, H$\beta$ (solid), and \ion{O}{i}\,8446
     (dashed) line profiles of spectrum ``E'' (left) and the
     {\sc AMBER}-simultaneous {\sc FEROS} spectrum during triple-peak phases.  The
     absorption on the blue side of \ion{O}{i}\,8446 is a Paschen
     line}
\label{Balm_OI}
    \end{figure}
%%%%%%%%%%%%%%%%%%%%%%%%%%%%%%%%%%%%%%%%%%%%%%%%%%%%%%%%%%%%%%%%%%

%%%%%%%%%%%%%%%%%%%%%%%%%%%%%%%%%%%%%%%%%%%%%%%%%%%%%%%%%%%%%%%%%%
   \begin{figure}[t]
   \includegraphics[viewport=37 46 543 769,width=9cm,angle=0]{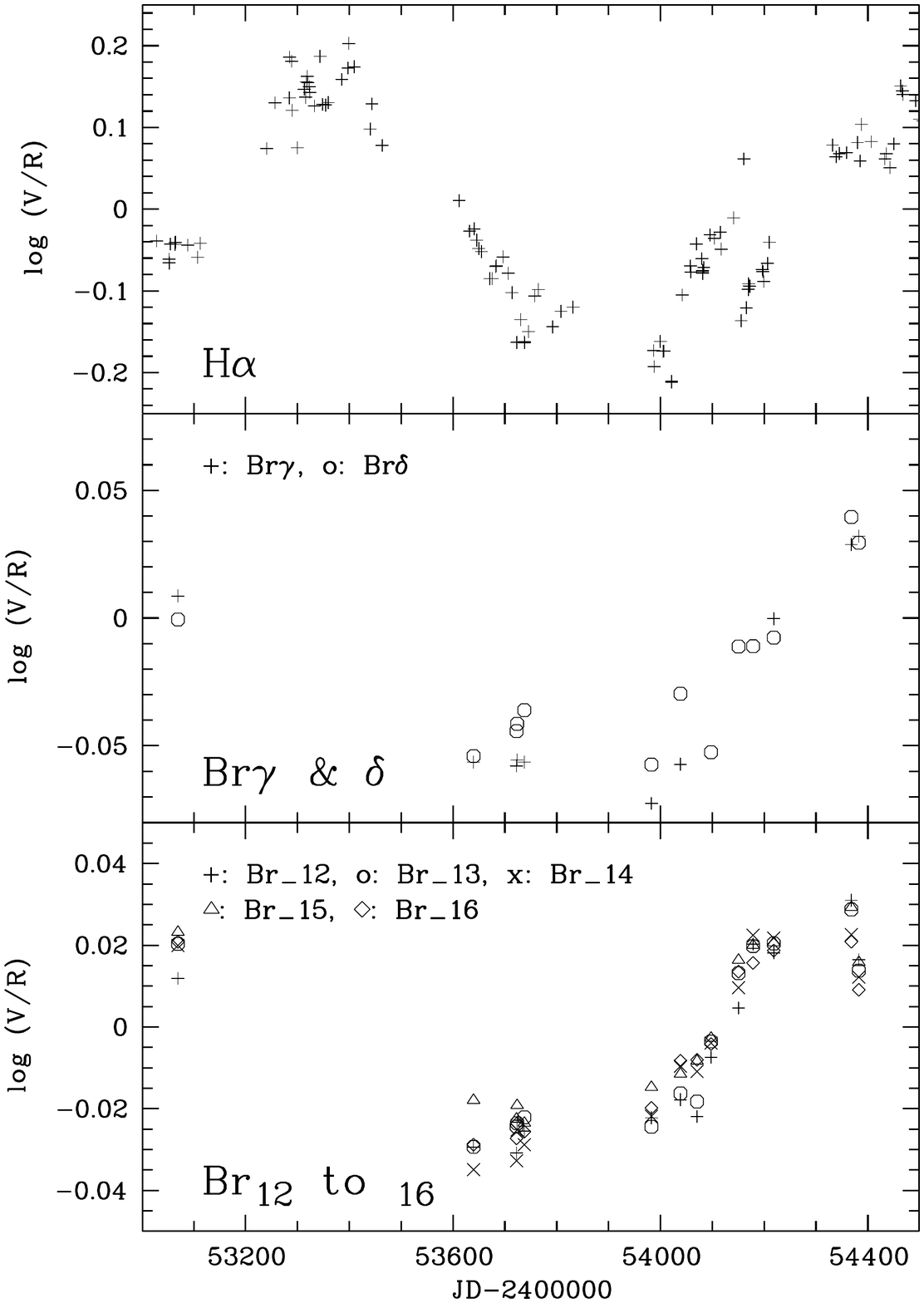}
   \caption{ $V/R$ variations of Br\,$\gamma$ and $\delta$ (middle
     panel) and Br\,12\,--\,16 (lower panel).  
     H$\alpha$ values are shown in the upper panel for comparison}
\label{IR_VtoR}
    \end{figure}
%%%%%%%%%%%%%%%%%%%%%%%%%%%%%%%%%%%%%%%%%%%%%%%%%%%%%%%%%%%%%%%%%%%%%

\subsection{Spectral evolution during the $V/R$ cycle}
\label{cycledesc}

The repeatability of the $V/R$ cycles for more than a decade 
justifies the introduction of the concept of an ephemeris and 
a phase so that observations can be compared between cycles.  
In the following, all phases refer to the formal cycle length of 
1429\,days from above and adopt the $V/R$ maximum of 
JD\,2\,450\,414 as the reference epoch.   The co-phased 
$V/R$ curve is given in Fig.~\ref{fig_cycle}.

In order to more completely characterize the disk variability than the
$V/R$ ratio can do, 6 optical reference spectra were selected.  Their
$V/R$ phases are marked in Fig.\ \ref{fig_cycle}. The spectra were sorted
such that the phase of equal peak height in presumably optically thin
lines like \ion{Fe}{ii}\,6318 has just finished in spectrum ``F'' and
is about to return after spectrum ``E''.  (Note that in Fig.\,\ref{fig_cycle} 
"F" appears at an earlier phase than "E".  But this is
only an artifact of the co-phasing of cycles with different length.
Phenomenologically, ``F'' is actually {\em following} ``E''.)

This classification is restricted to the visual spectral range.  
For the IR spectra, which are more limited in phase coverage 
and spectral resolution, a separate scheme is introduced in Sect.\
\ref{resIRspect}.

%%%%%%%%%%%%%%%%%%%%%%%%%%%%%%%%%%%%%%%%%%%%%%%%%%%%%%%%%%%%%%%%%%%%%%%
% 3pp evolutiion
  \begin{figure}[t]
   \includegraphics[width=9cm,angle=0]{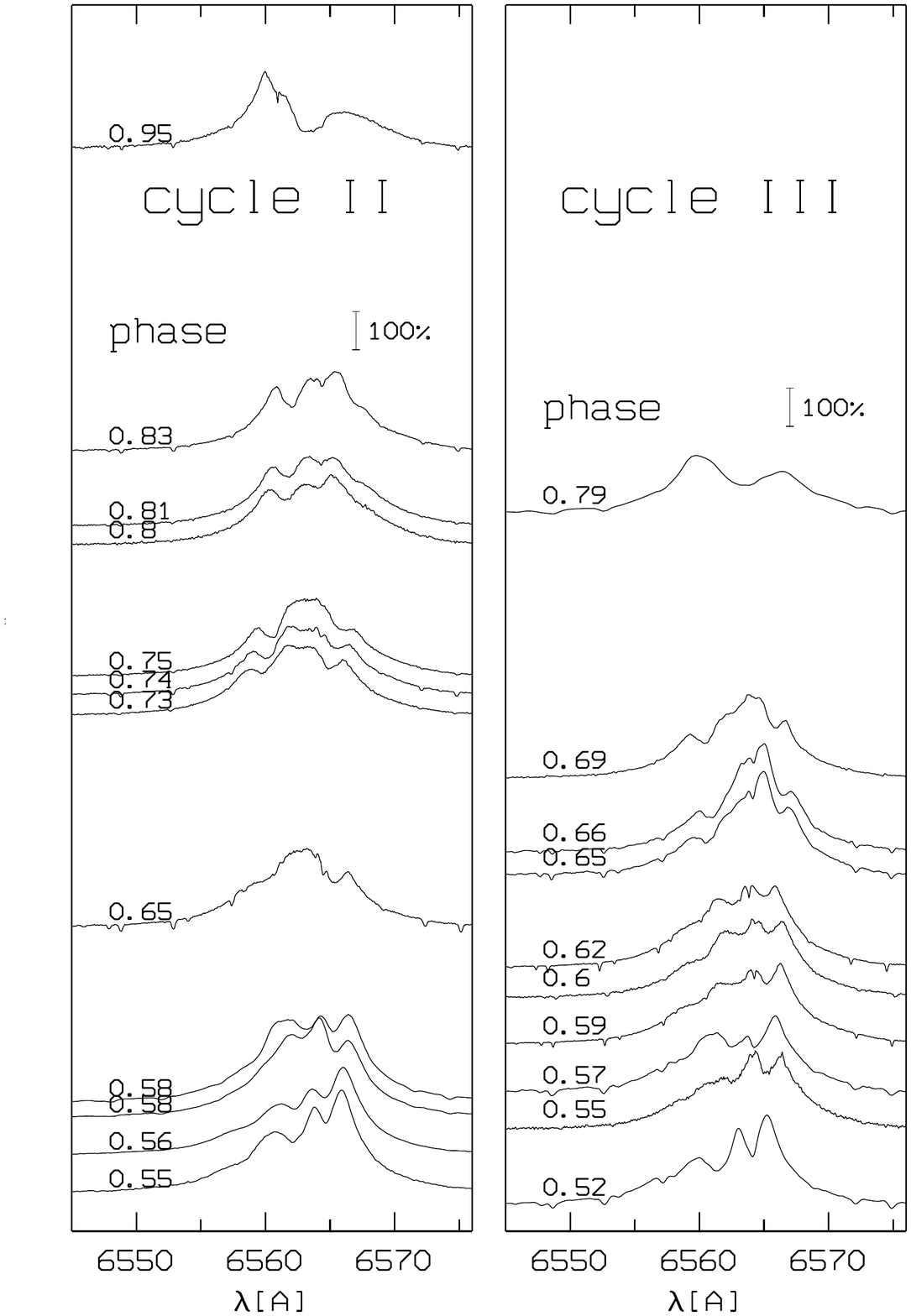}
   \caption{Evolution of the H$\alpha$ triple-peak profiles in the two
     most recent $V/R$ cycles of $\zeta$~Tau. The shifts between the spectra 
     correspond to differences in phase as indicated on the left side. 
     The phase is computed separately for each cycle, using different  
     starting dates (V/R maxima) and lengths as listed in Table~\ref{vtor_cycles}.
     The vertical bar indicates the 100\,\% continuum level
     }
    \label{3ppevol}
    \end{figure}
%%%%%%%%%%%%%%%%%%%%%%%%%%%%%%%%%%%%%%%%%%%%%%%%%%%%%%%%%%%%%%%%%%%%%%%%

The following subsections describe the main optical $V/R$ phases and
their transitions.  The definition of the $V/R$  phases is for H$\alpha$.
The variability of other lines may have significant offsets in phase,
as detailed below.

%%%%%%%%%%%%%%%%%%%%%%%%%%%%%%%%%%%%%%%%%%%%%%%%%%%%%%%%%%%%%%%%%
 \begin{figure}[t]
\parbox{8.8cm}{%
\parbox{8.8cm}{\includegraphics[viewport=41 54 568 
780,angle=270,width=8.8cm,clip]{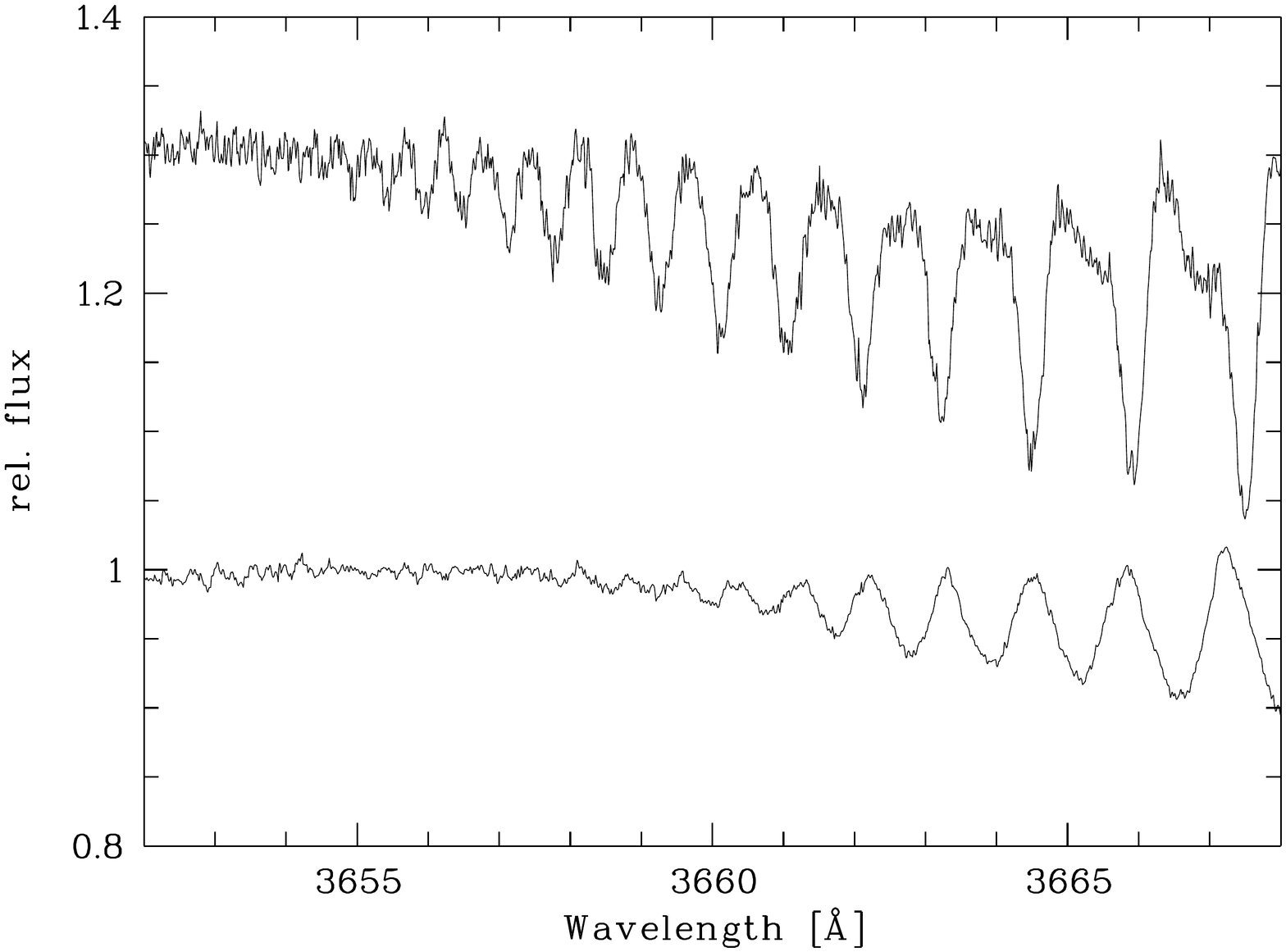}}%

\parbox{8.8cm}{\includegraphics[viewport=41 54 568 
780,angle=270,width=8.8cm,clip]{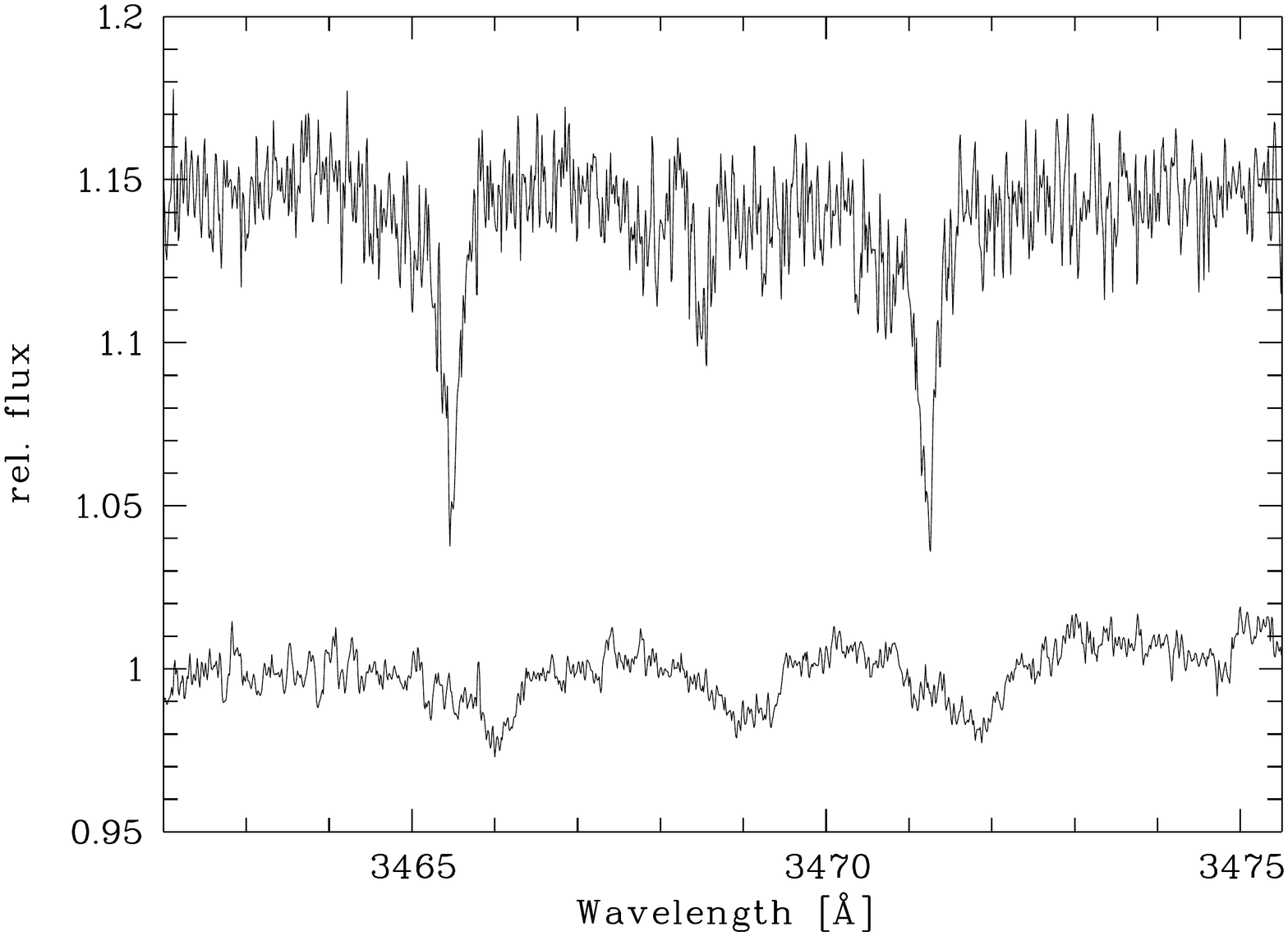}}%
}
\caption[xx]{\label{fig_blue}The region of the Balmer discontinuity
  (upper panel) and the narrow shell lines (lower panel) in the Balmer
  continuum in reference spectra ``A'' (lower plots) and ``C''
  (upper plots), see Sect.~\ref{vtor_ver}}
\end{figure}
%%%%%%%%%%%%%%%%%%%%%%%%%%%%%%%%%%%%%%%%%%%%%%%%%%%%%%%%%%%%%%%%%

\subsubsection{$V > R$} 
This phase is represented by the ``A'' profiles in Fig.\ \ref{splines}.  In all emission 
lines the violet peak is higher than the red one.  The shell absorption 
lines weaken and become narrower.  Both NLG and BLG are red-shifted
with respect to the systemic velocity.  However, there are differences:  The
red edge of BLG shell absorption is very steep, and the blue edge
rather asymptotically joins the continuum.  In the NLG, it is
the blue edge which is steeper than the red one.  All ionic species
are present but some low-ionization lines begin to 
weaken.  

\subsubsection{\label{vtor_ver} $V=R$, descending to $V < R$, and deep absorption} 
When the strength of the $R$ peak in H$\alpha$ begins to approach 
the one of the $V$ peak, the peaks in H$\beta$ are already of 
almost equal height, and \ion{Fe}{ii} lines even exhibit an inverted $V/R$ 
ratio (cf.\ spectra ``B'' and ``C'' in Fig.\ \ref{splines}).

Some of the low-ionization shell lines of the BLG category 
have completely vanished, while the NLG shell lines are at maximum
depth now.  The Balmer series is maximally visible up to H40/41. 
At other phases, the limit is reached around H34/H35 
(see Fig.~\ref{fig_blue}).

While pure BLG lines such as \ion{Mg}{ii}\,4481 are slightly
blue-shifted, pure NLG lines reside at zero velocity and are very
narrow.  Also shortwards of the Balmer discontinuity several shell
lines are far more clearly present than at other phases,
phenomenologically belonging to the NLG.  Their FWHM is only about 17
to 20\,km\,s$^{-1}$, i.e.\ about the thermal width.

Shell lines of \ion{Fe}{ii} and \ion{Si}{ii}, however, develop a
two-component structure related to both groups: a very narrow and deep
absorption core at zero velocity is superimposed on a broad and 
shallower absorption that has its own distinct core farther to the
blue.  The two cores can be clearly identified as parts of the BLG and
NLG characteristics, respectively.

\subsubsection{$V < R$} 
The ``D'' profiles in Fig.\ \ref{splines} illustrate this phase.
The red peak reaches its maximal strength.  Both NLG and BLG
lines are blue-displaced with respect to the systemic velocity.  The dual
NLG/BLG characteristics of some lines weakens and finally vanishes 
in that NLG cores disappear while the remainder of the profile 
evolves as if belonging to the BLG.

\subsubsection{$V=R$, ascending to $V > R$, and triple-peak structure} 
\label{3pp}
Spectra ``E'' and ``F'' in Fig.\ \ref{splines} offer a general overview
of this phase,
and Fig.\ \ref{3ppevol} illustrates the temporal evolution in H$\alpha$.  
The most intriguing property of this phase is that
the H$\alpha$ line emission strongly deviates from the classical Be
star profile in that it no longer shows clear double peaks with a
well-pronounced central depression.  Rather, the central depression is
filled in (or split) and three peaks of
similar strength co-exist.  Although the central depression does not get
filled in in the other Balmer lines, and they do not show multiple peaks,
their structure is more complex than in other phases.

The \ion{O}{i}\,8446 line, which is excited by
fluorescence of Ly$\beta$ and therefore usually assumed to present an
optically thin tracer of the hydrogen distribution, does not show any
sign of a triple peak (see Fig.~\ref{Balm_OI}).  Instead, the profile is 
double-peaked with a central depression.  Noteworthy is that the 
radial velocity of the red
peak of the \ion{O}{i} profile coincides with the local {\em minimum}
in the triple peaked H$\alpha$ and H$\beta$ profiles.

The shell absorption lines are strong and yet relatively broad.  Lowly
ionized species like \ion{Ti}{ii}, \ion{Cr}{ii}, or \ion{Mg}{ii}
prevail, but high-ionization species like \ion{Fe}{iii}
still produce noticeable absorption. 

Spectrum ``F'' (Fig.\ \ref{splines}) was obtained 
in 1991.  Since then, the shell spectrum has generally weakened. 
But there is no doubt that spectrum "F" is representative 
of the general $V/R$ pattern about this phase.  The latter includes 
that, between spectra ``E" and ``F", BLG shell absorptions 
shift from zero to moderately positive velocities.  Examples are
\ion{Mg}{ii}\,4481, \ion{Si}{ii}\,6347, and \ion{Fe}{ii}\,5317 in
Fig.\ \ref{splines}.  This is accompanied by a change in profile symmetry.  
While in ``E" there is a sharp blue edge, ``F" exhibits a sharp 
red edge.

In the NLG shell lines, however, the position of the line minimum 
as well as the symmetry change in exactly the opposite way, from 
low positive velocity to zero velocity in the same two observations.  
See, e.g., \ion{Fe}{iii}\,5156 and \ion{He}{i}\,6678 in Fig.\,\ref{splines}.  

There are very few pure emission lines, e.g.\ \ion{Fe}{ii}\,6319.  But
all of them, as well as lines such as H$\beta$, where emission still 
dominates over the shell absorption, already show the $V$ peak
higher than $R$.  That is, the $V/R$ phase of these lines is
ahead of the one of H$\alpha$.

Fig.\,\ref{3ppevol} illustrates the evolution of the triple peak
profiles in the two most recent $V/R$ cycles.  The triple-peak
interval in Cycle II lasted for almost 500 days, while in Cycle III it
was only 200 days.  As can be seen from Figs.\ \ref{figvr} and
\ref{fig_cycle}, the onset of the triple-peak interval is typically
fairly close to phase 0.5.  Much stronger cycle-to-cycle variations
are evident for its terminal phase.  In spite of the varying time
scale, however, the morphological evolution of the H$\alpha$ line
profile follows approximately the same pattern in all 
cycles.
 
%%%%%%%%%%%%%%%%%%%%%%%%%%%%%%%%%%%%%%%%%%%%%%%%%%%%%%%%%%%%%%%%%
\begin{figure*}[t]
\parbox{18.8cm}{%
\parbox{9.2cm}{\includegraphics[viewport=50 61 562 800,angle=270,width=9.2cm]{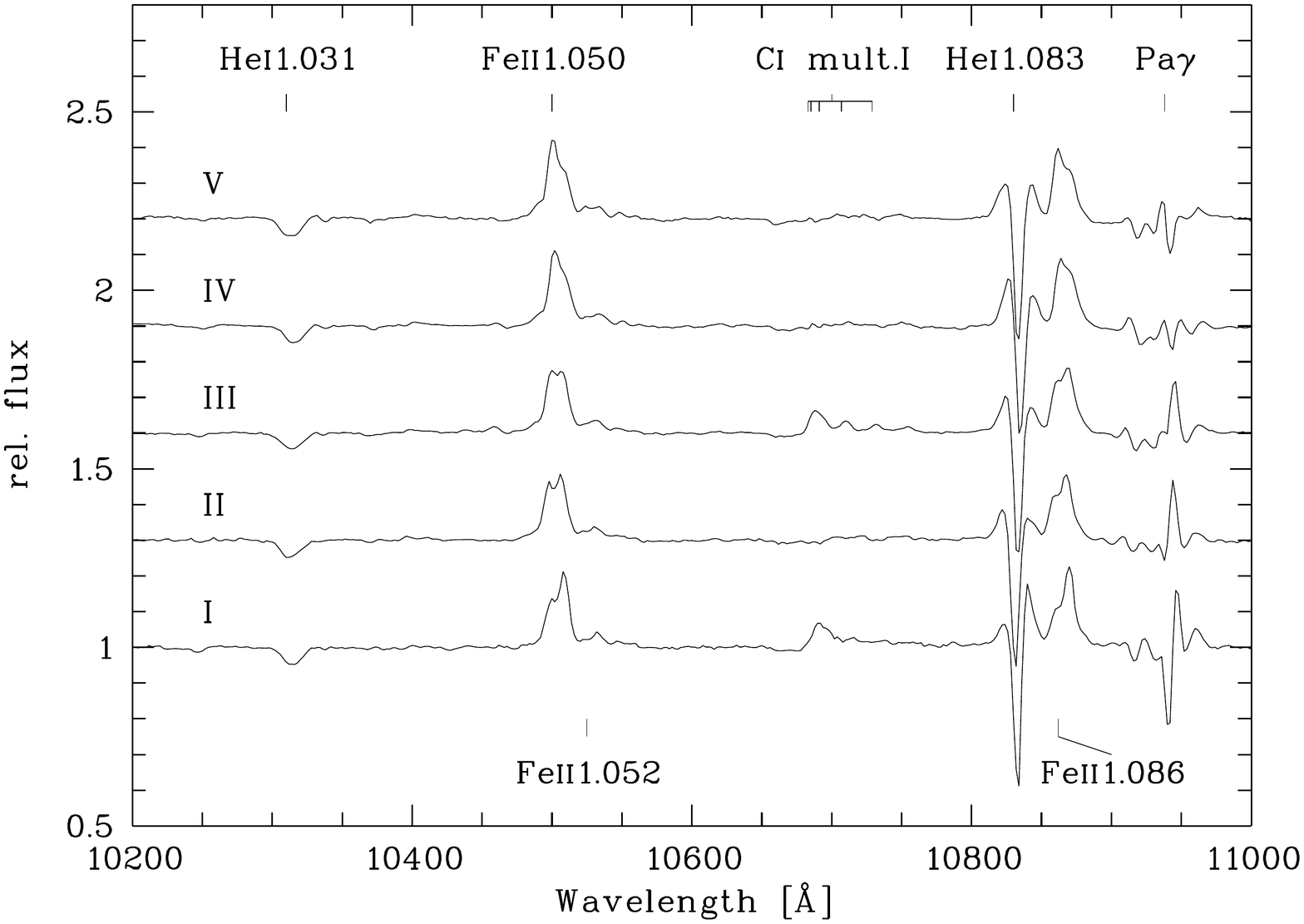}}%
\parbox{9.2cm}{\includegraphics[viewport=50 61 562 800,angle=270,width=9.2cm]{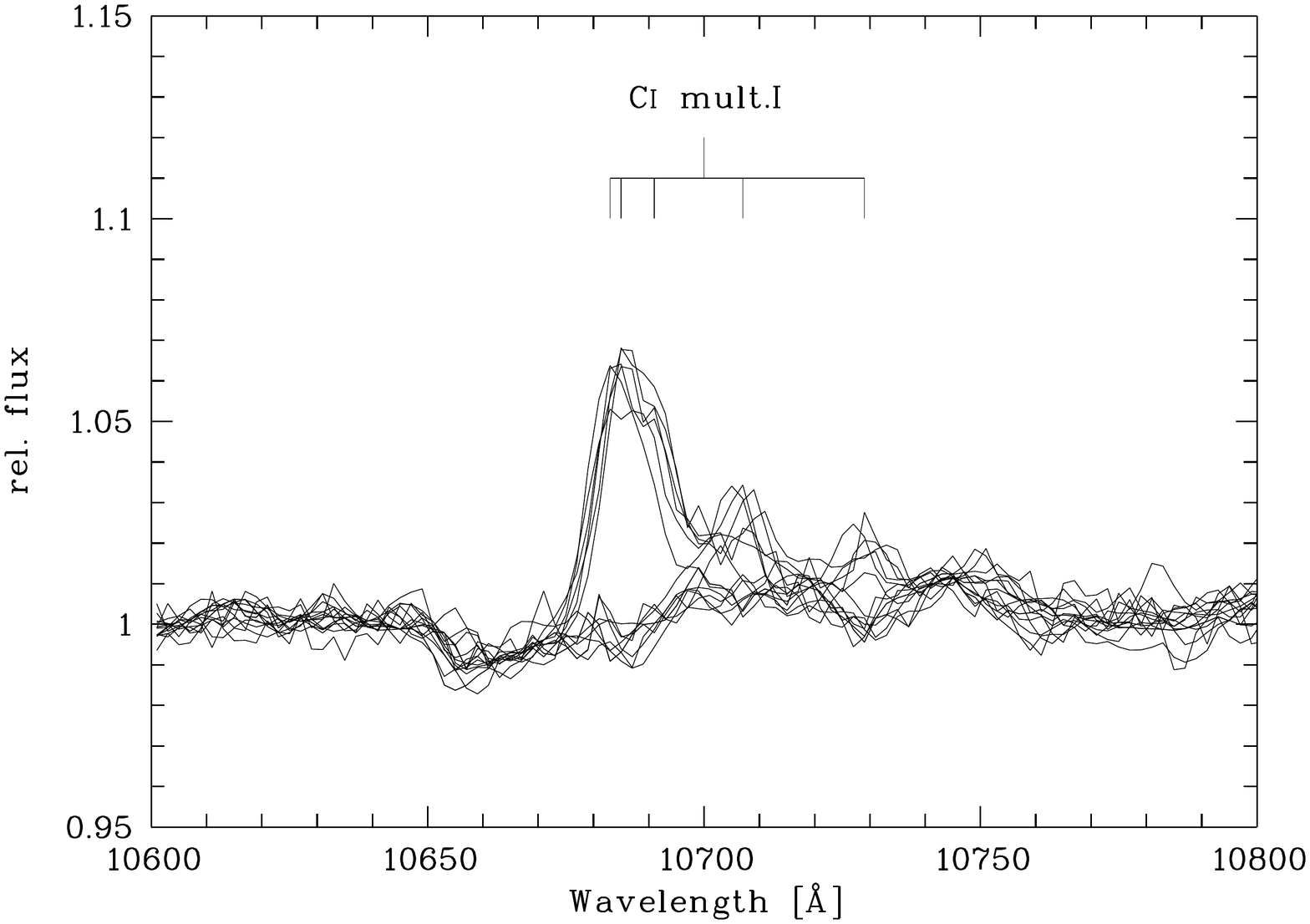}}%
}

\parbox{18.8cm}{%
\parbox{9.2cm}{\includegraphics[viewport=50 61 562 800,angle=270,width=9.2cm]{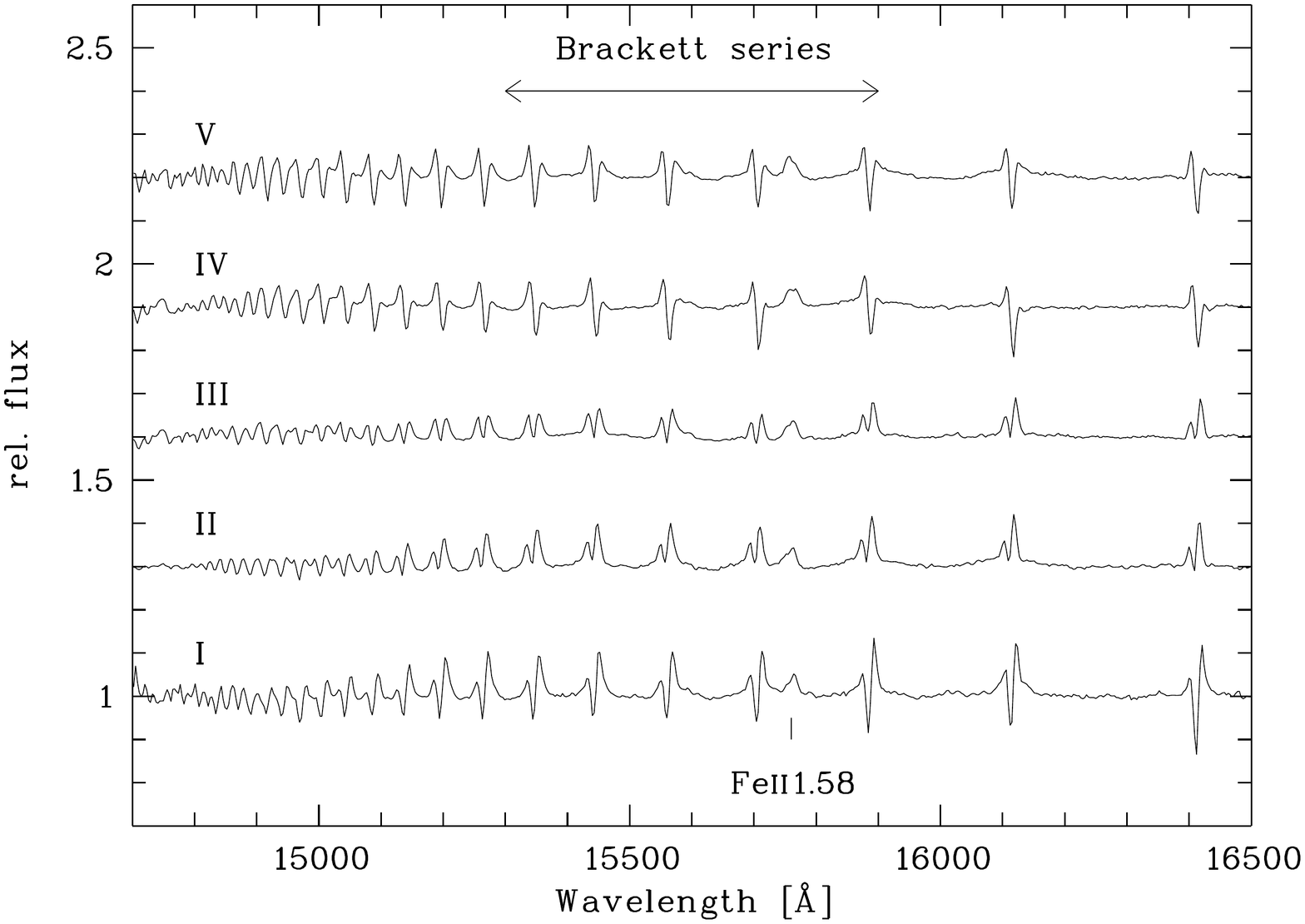}}%
\parbox{9.2cm}{\includegraphics[viewport=50 61 562 800,angle=270,width=9.2cm]{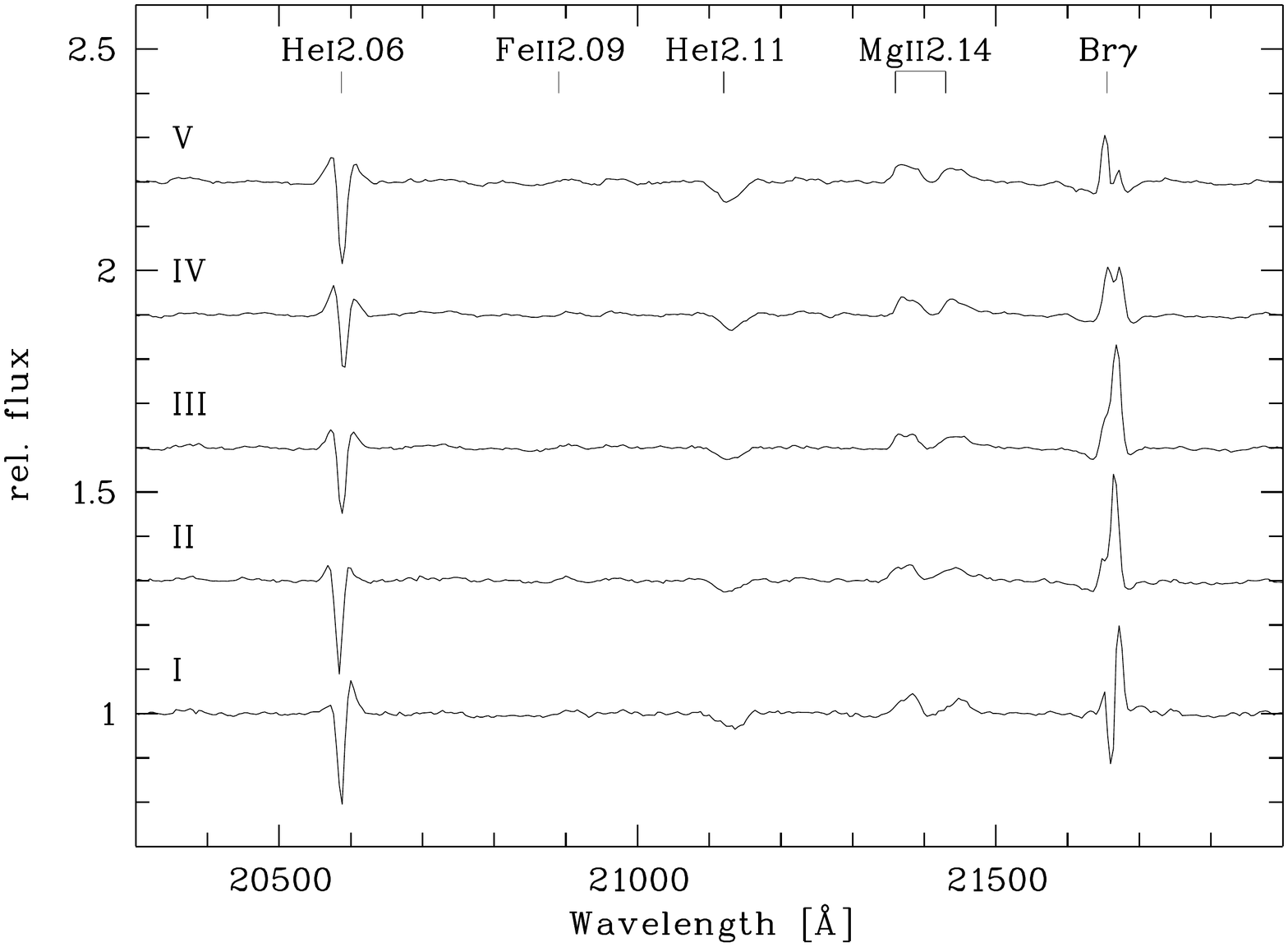}}%
}
\caption{Selected spectral windows in the J, H, and K bands.  
Observing dates of spectra 
I through V are given in Table \ref{vtor-IRspectra}; their phases in
the mean H$\alpha$ cycle are marked in Fig.\,\ref{fig_cycle}.
Pa\,$\gamma$ is blended with He\,{\sc i}\,16\,600 and two magnesium
lines.  Spectrum III is the one closest in time to the {\sc AMBER} 
observations.  The upper right panel combines all spectra of \ion{C}{i}
multiplet 1 in the J-band, illustrating its on/off behavior}
\label{IR_lp}
\end{figure*}
%%%%%%%%%%%%%%%%%%%%%%%%%%%%%%%%%%%%%%%%%%%%%%%%%%%%%%%%%%%%%%%%%

\section{Infrared  spectral variations}
\label{resIRspect}

\begin{table}
\begin{center}
 \caption[]{\label{vtor-IRspectra} Representative infrared spectra}
\begin{tabular}{ccc}
\hline\noalign{\smallskip}
\hline
  Spectrum &  Date [JD\,24...]     & Comment  \\
\hline 
I   & 53\,722 & 22 days after spectrum ``D" \\
II  & 53\,982 & just before triple peak \\
III & 54\,097 & contemporaneous to AMBER  \\
IV  & 54\,281 & just after triple peak \\
V   & 54\,382 & \\
\hline\noalign{\smallskip}
\end{tabular}
\end{center}
\end{table}

As for the optical domain, a number of phase-representative reference
IR spectra were selected to facilitate the description of the main
variability.  Their dates are marked as I through V in Fig.\,\ref{fig_cycle},
and listed in Table \ref{vtor-IRspectra}.

The IR spectra (see Fig.~\ref{IR_lp}) cover the JHK bands 
and the H$\alpha$ $V/R$ phases from about 0.3 through 0.8.  
Although of lower resolution, some of the conclusions already drawn 
from the visual spectra are confirmed:  The $V=R$ deep absorption 
phase has the same characteristics in the IR, as far as \ion{H}{i} and 
\ion{He}{i} are concerned.  In the triple-peak phase, the shell absorption 
gets shallower in  \ion{He}{i} and even disappears completely in
the higher-order Brackett lines, leaving a pure emission
profile although with a pronounced central depression.  All observed
Brackett lines are double-peaked at all times. But Br$\gamma$ 
and Br$\delta$ develop single-peak
profiles when the one in H$\alpha$ shows a triple-peak structure.  Also
metallic lines, in particular those of \ion{Fe}{ii} may show single
peak profiles, but at other phases.  But it must be suspected that this 
merger is only apparent due to the limited spectral resolving power.  

The IR lines provide the basis  for a more in-depth discussion 
of the phase lags between H$\alpha$ and other lines
(Sect.\ \ref{3pp}).  At some phases, they are well visible even within 
\ion{H}{i} emission of the Brackett series.  In Spectrum III of Fig.\
\ref{IR_lp} the lower
Brackett lines still show $V < R$ and, in \BrG, even $V \ll R$, but 
towards the end of the series $V/R$ reaches unity.  In a spectrum 
taken less than a month later, some of
the latter lines even appear with $V > R$ while \BrG\ looks unchanged.
The $V=R$ state is also reached in \ion{He}{i}, \ion{Fe}{ii}, and \ion{Mg}{ii}.  
These two particular observations (incl.\ Spectrum III) 
were obtained almost
simultaneously with the {\sc AMBER} data described below.  

The phase lag between higher- and lower-order Brackett lines is also
well visible in Fig.~\ref{IR_VtoR}.  Already the first IR observation
on JD\,2\,453\,069 shows that, while log$(V/R)$ is still below unity
for H$\alpha$, it is about unity for Br$\gamma$ and Br$\delta$, and
still larger for higher Brackett lines.  In general, the $V/R$ values
for both Br$\gamma$ and Br$\delta$ precede the $V/R$ curve of H$\alpha$,
but clearly lag behind the ones of the Br 12-16 lines.  The $V/R$
amplitude of the Br 12-16 lines is slightly lower than the ones of
Br$\gamma$ and Br$\delta$.  But it is lower by a factor 8-10 relative
to the one of H$\alpha$.

Contrary to H$\alpha$ (see Fig.~\ref{fig_cycle}), a distortion of
the $V/R$ curve around phase 0.6 due to triple-peak profiles cannot be
recognized in any IR lines.  Observations at higher spectral resolution 
are desirable to establish this definitively.  

A very conspicuous type of behavior without counterpart in the visible
range is associated with the infrared \ion{C}{i} lines (see upper
right panel of Fig.~\ref{IR_lp}).  IR multiplets 1 and 24 of this
species are seen in emission in six out of fourteen IR spectra (JD24:
53\,069, 53\,639, 53\,722, 53\,723, 54\,038, and 54\,070) and absent
in the others.  Since intermediate cases were not observed, this may
constitute some type of ``on/off'' behavior.  Contrary to the typical
circumstellar lines described above, these lines have a symmetric
single-peak profile and exhibit no variability other than their being
either present or absent.  No relation could be found of the presence
or absence of the \ion{C}{i} emission to the $V/R$ cycle or the
132.9735-d orbital period. It was also checked that the variations cannot be 
produced by an overlapping OH telluric emission \citep{2000A&A...354.1134R}. 

%%%%%%%%%%%%%%%%%%%%%%%%%%%%%%%%%%%%%%%%%%%%%%%%%%%%%%%%%%%%%%%%%
 \begin{figure*}[t]
\parbox{18.8cm}{%
\parbox{4.6cm}{\includegraphics[viewport=40 35 571 825,angle=270,width=4.6cm,clip]{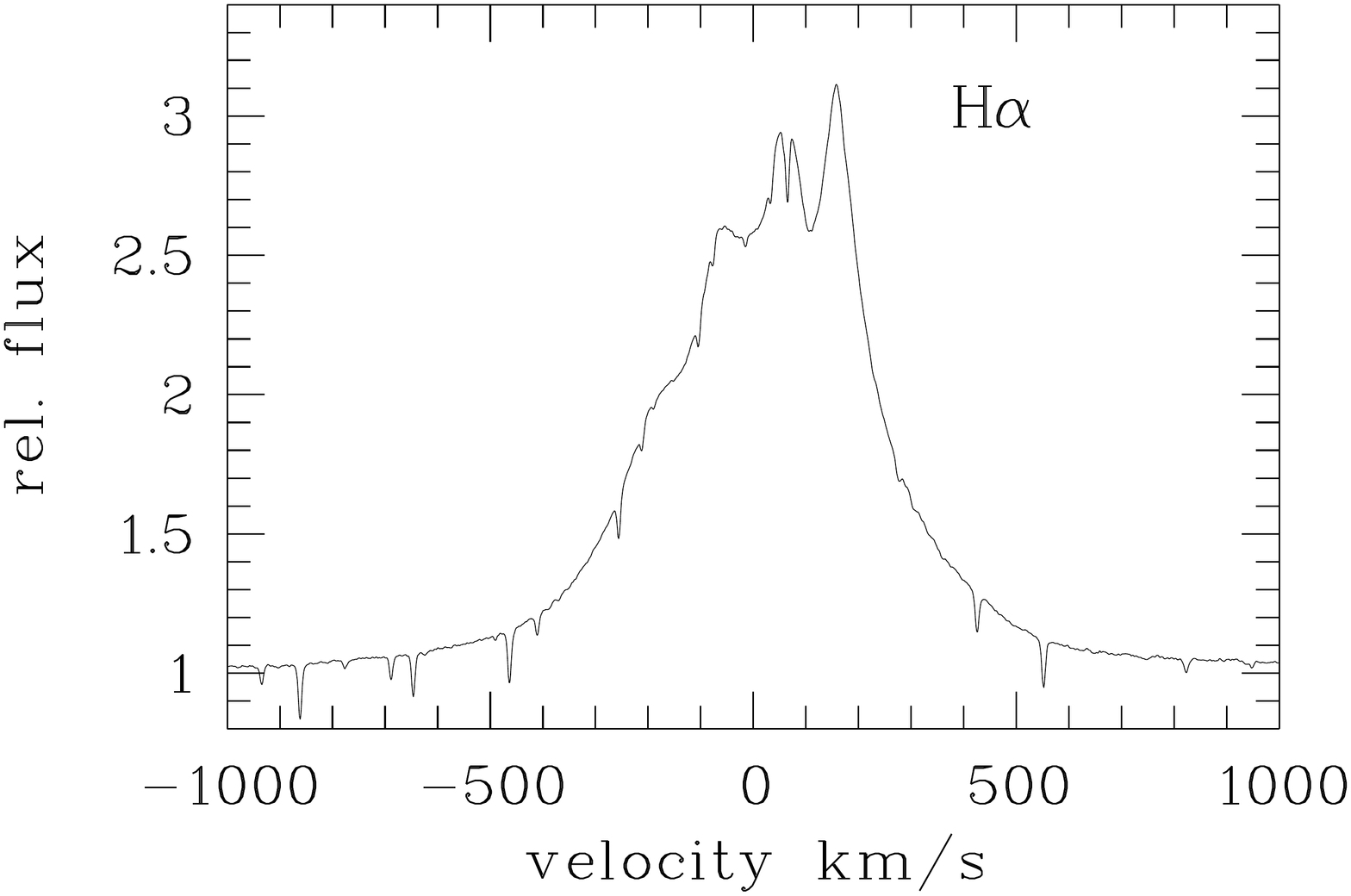}}%
\parbox{4.6cm}{\includegraphics[viewport=40 35 571 825,angle=270,width=4.6cm,clip]{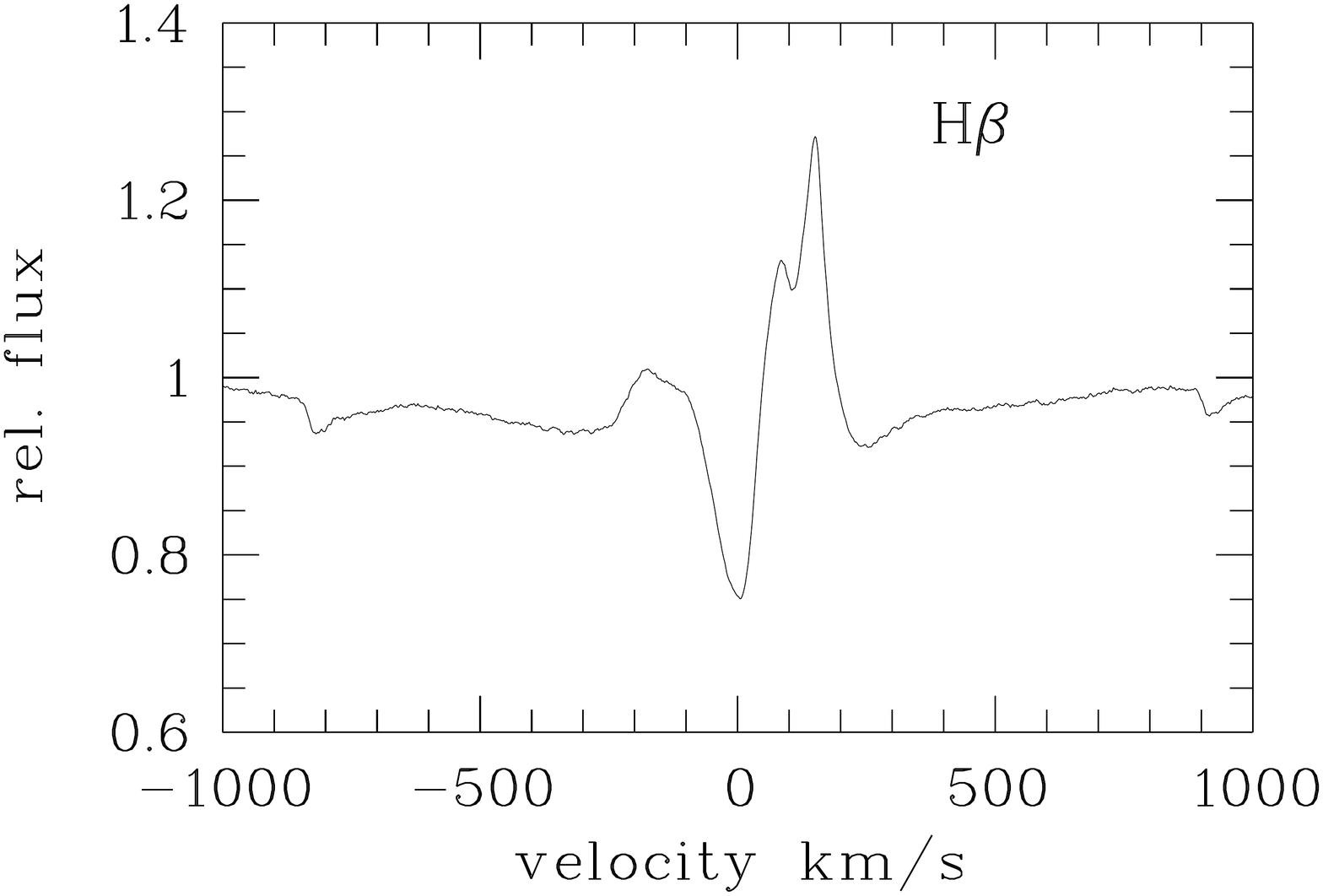}}%
\parbox{4.6cm}{\includegraphics[viewport=40 35 571 825,angle=270,width=4.6cm,clip]{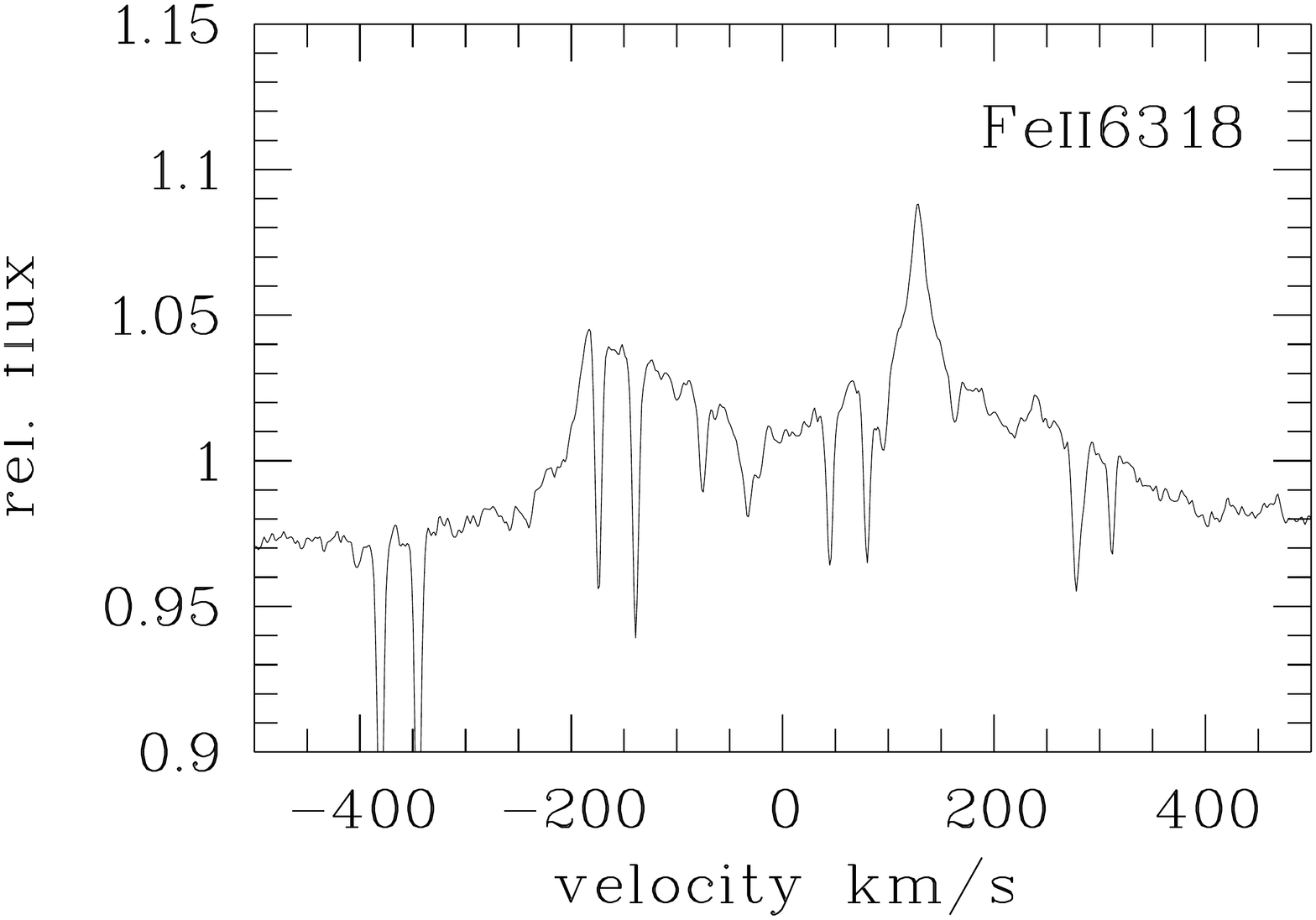}}%
\parbox{4.6cm}{\includegraphics[viewport=40 35 571 825,angle=270,width=4.6cm,clip]{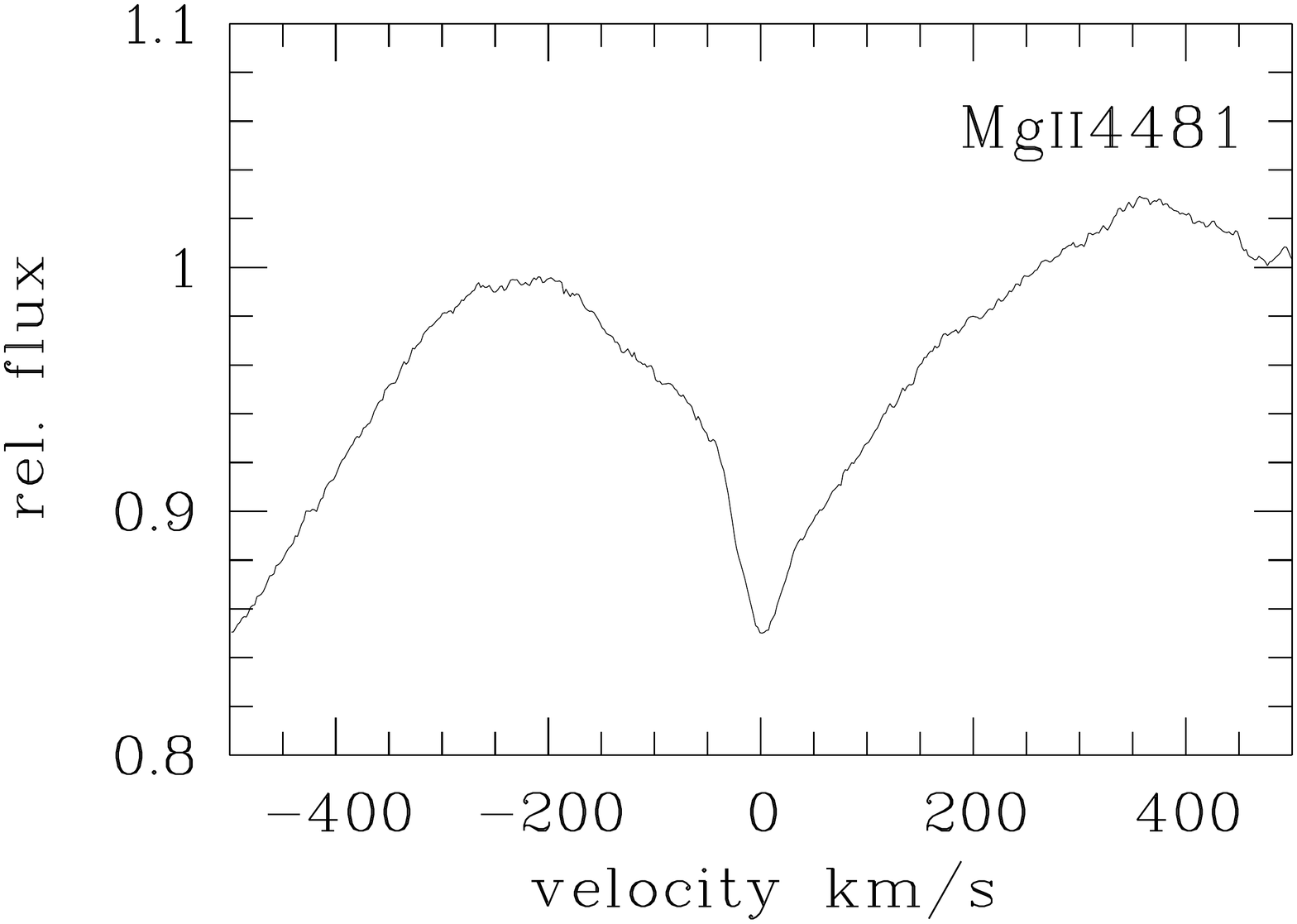}}%
                                        
\parbox{4.6cm}{\includegraphics[viewport=40 35 571 825,angle=270,width=4.6cm,clip]{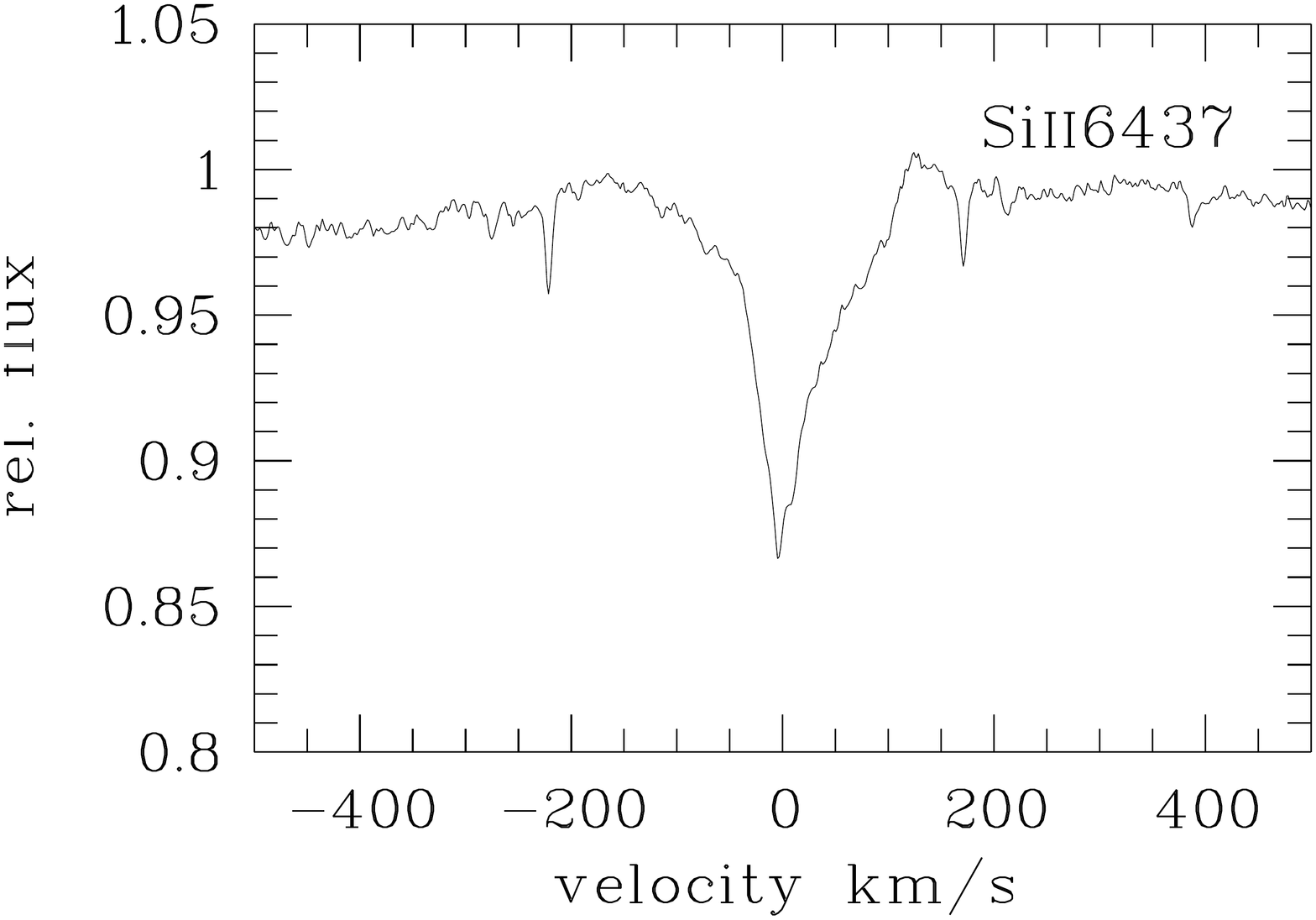}}%
\parbox{4.6cm}{\includegraphics[viewport=40 35 571 825,angle=270,width=4.6cm,clip]{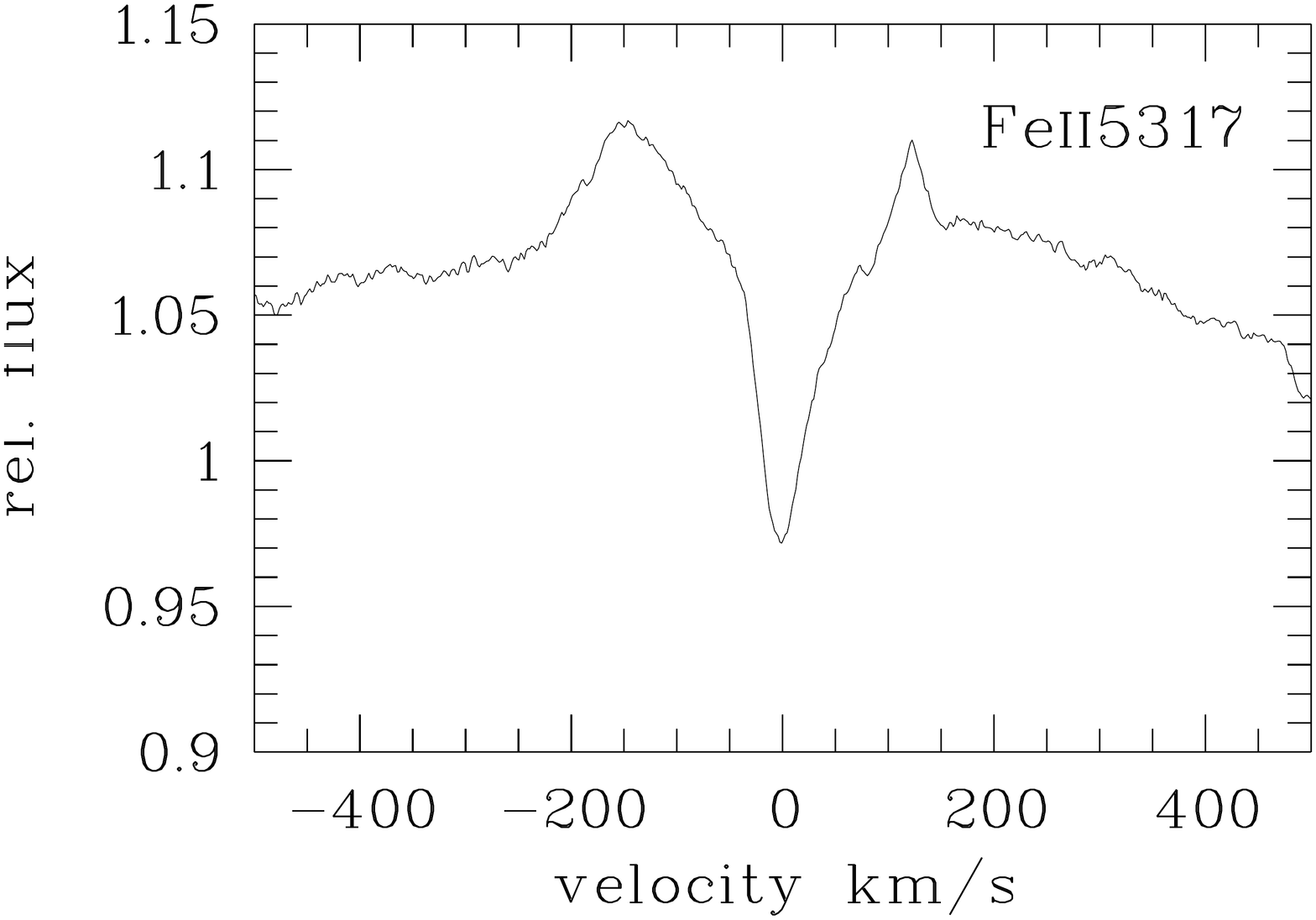}}%
\parbox{4.6cm}{\includegraphics[viewport=40 35 571 825,angle=270,width=4.6cm,clip]{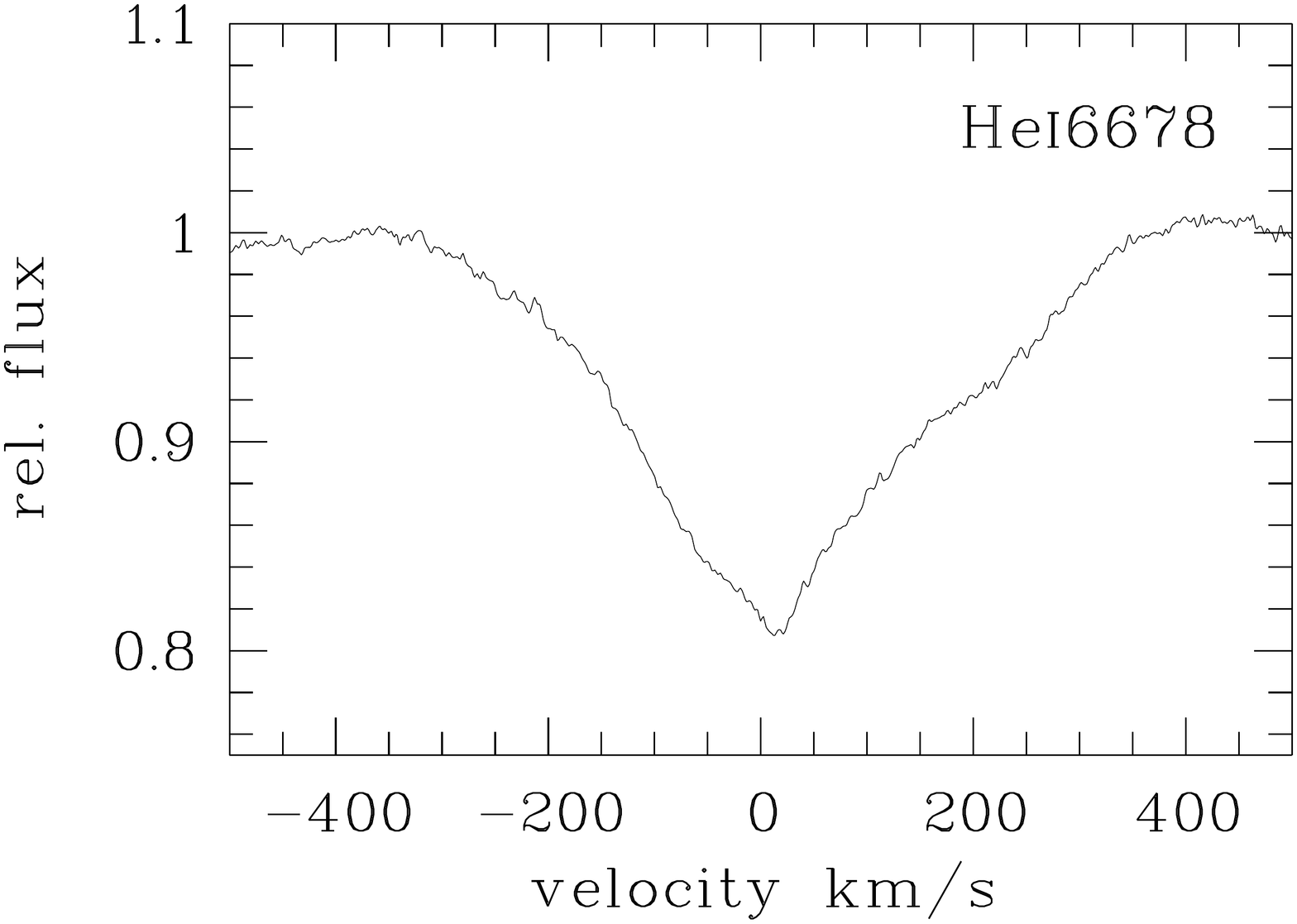}}%
\parbox{4.6cm}{\includegraphics[viewport=40 35 571 825,angle=270,width=4.6cm,clip]{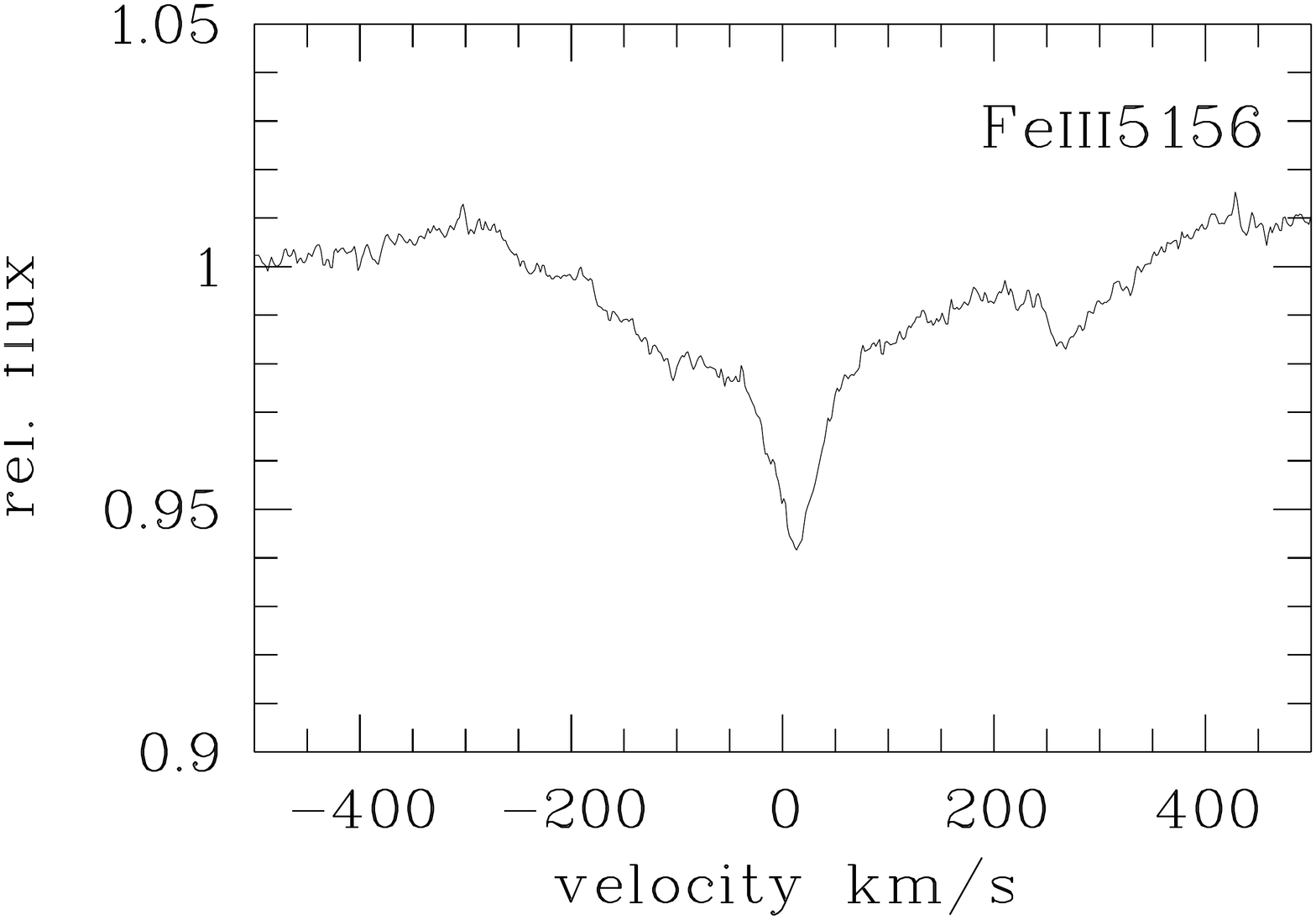}}%
}
\caption[xx]{\label{AMBER}Visual spectroscopic state of $\zeta$\,Tau
  during the {\sc AMBER} observations 
 % The spectral lines shown are the same as for Fig.~\ref{cycle}.
}
\end{figure*}
%%%%%%%%%%%%%%%%%%%%%%%%%%%%%%%%%%%%%%%%%%%%%%%%%%%%%%%%%%%%%%%%%
%%%%====================================================================

\section{Interferometric results}
\label{resinterfer}
\subsection{Results from {\sc AMBER}}

\begin{figure}[t]
  \centering
  \includegraphics[width=9.2cm]{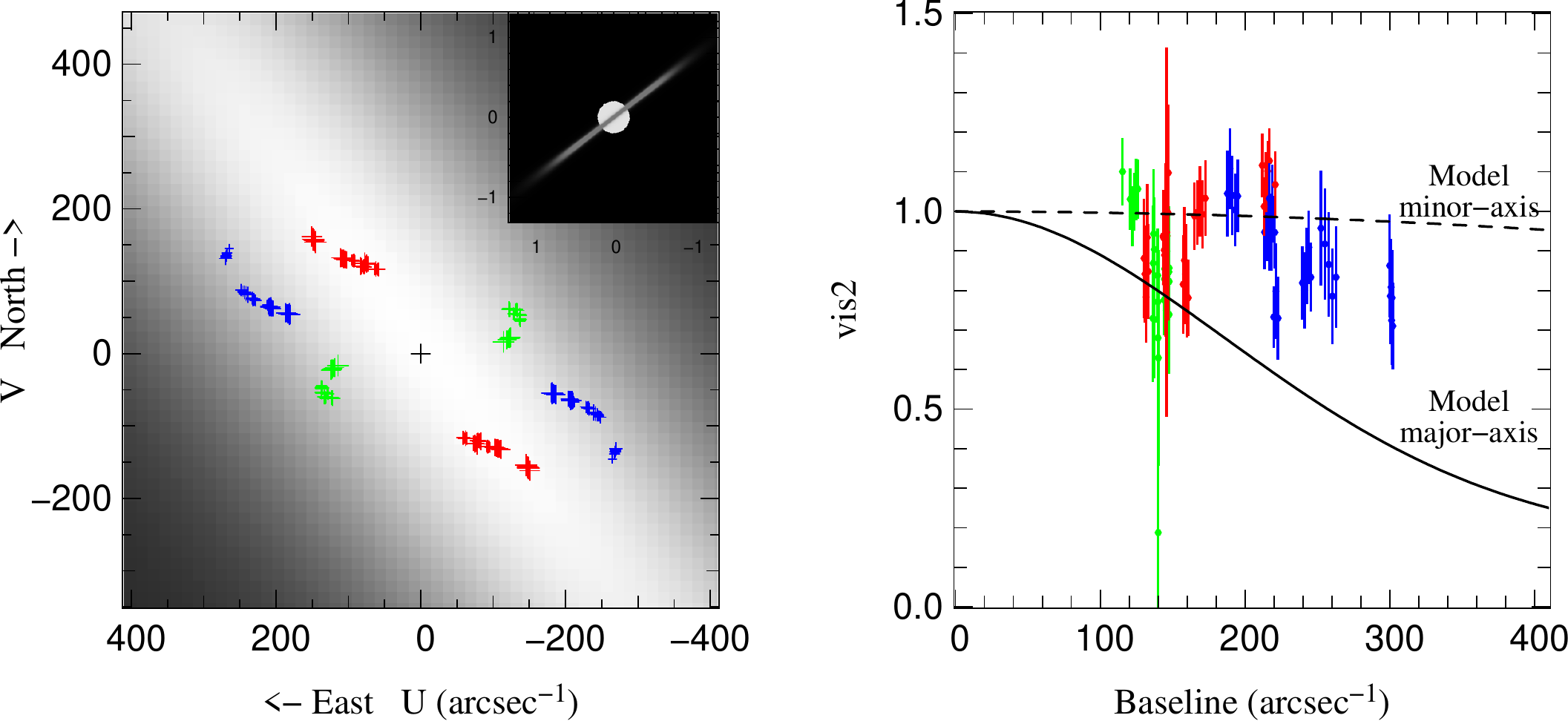}
  \caption{Schematic comparison of the VLTI/AMBER observations with the CHARA 
    model by \citet{2007ApJ...654..527G}.  
    In the colour plot available in the on-line version of the paper, the green, 
    red and blue colours correspond to the UT\,3-4. UT\,1-3 and UT\,1-4 baselines, 
    respectively.
    Left panel: Coverage of the $uv$ plane by {\sc AMBER} observations 
   of Dec 12, 2006. The sub-frame in the upper right corner shows the CHARA model 
   projected on the sky (before the transformation). The sub-frame covers the 
   region of about 3mas~x~3mas and corresponds to the K-band continuum. The squired 
   visibility of the {\sc AMBER} observations is linearly proportional to the size of
   symbols, for the CHARA model (in the sub-frame and transformed  in the background) 
   to the brightness. 
   Right panel: Absolutely calibrated continuum squared visibilities as a function 
   of the baseline length. Squared visibilities of the major and minor disk axes 
   derived from the CHARA model are overplotted.
  }
  \label{fig:continuum_size}
\end{figure}

In Fig.~\ref{fig:continuum_size}, the left panel shows the coverage of the $uv 
$ plane by our {\sc AMBER} observations. Following the ESO/VLTI  
documentation, the following baseline position angles were accepted: U1- 
U3\,=\,32.369, U1-U4\,=\,60.396 and U3-U4\,=\,110.803 deg. The right  
panel shows the absolutely calibrated squared visibilities in  
continuum.  Because the transfer function is very uncertain, the final  
precision is rather poor, errors are as high as 20-30\%. The model 
visibilities for baselines along the major and minor axis  
of the $\zeta$~Tau model derived from  the
CHARA K'-band observations by \citet{2007ApJ...654..527G} are over- 
plotted. The {\sc AMBER} continuum observations  -- in spite of the large scatter  
-- are compatible with the CHARA model. However their accuracy and the  
fact that \zetaTau{} is only marginally resolved by AMBER/VLTI in the  
K-band continuum prevent us to test the model in more detail.

\begin{figure}[t]
  \centering
  \includegraphics[width=9.2cm]{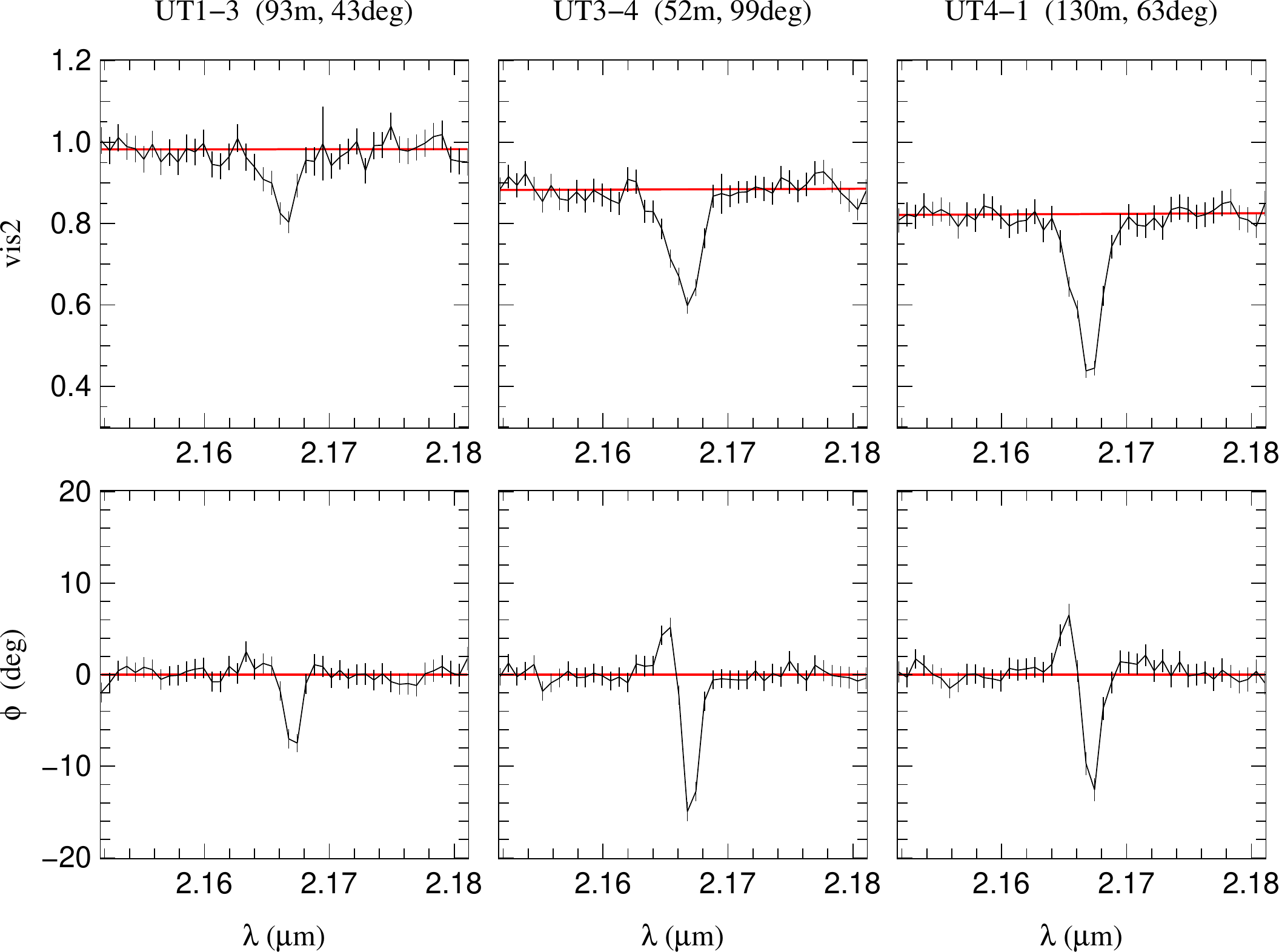}
  \caption{{\sc AMBER} visibilities and phases around $2.18\,\mu$m 
   normalized to the model of \citet{2007ApJ...654..527G}. 
   }
  \label{amber_res}
\end{figure}

\begin{figure}
\vspace{0.3cm}
  \centering
  \includegraphics[width=9.2cm]{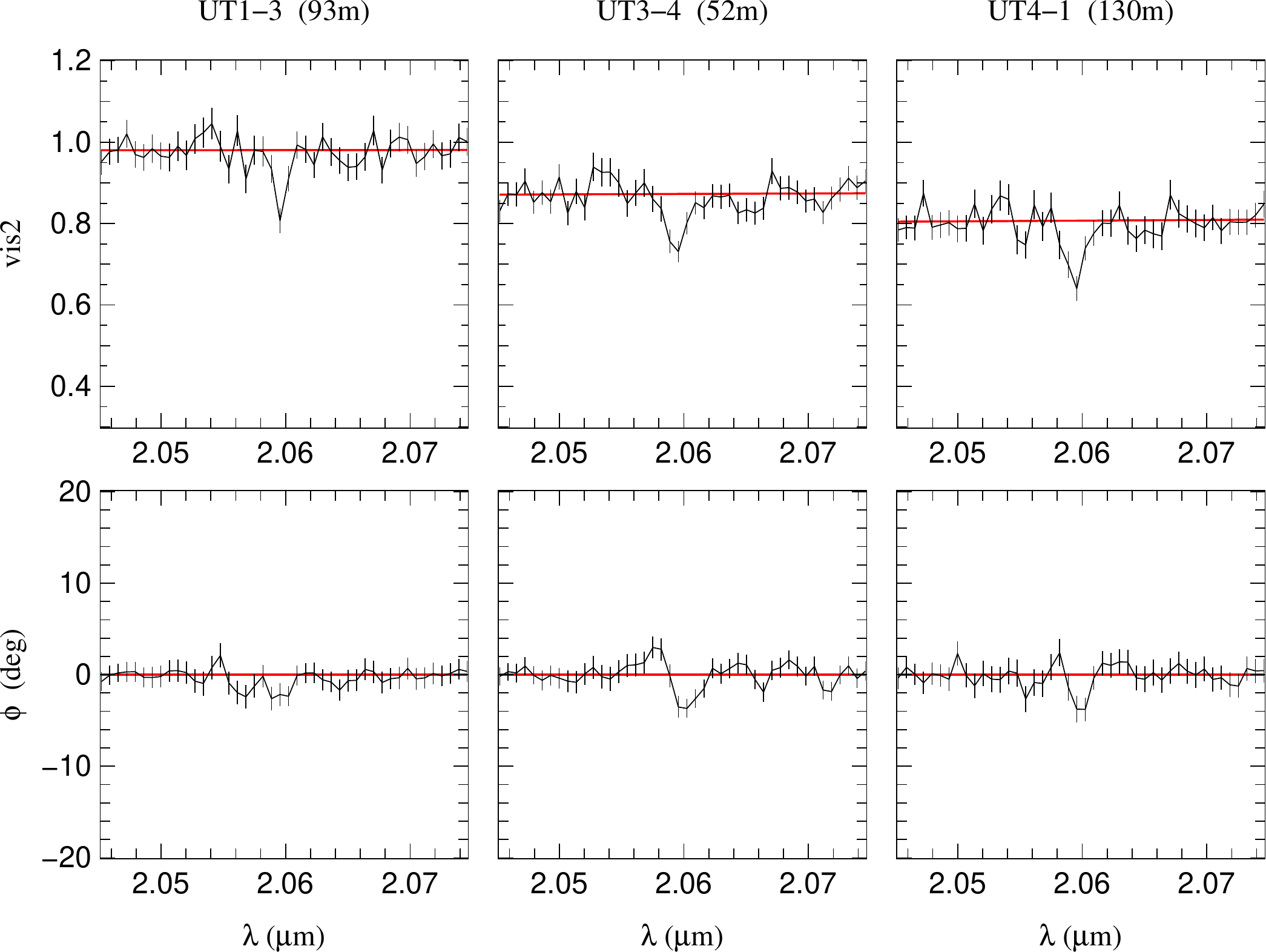}
  \caption{Same as Fig.~\ref{amber_res} but around $2.06\,\mu$m. The
    signal is significantly weaker but the He\,{\sc i} line is still
    detected}
  \label{amber_res_2}
\end{figure}

{\sc AMBER} spectro-interferometric measurements were combined with the CHARA 
model in order to derive visibilities and phases in Br\,$\gamma$  and   
\ion{He}{i} 2.06\,$\mu$ emission lines (see Sec.~\ref{datainterfer}).  
The differential values are  shown in Figs.~\ref{amber_res} and 
Figs.~\ref{amber_res_2}  for the regions around Br$\gamma$ and 
\ion{He}{i}~2.06\,$\mu$, respectively. In all baselines, the signal is much  
stronger in the Br$\gamma$ line (2.17\,$\mu$m), but it is also clearly  
detected in the \ion{He}{i} line (2.06\,$\mu$m). The drops in visibility   
prove that the Br$\gamma$ and \ion{He}{i} emitting regions are more  
extended than the continuum emitting region. Moreover, an asymmetry in  
the visibility profile is clearly seen, with the visibility in the red  
part being lower than in the blue part. Fitting these visibility  
profiles across the lines requires proper modeling. Such a work is  
presented in paper II.

Apart from the visibility signature, also a phase effect was found in  
all baselines for Br$\gamma$, indicating that the photocenter across  
the line is shifted with respect to the photocenter in the continuum.  
To allow a geometrical representation of the AMBER phases, the   
differential-phases $\phi(\lambda)$ across the \BrG{} line were
converted into 2D astrometric shifts $\vec{p}(\lambda)$ by inverting
the well-known formula for marginally resolved interferometric
observations \citep{2003A&A..400..795L}:
\begin{equation}
\phi(\lambda) = -2\pi \cdot \vec{p}(\lambda) \cdot \frac{\vec{b}} 
{\lambda}
\end{equation}\vspace{0.07cm}
\noindent where $\vec{b}$ is the interferometric baseline vector  
projected onto the sky. Uncertainties were propagated to the  
astrometric vector $\vec{p}$  by standard formulas. Neither the uncertainty in 
the baseline length nor in the spectral calibration were taken  
into account, because they affect all spectral bins in the same  
way. 

Figs.~\ref{amber_res} and \ref{amber_res_2} 
prove the presence of phase differences of the V and R emission 
components with respect to the continuum. 
The observed (differential) phases are the flux weighted vector sum of the
source phases of the continuum and line components. With a physical model of
the disk and radiative transfer, the fluxes are known and the phase
differences between the continuum and line emitting regions can be
translated into their angular separation.
Such a model is presented 
in Paper II.  The model-independent conclusion is that the line emitting 
regions and the continuum source lie in one common plane but that the V 
and the R components arise from different locations that do not coincide 
with the continuum source.  

\begin{figure}
  \includegraphics[width=9.2cm]{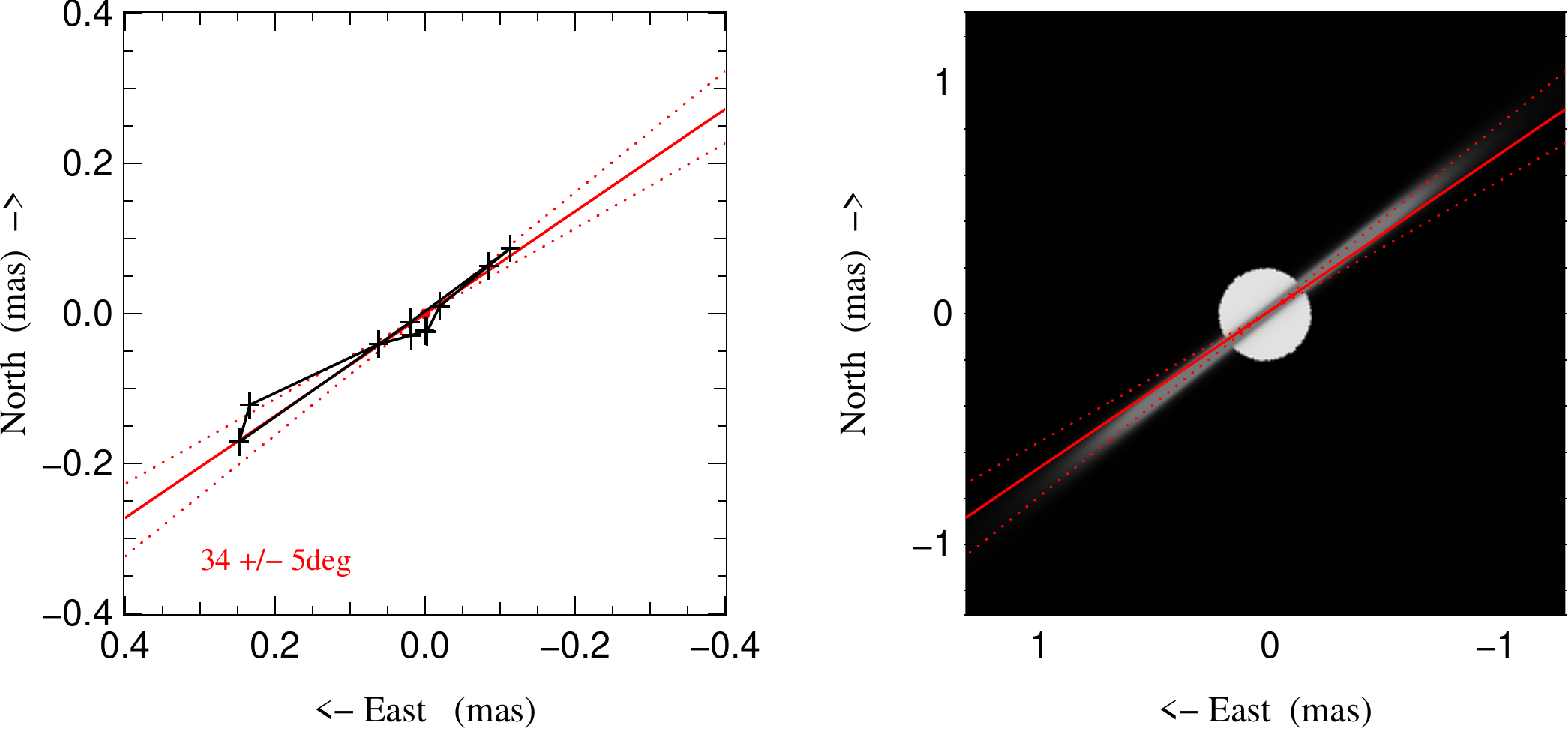}
   \caption{Left: Photocenter shifts derived from the {\sc AMBER} relative phases
     across Br$\gamma$.  The maximum shift is about 0.4\,mas within the plane of
     the circumstellar disk, while no significant offset perpendicular to 
it can be found (black line). Right: The position angle derived from our 
differential data overplotted on the model of \citet{2007ApJ...654..527G}.
}
    \label{photocenter}
\end{figure}

Fig.~~\ref{photocenter}, left panel shows the computed relative offsets  
in angular units. The phases across the Br\,$ 
\gamma$ line is converted into 2D photocenter shift in the plane of the  
sky. The photocenter is displaced toward the NW direction in the blue  
part of the line, and toward the SE direction in the red part of the  
line. This is the clear signature of the rotating disk emitting the 
Br$\gamma$ line. The position angle of the displacement is perfectly 
compatible with the position  
angle of the CHARA model based on measurements made in the K'  
continuum, as shown in the right panel of Fig.~\ref{photocenter}.  
Moreover, the amplitude of the photocenter displacement is  
significantly larger in the red wing ($280\,\mu$as, SE) than in the  
blue wing ($120\,\mu$as, NW). The simplest explanation is an asymmetry  
in the Br$\gamma$-emitting material, with the SE part being brighter  
and/or more extended than the NW part. Again, modeling of this signal  
in the framework of the one-armed disk-oscillation model is developed  
in Paper II.

%%%%%%%%%%%%%%%%%%%%%%%%%%%%%%%%%%%%%%%%%%%%%%%%%%%%%%%%%%%%%%%%%
%%%%%%%%%%%%%%%%%%%%%%%%%%%%%%%%%%%%%%%%%%%%%%%%%%%%%%%%%%%%%%%%%

\subsection{Optical and IR spectra accompanying the {\sc AMBER} observations}

When {\sc AMBER} observed $\zeta$\,Tau, simultaneous {\sc FEROS} 
optical spectra as well as quasi-simultaneous IRTF spectroscopy were 
also obtained.  Fig.~\ref{AMBER} depicts representative visual line profiles.  
Further {\sc FEROS} observations were made throughout December 2006, 
which differ but little from the one at the time of the VLTI observations.   
In the IR, Spectrum III (see Fig.~\ref{IR_lp}) comes closest (12 days before) 
in time to the {\sc AMBER} observations. A next spectrum, obtained 
16 days after {\sc AMBER} data, is very similar.  Therefore, the low variability 
in both optical and IR spectra suggests that the selected spectra represent well the 
dynamics of the disk at the time of the {\sc AMBER} observations.  

As IR spectrum
III shows (Fig.~\ref{IR_lp}), the emission is almost symmetrical in
\ion{He}{i}\,2.06\,$\mu$m, so that the much lower amplitude of the phase
signature in the same line is in good agreement with the spectroscopic
data. The distribution of the \ion{He}{i}\,2.06\,$\mu$m emitting
material is symmetric with respect to both systemic velocity and
location of the continuum source.

The situation is different for Br$\gamma$ but the reasoning is the
same: The closure phase is undoubtedly non-zero for this line and
proves a spatial asymmetry in the brightness distribution of the
system (continuum vs.\ line-emitting region).  At the same time, the 
$V/R$ value of Br$\gamma$ is far from unity.  The combination of these 
two asymmetries places the bulk of the Br$\gamma$-emitting region to 
the receding sector of the disk. 

The scheduling of the {\sc AMBER} observations was very successful in
that they could be made during a triple-peak phase of the H$\alpha$
emission.  However, with single-epoch data it is difficult to say
whether they harbor any related hidden anomaly.  For a discussion of the 
azimuth (in a not point-symmetric model) of the region of formation 
of this spectral structure see Sect.\ \ref{3peak}.  

%%%%%%%%%%%%%%%%%%%%%%%%%%%%%%%%%%%%%%%%%%%%%%%%%%%%%%%%%%%%%%%%%

%%%%%%%%%%%%%%%%%%%%%%%%%%%%%%%%%%%%%%%%%%%%%%%%%%%%%%%%%%%%%%%%%

%%%%%%%%%%%%%%%%%%%%%%%%%%%%%%%%%%%%%%%%%%%%%%%%%%%%%%%%%%%%%%%%%
%%%%%%%%%%%%%%%%%%%%%%%%%%%%%%%%%%%%%%%%%%%%%%%%%%%%%%%%%%%%%%%%%
%

\section{Polarimetric results}
\label{respolar}

%%%%%%%%%%%%%%%%%%%%%%%%%%%%%%%%%%%%%%%%%%%%%%%%%%%%%%%%%%%%%%%%%

%%%%%%%%%%%%%%%%%%%%%%%%%%%%%%%%%%%%%%%%%%%%%%%%%%%%%%%%%%%%%%%%%
\begin{table}
\begin{center}
  \caption[]{\label{table_pola} Mean polarimetric properties, taking
    into account only HPOL data after JD\,2\,490\,000}
\begin{tabular}{ccccc}
  \hline\noalign{\smallskip}
  \hline
Spectral & \multicolumn{2}{c}{Pol.~degree}  & \multicolumn{2}{c}{Pol.~angle} \\
Band & Mean [\%] & $\sigma$  & Mean [$\deg$]& $\sigma$ \\
\hline 
$U$ & 0.99 & 0.14 & 32.3 & 3.5 \\
$B$ & 1.59 & 0.08 & 31.6 & 1.6 \\
$V$ & 1.46 & 0.08 & 31.3 & 1.5 \\
$R$ & 1.30 & 0.06 & 31.3 & 1.2 \\
$I$ & 1.18 & 0.06 & 31.5 & 1.5 \\
\hline\noalign{\smallskip}
\end{tabular}
\end{center}

\end{table}
%%%%%%%%%%%%%%%%%%%%%%%%%%%%%%%%%%%%%%%%%%%%%%%%%%%%%%%%%%%%%%%%%

Since the $V/R$ cyclicity before JD\,2\,490\,000 is not well
constrained by the observations and, as described in Sect.~\ref{haew},
$\zeta$\,Tau reached a stable disk state only after that date, 
only polarization measurements obtained after JD\,2\,490\,000 
were taken into account for the computation of the mean 
values shown in Table~\ref{table_pola}.

The polarization angle (PA) is the same in all bands within both the
individual uncertainty and the standard deviation.  Although the
standard deviation of the PA measurements is slightly
higher than the uncertainty estimated for a single measurement, this
difference is not large enough to conclude a variable polarization
angle.  Even after inclusion of all data, i.e.\ also the observations 
taken before the $V/R$ cycle is well defined, the values for the PA
in Table~\ref{table_pola} would change only in the last digit, if at 
all, so that the conclusion about a non-changing angle is valid over the
entire period of observation.

\begin{figure}
   \includegraphics[width=9cm,viewport=45 42 543 656,angle=0,clip]{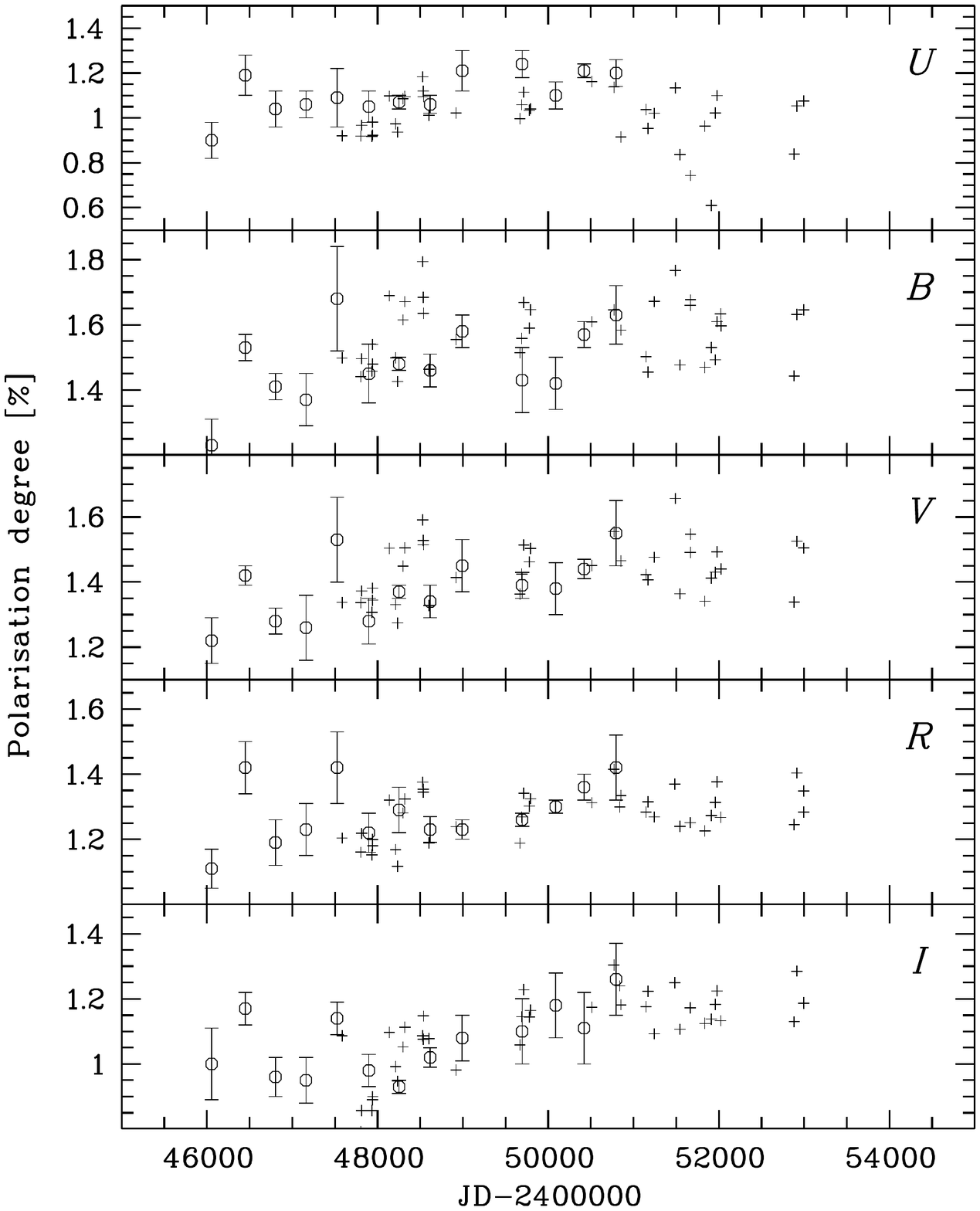}
   \caption{\label{fig_poladeg}Temporal behaviour of the polarization
     degree. Data from \citet{1999PASP..111..494M} are shown as
     circles. The uncertainty of the HPOL data (+) is about the size
     of the symbols, i.e.\ 0.01\,\%  }
\end{figure}

Other than the angle, however, the polarization degree does vary. The
variability shown in Fig.~\ref{fig_poladeg} can be disentangled into
two components.  A slow and secular change is visible well in the
$VRI$ and, though somewhat noisy, also in the $UB$ bands in
Fig.~\ref{fig_poladeg}. This behaviour is in agreement with the
general evolution of the H$\alpha$ disk as described in
Sect.~\ref{haew}: The mean polarization degree is lower before
JD\,2\,448\,000, then rises until JD\,2\,450\,000, and after that
remains stable again.  Superimposed on that pattern is short-term
variability.  Unfortunately, the data do not cover the timescales
required to properly constrain the short-term component.  However, the
observed behavior is not in contradiction with polarimetric
variability produced by a series of discrete and individually rather
minor ejections of matter into the immediate stellar environment, as
already suggested in Sect.~\ref{diskev}.

In summary, neither the variability of the polarization degree nor the
polarization angle, which is stable, can be connected to the cyclic
$V/$R variability. No modulation due to the orbital period could be detected.

%%%%%%%%%%%%%%%%%%%%%%%%%%%%%%%%%%%%%%%%%%%%%%%%%%%%%%%%%%%%%%%%%
%%%%%%%%%%%%%%%%%%%%%%%%%%%%%%%%%%%%%%%%%%%%%%%%%%%%%%%%%%%%%%%%%

\section{Synopsis and discussion}
\label{discuss}

\subsection{$\zeta$\,Tau spectral type}
Table~\ref{table-lines} lists several near-UV \ion{He}{i} lines
as having purely photospheric profiles.  They offer the
opportunity to shed some light on the various discrepant spectral
types that have been published for $\zeta$\,Tau.  A comparison of the 
high-quality UVES data to several early B-type stars in the 
UVES POP-library \citep{2003Msngr.114...10B} showed unambiguously 
that the observed strength of these lines is inconsistent with an
effective temperature corresponding to a 
spectral type of $\zeta$\,Tau later than B2. However, an exact re-determination 
of the spectral type is beyond the scope of this study.

\subsection{Binarity}
With $\zeta$\,Tau having an unseen companion, any variability must be
checked for a relation to the 132.9735\,d orbital period established
by \citet{1984BAICz..35..164H}.  However, the strength of the disk
during the observations implies that the vast majority of the spectral
lines are contaminated by line emission primarily varying on a
different time scale.  The other lines are mostly too shallow for
radial-velocity measurements that would permit a verification of
\citeauthor{1984BAICz..35..164H}'s amplitude of 
$K$\,$\approx$\,10\,km/s to be attempted, which was derived from
observations at more $V/R$-quiescent epochs. The photospheric \ion{He}{i} 
lines are all in the blue spectral region, which is covered only by
a small number of spectra used in this study. 

It  follows that the used dataset of optical spectra, focused on the H$\alpha$ V/R
variations is not suitable for a refinement of orbital parameters. As a compromise 
only a check of the period using the He{\sc i}\,4026 line was performed.  Radial 
velocities were measured by fitting Gaussians to the line core.  Phasing this data with
\citeauthor{1984BAICz..35..164H}'s ephemeris gives a plot similar to
his Fig.\,~1a but with higher errors and a few outliers.  A
time-series analysis finds the second strongest
peak corresponding to a period of $131.6\pm0.5$\,d, in reasonable
agreement with \citeauthor{1984BAICz..35..164H}.  The strongest peak
present is a nearby alias of this one, probably enhanced by noise
effects.  No other quantity investigated in this work shows a
significant modulation with the orbital period.

Even over the given large spectral range, lines due to a secondary, 
either hot or cool, were not found.   In particular, no trace of 
\ion{He}{ii}\,4686 was detected that would have
indicated a hot subdwarf-like companion.  

Re-analysis of the publicly available photometric data \citep[][and data
from the HIPPARCOS-mission]{1995A&AS..112..201G,1997A&AS..125...75P, 1997JAD.....3....5H}
did not confirm the presence of photometric eclipses
suggested by \citet{1988HvaOB..12...15B}.  Rather, the fainting at
phase zero they observed seems to be an event unrelated to the orbital
phase.  There are more such fadings but at random phases 
and none of them at phase zero.  

\subsection{\label{disk_orient} Disk orientation on the sky}

In an axi-symmetric circumstellar disk the PA is
in general perpendicular to the disk plane\footnote{For very optically
thick disks, the polarization can be parallel to the disk plane.}
and thus parallel to the disk position angle ($\chi$).

The spectropolarimetric monitoring of $\zeta$~Tau from the PBO
observatory shows that the polarization angle has been remarkably
constant from at least 1989 to 2004 (see Sect.~\ref{respolar}). The
average $V$-band polarization angle is ${\rm
 PA_{PBO}}=31.3~\pm~1.5~\degr$ with an individual measurement
uncertainty of $1\deg$.  The data from Limber Observatory, taken
between 1984 and 1997 \citep{1999PASP..111..494M}, gave very similar
results: ${\rm PA_{Limber}}=32.5 \pm 1.1\degr$, and a typical
individual uncertainty of $0.5\degr$.

Further, there are three independent measurements of the position
angle from interferometric studies.  \citet{1997ApJ...479..477Q}
determined from a 2D Gaussian fit for data from the Mark III
interferometer a value of $\chi_{\rm Mark III}=31 \pm4\degr$. Using
data from NPOI, \citet{2004AJ....127.1194T} obtained $ \chi_{\rm
  NPOI}=28\pm4\degr$, while
\citet{2007ApJ...654..527G} report $\chi_{\rm CHARA}=37
\pm2\degr$ from their CHARA data.
Finally, the VLTI data of Sect.~\ref{resinterfer} show 
that $\chi_{\rm AMBER} = 32 \pm 5\degr$

The orientation of the disk on the sky, as indicated by all these
measurements, has not changed since 1984, and in particular there
seems to be {\em no} binary phase-locked variability throughout the $V/R$
cycle, at least within the individual measurement uncertainties.  The
stability of the disk position angle is typical for Be stars, the only
few counter-examples ever found are suspected to undergo precession of
a non-equatorial disk \citep{1998A&A...330..243H,2007ASPC..361..267H}.

\subsection{$V/R$ cycle length}
The analysis of the up to now most comprehensive observational 
data set of $\zeta$~Tau confirmed that its $V/R$ variations follow a cycle
with relatively stable amplitude and length of 1405~-~1430 days duration 
during the present V/R variable phase starting at the beginning of nineties.
The length is different and much more stable than during the previous
V/R active phase in  1955~-~1980.

Table~\ref{vtor_cycles} shows that the cycle length varies and 
suggests a correlation between the cycle length and the duration of the 
triple-peak epoch.  In fact, the differences in cycle length seem to be 
dominated by the duration of the triple-peak phase.  For instance,
the lengths of Cycles II and III differ by 291 days.  But after subtraction 
of the duration of the triple-peak epochs, the remainders of the cycles 
are of the same length (1033 and 1027 days, respectively).
The lack of an explanation of the variation of the duration of
triple-peak phases may limit models for the basic $V/R$ activity as
well.  Fortunately, at less than $\pm 10$\,\%, the quantitative effect
is small.

The well-observed $V/R$ maxima in H$\alpha$ have been decreasing 
through Cycles I to III.  This may be indicative of a decrease also of 
the amplitude of the perturbation causing the $V/R$ variations.  

\subsection{$V/R$ phase differences between emission lines}
The  relative shifts in the $V/R$ variations of Balmer, Brackett,
\ion{He}{i}, and metal emission lines should probe the disk at
decreasing distances from the central star and so place considerable 
constraints on any  explanation of the cyclic $V/R$ cyclic of $\zeta$~Tau
(and other Be stars).   The phase lag between H$\alpha$ and the higher
Brackett lines is of the order of $\Delta\phi \approx 0.25$ 
(see Fig.~\ref{IR_VtoR}).  Unfortunately, any quantitative modeling is 
somewhat hampered by the medium spectral resolution of both the 
interferometry and IR-spectroscopy.

From a subset of the present IR spectra, \citet{2007ApJ...656L..21W} 
derived the opposite $V/R$ phase relation between IR lines and H$\alpha$.  
The larger phase and wavelength coverage of the present study 
appears to exclude this possibility.  

\subsection{Triple-peak H$\alpha$ profiles}
\label{3peak}
The present compilation of spectra maps the evolution of the 
triple-peak profiles in Cycles III and IV in much detail.  Qualitatively, 
it follows a very similar pattern in either cycle.   But the duration of 
the triple-peak phases is variable and may last from 200 to 500 days. 
They are not phase-locked to the companion star.  However,
this does not invalidate the small-number statistics produced by 
\citet{2006A&A...459..137R}, who report that triple-peak line 
profiles only occur in shell stars that are both 
$V/R$ variable {\em and} multiple.

As is apparent from Fig.\,\ref{splines}, the triple-peak profile in
H$\alpha$ is accompanied by a somewhat disturbed profile in H$\beta$. 
But other non-hydrogen emission lines in the visual range are largely
unaffected.  Although the lower resolution of the IR spectra
leaves some space for undetected line profile deformations, IR 
emission lines, too, (including those of \ion{H}{i}) do not seem 
to undergo related variations.  

The absence of triple peaks from optically thin emission lines and in
particular the stability of \ion{O}{i}\,8446, which probes the
H$\alpha$-forming region due to Ly$\beta$ resonance pumping, suggest
that the triple-peak feature does not probably correspond to an actual
density structure.  Triple peak profiles have so far only been
reported in shell stars.  This detail, too, would not easily be
explained by real density enhancements.  

This gives rise to the speculative conjecture about a change in the
local escape probability due to distortions of the local velocity
field, which of course would have an effect only on optically thick
lines, like H$\alpha$.  Another possibility, guided by the finding by
\citet{2007ASPC..361..274S} that triple-peak emission lines occur in
binary Be stars, might be some resonance between the orbital motion of
the companion stars and the orbital motion of gas in the outer disk.

H$\alpha$ and H$\beta$ are not the only optically thick emission
lines, though: It is commonly accepted that the Paschen and Brackett
lines as well as some strong IR \ion{Fe}{ii} emission lines form
closer to the star than the first few lines of the Balmer series.
Since only H$\alpha$ and H$\beta$ show some form of triple peak, this
may point at the outer regions of the disks as the locus of the
physical variability.  Such a region  was on the line of sight at 
the time of the {\sc AMBER} observations. At that time, the bisector line of
the H$\alpha$ emitting region was close to the line of sight, whereas 
the bisector lines of
the regions of formation of emission lines requiring higher proximity
to the exciting central star passed the line of sight up to one year
earlier.

\subsection{Narrow- and broad line shell absorptions}
\label{BNLGdiscuss}
As described in Sect.\,\ref{shellabs}, 
NLG lines are formed in a broad diversity of ionic 
species and over a wide range of ionization and/or excitation 
potentials beginning with \ion{O}{i} and \ion{Na}{i} and 
extending to \ion{Fe}{iii}.  Conversely, BLG characteristics are 
exclusively associated with transitions expected from 
comparatively cool conditions; examples are \ion{Mg}{i} or 
\ion{O}{i}. The simultaneous, let alone long-term, presence of 
both \ion{Fe}{iii} and \ion{Mg}{i} shell absorption is very unusual.
A search of the large FLASH, HEROS, and FEROS 
database of Be spectra \citep{2003A&A...411..229R} furnished no 
second example.

The occurrence of lines with persistent mixed BLG and NLG 
characteristics suggests that the circumstellar disk does not have 
a simple and smoothly varying structure.  
Rather, NLG and BLG components seem to form in two spatially 
separated regions with distinct physical conditions.  Simple 
common-sense considerations lead to conflicting conclusions 
about the location of these regions relative to the central star.  

On the one hand, their higher excitation would place the NLG  
closer to the star and the lower-excitation BLG in the outer disk.  
On the other hand, the line {\em profiles} point to the
opposite, provided that the disk dynamics is crudely Keplerian:  
The narrowness of the NLG, which in some UV lines 
hardly exceeds the thermal width, as well as their lower  RV
amplitudes suggest, then, that the region of formation of NLG lines 
is relatively far from the central star.  The broader BLG lines would 
form in the inner disk, where the velocity ranges are larger.  

Apart from their width, NLG and BLG lines also 
differ in the variation of their equivalent width with $V/R$ phase. 
While BLG lines are strongest in the second
half of the cycle, NLG lines are strongest around phase 0.25.

The radial-velocity variations of both groups have comparable 
amplitude through the $V/R$-cycle (see Fig.~\ref{splines}).  
However, there is a phase difference of about $\Delta\phi\approx0.25$ 
with the NLG trailing the BLG. 

%%%%%%%%%%%%%%%%%%%%%%%%%%%%%%%%%%%%%%%%%%%%%%%%%%%%%%%%%%%%%%%%%

\section{Conclusions}
\label{concl} 

$\zeta$\,Tau is a Be-shell star.  Its spectral type is B2 or earlier
one.  Both interferometry and polarimetry demonstrate that its
circumstellar disk is flat and seen edge-on.  The assumption that {\em
all} Be-shell stars are observed edge- and equator-on has been vital
for the analysis  by \citet{2006A&A...459..137R} of the fractional critical 
rotation rates of Be stars.  

The plane of the disk has remained stable for decades.  Warping or
tilting, as diagnosed in other Be stars by 
\citet{1998A&A...330..243H}, may be present but was not detected.  
The disk seems to have been persistent for about a century.  But 
the strength of emission and shell absorption lines has varied 
on time scales of years to decades.   

The photocenter of the Br$\gamma$ line emission lies in the plane of
the disk but is offset from the continuum source.  The same could not
be diagnosed from the weaker \ion{He}{I} 2.06\,$\mu$m emission line.
This may be a data-quality issue or result from an alignment, at the
time of the observations, of the photocenter of the \ion{He}{I} line
with the continuum source.

The Br$\gamma$ result may also resolve a general ambiguity in the
interpretation of $V/R$-asymmetric emission lines.  From the spectra
alone one cannot decide whether the $V/R$ asymmetry is only in
velocity or also in configuration space.  The VLTI observations show
that the latter is true in $\zeta$ Tau.

Like some other, but by far not all, Be stars, $\zeta$ Tau is a binary.
The companion remains undetected at optical and IR wavelengths.  An
intensive search in a dense series of multi-wavelength observations of
a very broad range of electromagnetic observables did not furnish any
effect of the companion on the long-term variability or orientation of
the disk.  The variability of the circumstellar disk can, therefore,
be considered intrinsic to the disk itself and the central B-type
star.

The most prominent spectroscopic signature of the disk activity 
is the cyclic $V/R$ variability.  Its amplitude can for decades drop 
below the level of easy detectability.  Pronounced $V/R$ variations
were resumed about 1992.  Since 1996, three complete $V/R$ cycles 
were observed.  Their similarity in length and amplitude has 
enabled a detailed phenomenological description of the variability.  

The formal mean cycle length is 1405\,-\,1430 days, with cycle-to-cycle
variations of less than $\pm 10$\,\%.  The basic log($V/R$) curve for
H$\alpha$ is smooth and symmetric.  However, about the $V/R$ minimum a
perturbation develops, which has been qualitatively the same in all
three of the most recent cycles.  The H$\alpha$ emission profile
develops a comb-like structure at its top, which consists of three
small peaks separated by equally small depressions.  The duration of
this phase varies from cycle to cycle, and the changes may be the main
reason of the variation in length of the main cycle.  In the available
observations, triple-peak profiles are restricted to H$\alpha$, with
traces showing up in H$\beta$.  Therefore, the underlying perturbation
may be confined to the outer disk.

All emission lines partake in the $V/R$ variability.  But there 
are line-specific shifts in phase of up to 25\% of the cycle length. 
H$\alpha$ $V/R$ lags behind Br$\gamma$, which trails the higher
Brackett lines.  Metallic lines typically behave like the higher
Brackett lines.  This supports the notion that the phase lag 
increases with decreasing density and increasing distance 
from the central star.  

The shell absorption lines crudely fall into two different groups
mainly distinguished by the line widths.  Low-ionization species and
low-excitation lines are broader than the ones involving higher
energies.  Some lines also share the behavior of both groups so that
each group may predominantly arise from different locations.  If the
disk motions are basically Keplerian and density and temperature
mainly drop with distance from the central star, the grouping provides
seemingly contradictory diagnostics of these locations relative to the
central star.  Both groups undergo similar radial-velocity variations
but the broad lines lead in phase by about one-quarter of a $V/R$
cycle.

The only variability apparently unrelated to both the $V/R$ cycle and
the binarity is the occasional presence of infrared \ion{C}{i}
emission.  

Discrete major mass loss events were not observed.  But 
observations of other Be stars and the relative constancy of the 
total emission strength suggest that minor ones have likely 
taken place.  They may provide a partial explanation of 
not fully repetitive variations.  

The complexity of the observations and especially the various cyclically
repeating phase differences show that the underlying perturbations are
not both point-symmetric {\em and} radially monotonic. Consequently, 
the perturbations are hardly reconcilable with  roughly
spherically symmetric models as, e.g., the one of \citet{1987LNP...274...96D}.
Major outbursts might temporarily lead to such a situation but 
the cyclic variability would probably require them, too, to be 
cyclic.  A perturbation that propagates in a spiral-like pattern 
has a physical foundation in global $m$=1 disk oscillations, which 
will be developed into a fully self-consistent model in Paper II 
(Carciofi et al., this volume).  

\begin{acknowledgements}

The polarimetric observations were supported in part by NASA via
contract NAS5-26777 with the University of Wisconsin and grants
NAG5-3447 and NAG5-8054 to the University of Toledo. We thank Ken
Nordsieck for access to HPOL.  We also thank the many members of the
PBO observing and data reduction teams over the years, with special
thanks to Marilyn Meade and Brian Babler, for assistance with data
acquisition and reduction.

This work has made use of the BeSS database, operated at GEPI, Observatoire 
de Meudon, France: http://basebe.obspm.fr.  We would like to thank the amateur 
observers who obtained valuable spectra and generously made them available at 
the BeSS web page, namely to E.~Barbotin, C.~Buil, J.~Guarroflo, B.~Mauclaire,
J.~Ribeiro, J.~Terry, O.~Thizy and V.~Desnoux.

We thank Dr. D.~Gies who provided us with their model derived from the 
CHARA observations, and the unknown referee for his/her very constructive 
comments and suggestions. 

We thank the observers and staff at Ritter Observatory, especially
Nancy Morrison, for their assistance in providing data used in this
paper.  Observations at Ritter Observatory are supported by the NSF
under the PREST program, grant AST04-40784. JPW is supported by a NSF
Astronomy \& Astrophysics Postdoctoral Fellowship under award
AST-0802230.  The observations at Ond\v{r}ejov Observatory were supported 
by the Grant Agency of the Academy of Sciences of the Czech Republic (grants
AA3003001,  AA 3003403).  The Heros@Ond\v{re}jov monitoring project was part of a 
joint project supported by the German Bundesministerium f\"{u}r Bildung and
Forschung and the Ministry of Education of the Czech Republic (TSE-001-009, 
436 TSE 113/18 and 41).  This work was also supported by FAPESP grant 04/07707-3 to A.C.C.
\end{acknowledgements}
\bibliographystyle{aa} % style aa.bst
\bibliography{zetaTau.bib} % your references Yourfile.bib
\end{document}